\documentclass[
      twocolumn,
      twocolappendix,
      amsmath
      ]{aastex702}

\graphicspath{{./}{apjs_fig/}}

\begin{document}

\title{Correlated Galactic Confusion Foreground for the LISA-Taiji-TianQin Network and Its Impacts on  Resolvable-Source Analysis}

\shorttitle{Correlated Galactic Confusion Foreground for the LISA-Taiji-TianQin Network}
\shortauthors{Du, Xu, \& Luo}
% \submitjournal{ApJS}

%% ORCID iDs are given through the optional argument of \author, exactly as in
%% sample702.tex:  \author[orcid=0000-0002-9072-1121]{Greg Schwarz}
%% The class turns this into a hyperlink to http://orcid.org/<iD> and prints the
%% ORCID icon (orcid-ID.png) after the author name. The value must be four
%% groups of four digits (or X as the last character); leaving it empty silently
%% drops both the icon and the link, and a non-numeric placeholder
%% (e.g. XXXX-XXXX-XXXX-XXXX) stops the compilation entirely.

%% Corresponding author: as in the published ApJS papers (e.g.\ Ma et al. 2026),
%% the corresponding author is flagged by appending his Linked e-mail address to
%% his affiliation, after a semicolon. Keep the \email{} line as well: AASTeX v7
%% requires one for every author.
%% NOTE: AASTeX merges affiliations by an EXACT match of the affiliation text.
%% Appending the e-mail to only one author's copy of an affiliation therefore
%% creates a second, separate entry and the institute gets printed twice (Nos. 1
%% and 2). Writing the very same string (i.e. this macro) for every author who
%% shares the institute keeps it a single entry, exactly as in the published
%% ApJS papers, where the corresponding author's e-mail is appended to the
%% (shared) first affiliation.
\newcommand{\IMechAffil}{Center for Gravitational Wave Experiment, National Microgravity Laboratory, \\ Institute of Mechanics, Chinese Academy of Sciences, Beijing 100190, China}
\newcommand{\IMechAffilCorr}{\IMechAffil; \href{mailto:duminghui@imech.ac.cn}{duminghui@imech.ac.cn}}

\author[orcid=0000-0003-2155-3280]{Minghui Du}
\affiliation{\IMechAffilCorr}
\email{duminghui@imech.ac.cn}

\author[orcid=0000-0002-3543-7777]{Peng Xu}
\affiliation{\IMechAffilCorr}   % same string as for Du -> merged into affiliation 1
\affiliation{Taiji Laboratory for Gravitational Wave Universe (Beijing/Hangzhou), \\ University of Chinese Academy of Sciences (UCAS), Beijing 100049, China}
\affiliation{Key Laboratory of Gravitational Wave Precision Measurement of Zhejiang Province, \\ Hangzhou Institute for Advanced Study, UCAS, Hangzhou 310024, China}
\affiliation{Lanzhou Center of Theoretical Physics, Lanzhou University, Lanzhou 730000, China}
\email{xupeng@imech.ac.cn}

\author[orcid=0000-0002-9533-8025]{Ziren Luo}
\affiliation{\IMechAffilCorr}   % same string as for Du -> merged into affiliation 1
\affiliation{Taiji Laboratory for Gravitational Wave Universe (Beijing/Hangzhou), \\ University of Chinese Academy of Sciences (UCAS), Beijing 100049, China}
\affiliation{Key Laboratory of Gravitational Wave Precision Measurement of Zhejiang Province, \\ Hangzhou Institute for Advanced Study, UCAS, Hangzhou 310024, China}
\email{luoziren@imech.ac.cn}

%%-----------------------------------------------------------------------
\begin{abstract}

The millihertz gravitational-wave sky is expected to be observed  by a network of space-based detectors (LISA, Taiji, and TianQin) in the 2030s.
A dominant noise in this band is the confusion foreground produced by $\mathcal{O}(10^7)$ unresolved Galactic binaries (GBs).
Because the same population projects onto the time-delay interferometry (TDI) channels of the different detectors, the foreground is necessarily correlated across the network.
In this paper, we construct the full foreground noise covariance matrix for the LISA-Taiji-TianQin network from a catalogue of $\sim 3 \times 10^7$ GBs through numerical simulation, and we  derive an analytic model of the cross-detector foreground coherence that provides a cross-check and physical interpretation of the numerical results.
We derive  the overall sensitivity of the detector network based on this covariance matrix,  and  further characterize the frequency- and time-dependent cross-detector foreground correlations.
Taking massive black hole binaries (MBHBs) and GBs as representative transient and continuous sources, we further  compare the block-diagonal (\textit{i.e.}, neglecting the cross-detector foreground correlation) and full-covariance noise models in terms of their impacts on the signal-to-noise ratio (SNR), parameter uncertainties, and Bayesian posteriors.
The impact is confined to specific signal regimes, affecting primarily
high-mass ($\mathcal{M}_c \sim 10^7\,M_\odot$) MBHBs   and  low-frequency ($f_0 \lesssim 2$~mHz) GBs, with SNR relative differences of up to $\sim 30 \%$ and $\sim 10 \%$, respectively. 
Besides, no statistically significant parameter bias arises under either noise model, as verified by probability--probability  tests. 
Parameter estimations for MBHBs and GBs are performed using the network analysis pipelines implemented in the \texttt{Triangle-BBH} and \texttt{Triangle-GB} codes.  
Both code repositories, together with the simulated foreground data and network sensitivities,  are publicly released for diverse scientific investigations.
\end{abstract}
%%-----------------------------------------------------------------------


\section{Introduction}
\label{sec:intro}
%=====================================================================

The millihertz gravitational wave (GW) band will provide new observational avenues for probing  astrophysics, cosmology, and fundamental physics.
Under optimistic projections, the 2030s might witness the simultaneous operation of three space-based GW detectors in the millihertz band: LISA~\citep{AmaroSeoane2017}, led by European Space Agency, and Taiji~\citep{HuWu2017}, TianQin~\citep{Luo2016}, both developed in China.
LISA and Taiji follow heliocentric orbits with arm lengths of $2.5\times 10^9$~m and $3\times 10^9$~m,  and are designed to trail or lead the Earth by $20^\circ$, respectively, while TianQin follows a geocentric orbit with an arm length of $\sqrt{3}\times 10^8$~m and the norm of its constellation fixed toward the verification binary HMCnc.
Operating together, the three detectors form a network with which joint observation accumulates the network signal-to-noise ratio (SNR) over all individual detectors, boosting sensitivity and thereby expanding the reach of population-level astrophysics studies, \textit{e.g.}, constraining the formation channels of supermassive black holes~\citep{ShenHanSCPMA}. 
Together with  SNR enhancement, different orientations of the individual detectors within this network break degeneracies in  the source's extrinsic parameters. 
This yields precise localisation~\citep{Ruan:2020smc,Wang2021,Gao2024}, which in turn underpins both standard-siren and dark-siren cosmological analyses~\citep{Wang:2021srv,Wang:2020dkc,Jin:2023sfc,Zhan:2025jqg}. 
Moreover, by cross-correlating  data from  multiple detectors, the network helps separate stochastic GW signals from instrumental noise,   thereby facilitating credible detection and analysis of the stochastic GW background (SGWB), including its anisotropy and polarization content~\citep{Zhao:2024yau,Li:2025pde,ChenLiuZhang2024}. 
This capability unlocks crucial observational probes into early-universe phase transitions~\citep{Huang:2025uer}, cosmic strings~\citep{Wang:2023ltz}, and primordial black holes~\citep{Yang:2022cgm}, \textit{etc.}. 
Looking ahead, the  network may also provide a crucial consistency check for  global-fit analyses that aim to  resolve the full population of sources~\citep{Zhang:2022wcp,Littenberg2023,Strub:2024kbe,Katz:2024oqg,Deng:2025wgk}.

To ensure the reliability of data analysis, both the noise and the signal components in the time-delay interferometry (TDI)~\citep{1999ApJ...527..814A,Tinto2021} data stream of space GW detectors must be modeled correctly. 
After the primary noise contributions are suppressed by TDI~\citep{Tinto2021,PhysRevD.103.123027} and related procedures (\textit{e.g.}  calibration and subtraction of tilt-to-length noise~\citealp{PhysRevD.106.042005}), the resulting TDI data streams primarily consist of two noise components.
The first is instrumental noise, which is  reasonably  regarded as statistically independent across detectors.
The second is the Galactic confusion foreground produced  by  $\mathcal{O}(10^7)$ unresolved compact binaries (mostly double white dwarfs). 
Near  1~mHz, 
it dominates over instrumental noise by about an order of magnitude for Taiji and LISA~\citep{Barack:2004wc,Liu2023}, while being slightly below  the instrumental noise for TianQin~\citep{Huang:2020rjf}. 
Crucially, the confusion foreground exhibits a fundamental distinction from instrumental noise: because the same Galactic binary (GB) population projects simultaneously onto the TDI data of all detectors in the network, the foreground is necessarily correlated across detectors.
This gives rise to  non-zero cross  spectral density (CSD) between detector pairs, in addition to the power spectral density (PSD) within each detector.  
Overlooking this cross-detector correlation  
amounts to misspecify  noise covariance model  in data analysis, and hence risking  introducing systematic parameter biases and  compromising the credibility of the  scientific objectives outlined above~\citep{Littenberg:2014oda,Romero-Shaw:2022ctb}.

Despite this physical inevitability, the confusion foreground for  space detector network, especially its  cross-detector correlation  has  received only limited attention.  
Rigorous modeling of this correlated foreground remains an open problem, and 
its potential implications for analyzing  the numerous population of resolvable  LISA, Taiji, and TianQin sources  remain unexplored. 
For instance, \cite{Wu2023} investigated the subtraction of resolvable GBs  for the LISA-Taiji-TianQin network, using a 1\% subset of the LISA Data Challenge (LDC) GB catalogue~\citep{Baghi:2022ucj} and a simplified detector response formalism.
While this work offers quantitative comparisons of GB detectability under various network orbital configurations, the total confusion noise was not evaluated  from  the full population with more realistic  TDI response, 
and the foreground correlations were  not taken into account.
In  the context of SGWB detection,  \cite{Liang:2024tgn} developed a cross-correlation  framework for the LISA-TianQin network, treating the confusion foreground as a sky-averaged spectrum whose cross-detector correlation is encoded through the time-dependent overlap reduction function (ORF). 
However, for the analysis of resolvable sources such as bright GBs and massive black hole binaries (MBHBs), 
existing  studies using space detector network   either do not consider the foreground, or setting the  cross-detector foreground correlation  to zero (\textit{e.g.}, \citealp{Gao2024,Zhang:2022wcp,Ruan:2019tje,Wang2021}).  
Building on these efforts, 
the present work aims to address these  gaps  by  modeling,  simulating,  and analyzing  this correlated confusion foreground for the   LISA-Taiji-TianQin network, with particular emphasis on  its implications for  the inference  of resolvable GW sources.

We begin by constructing the full foreground covariance matrix  for the LISA-Taiji-TianQin network. 
The  confusion foreground is  obtained from  a catalogue of $\sim 3 \times 10^7$ GBs from the LDC Radler dataset~\citep{Baghi:2022ucj}   and using  an iterative bright-source subtraction procedure adapted  from \cite{Karnesis:2021tsh}, under  two choices for the network SNR threshold (7 and 10). 
The complete set of foreground PSDs and CSDs is computed from the resulting data, and an analytic model of the cross-detector coherence is derived and cross-validated against the  numerical simulation,   providing   physical insight into the frequency and time dependence of the correlations. 
Building on the   covariance matrix that combines  foreground and instrumental noises, we compare the block-diagonal (neglecting cross-detector foreground correlations)  and full-covariance noise models (incorporating cross-detector foreground correlations) in terms of  their impacts on the analysis  of MBHBs and GBs. 
The comparison encompasses  SNR, Fisher information matrix (FIM) parameter uncertainties, and the full Bayesian posterior bias under noise-model misspecification, which is  illustrated through  both the posterior distrubiton of individual  sources and  probability--probability (P-P) tests across  ramdomly  generated populations.   
To facilitate broader scientific exploration based on  the space detector network, 
we publicly release the network analysis codes  for GBs and MBHBs implemented in  \texttt{Triangle-GB}~\footnote{https://github.com/TriangleDataCenter/Triangle-GB}  and \texttt{Triangle-BBH}~\footnote{https://github.com/TriangleDataCenter/Triangle-BBH},  as part of  the Taiji Data Challenge~\citep{Du:2025xdq} toolkit,   along with   the simulated foreground data and  network sensitivity curves~\footnote{https://zenodo.org/records/22694654}.

Besides, the relative orbital configuration of the detectors is another degree of freedom that has been explored in the literature. 
\cite{Wang2021} introduced a family of alternative LISA-Taiji networks, in which the Taiji constellation is deployed with different relative inclinations and orbital phases with respect to LISA.  
In this work we likewise consider different orbit realizations, with the results for the ``default'' configuration presented in the main text, and another representative configuration in the appendix, to demonstrate the dependence of the foreground correlation on the relative orbital geometry and to test the generality of our theoretical framework.

The paper is organized as follows.
Section~\ref{sec:theory} establishes the theoretical framework, covering the single-link GW response and TDI combinations, the network data model and  its instrumental and confusion noise properties, the construction of the network covariance matrix, matched-filtering statistics, sensitivity, and an analytic estimate of how foreground correlations affect the total SNR.
Section~\ref{sec:method} presents the methodology of our investigation, including the detector orbit and noise configurations, the two noise models, the simulation of the Galactic confusion foreground through iterative subtraction, and the network Bayesian inference framework for MBHBs and GBs.
Section~\ref{sec:results_discussion} shows and discusses  the results for foreground PSDs, CSDs, and network sensitivity, validates the analytic foreground model against simulations, and compares the block-diagonal and full-covariance noise models for both MBHB and GB analyses.
Section~\ref{sec:conclusion} summarizes the main conclusions and outlines future developments.
Appendices~\ref{app:GB} and~\ref{app:MBHB} detail the frequency-domain TDI responses for GBs and MBHBs, with the latter also including the network heterodyned likelihood. Appendix~\ref{app:foreground} derives the foreground CSD and characterizes its time dependence, while Appendix~\ref{app:configII} presents the results for the alternative orbit configuration.

Throughout this work we assume a 4-year  effective science operation duration  for all three detectors, neglecting  TianQin's  ``3-month-on, 3-month-off'' operation mode~\citep{Luo2016}. 
This simplification enables  fast frequency-domain analysis  without appreciably affecting our conclusions on the foreground correlation, since it is  dominated by the LISA-Taiji pair.

%=====================================================================
\section{Theoretical Framework}
\label{sec:theory}
%=====================================================================

\subsection{Models for signal and noise in TDI data streams}

\begin{figure*}[t]
\centering
\includegraphics[width=0.75\textwidth]{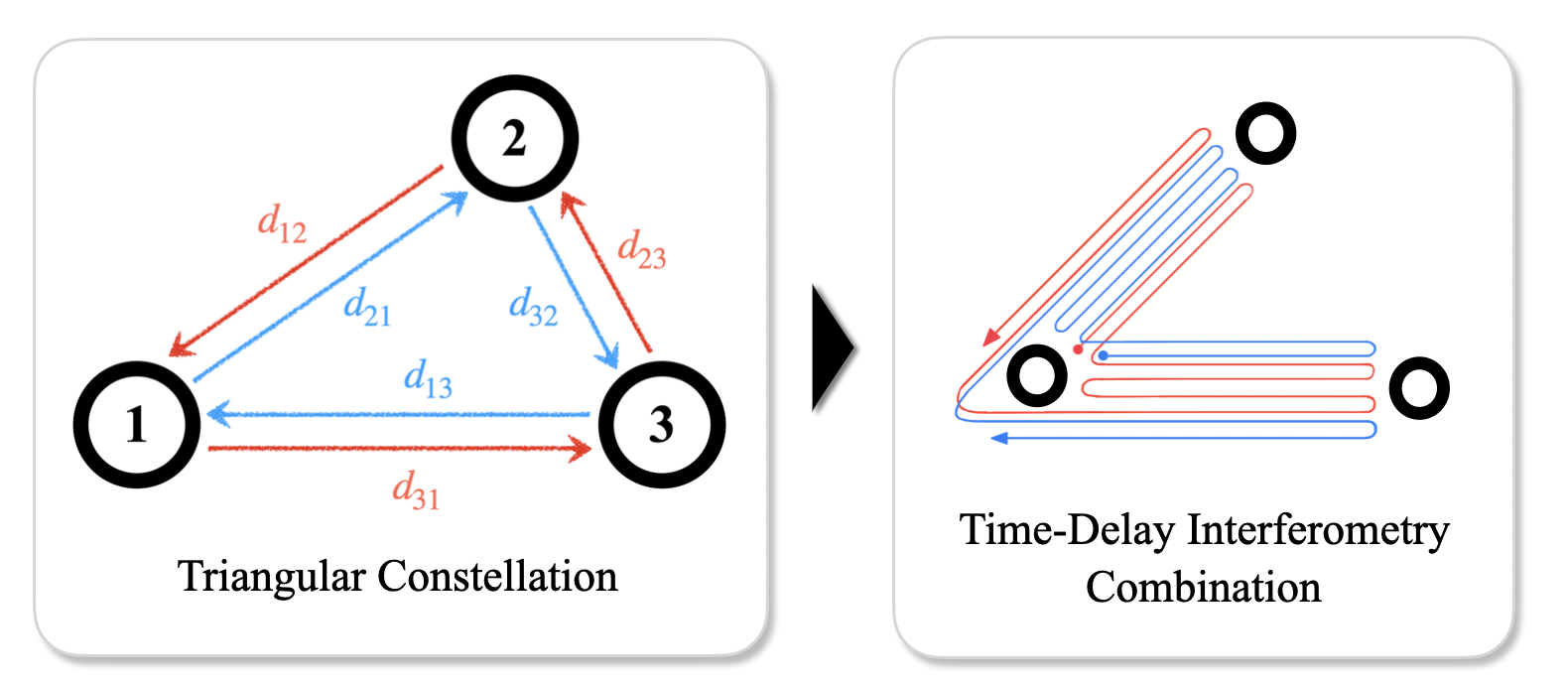}
\caption{Schematic illustrations for  a typical  space-based GW  detector and  TDI combination.
Left: the near-equilateral triangular constellation of a space-based detector, with the three SCs  numbered $1$, $2$, and $3$ and  six inter-SC laser links labeled $\{12, 23, 31, 21, 32, 13\}$.
Right: principle  for the second-generation Michelson-$X$ TDI combination, in which the single-link measurements are delayed and combined to cancel the dominant laser frequency noise.}
\label{fig:constell_tdi}
\end{figure*}

After the raw laser interferometric measurements are downlinked to the ground, they undergo a series of pre-processing steps that standardize the data format and mitigate primary noise sources (laser frequency noise, clock noise, optical bench jitter, tilt-to-length coupling, \textit{etc.})~\citep{Olaf_phdthesis}, producing the TDI data product used for extracting GW signals.
In Figure~\ref{fig:constell_tdi}, the left panel  schematically illustrates the near-equilateral triangluar  constellation  shared by LISA, Taiji and TianQin, together with the  indexing of  spacecrafts (SCs) and laser links. 
The right panel  sketches the  principle of  TDI  (taking the second-generation Michelson-$X$ combination as an example, whose  explicite expression is provided below). 
% For the purpose of GW signal analysis, 
For an individual detector, 
our modeling  starts from the fractional frequency shift  measurement $\eta_{ij}$ along laser link SC$_j \to$ SC$_i$ ($i, j \in \{1, 2, 3\}$), which 
captures the  imprints of both GW and non-GW effects.
Assuming the primary noises can be ideally suppressed via TDI and all time stamps have been perfectly synchronized to a unified reference frame (\textit{e.g.}, Barycentric Coordinate Time), $\eta_{ij}$ is expressed as the combination of single-link GW response $y_{ij}$ and  ``secondary'' instrumental  noise $n_{ij}$:
\begin{equation}
\eta_{ij}(t) = y_{ij}(t) + n_{ij}(t). 
\label{eq:eta_def}
\end{equation}
We adopt a two-component noise model:
\begin{equation}
n_{ij}(t) = N_{ij}(t) + \delta_{ij}(t) + \mathbf{D}_{ij}\,\delta_{ji}(t), 
\label{eq:noise_def}
\end{equation}
where $N_{ij}$ denotes the optical metrology system (OMS) noise of  laser link SC$_j \to$ SC$_i$, and $\delta_{ij}$ denotes the test-mass  acceleration (ACC) noise associated with the test-mass carried by 
SC$_i$ and facing SC$_j$.
The delay operator $\mathbf{D}_{ij}$ is defined by its action on an arbitrary time series $g(t)$:
\begin{equation}
\mathbf{D}_{ij}\,g(t) \equiv g\!\left[t - d_{ij}(t)\right], 
\label{eq:delay_op}
\end{equation}
with  $d_{ij}$ representing  the laser propagation time from SC$_j$ to SC$_i$.

Under the condition that the SC  motion is much slower than the speed of light, the single-link GW response  reads~\citep{Estabrook1975}
\begin{equation}
\begin{aligned}
y_{ij}(t) = \frac{1}{2\bigl[1 - \hat{\mathbf{k}}\cdot\hat{\mathbf{n}}_{ij}(t)\bigr]}
\Bigl[ &H_{ij}\!\left(t - d_{ij}(t) - \frac{\hat{\mathbf{k}}\cdot\mathbf{R}_j(t)}{c}\right) \\
     &- H_{ij}\!\left(t - \frac{\hat{\mathbf{k}}\cdot\mathbf{R}_i(t)}{c}\right) \Bigr]. 
\end{aligned}
\label{eq:single_link_td}
\end{equation}
$\mathbf{R}_i$ and $\mathbf{R}_j$ are  the positions of the two SCs  in the Solar-System-barycenter (SSB) frame, $\hat{\mathbf{n}}_{ij}$ is the  unit vector along the laser link. 
$\hat{\mathbf{k}}$ stands for  the GW's  wave vector, which is related to the source's ecliptic longitude $\lambda$ and   latitude $\beta$  as 
\begin{equation}
      \hat{\mathbf{k}} =  -\left[ \cos \beta \cos \lambda, \cos \beta \sin \lambda, \sin \beta \right]. 
\end{equation}
The orbital motion of the detector enters through the time dependence of $\mathbf{R}_i$, $d_{ij}$, and $\hat{\mathbf{n}}_{ij}$.
The function $H_{ij}$ projects the GW tensor onto direction $\hat{\mathbf{n}}_{ij}$ as:
\begin{equation}
H_{ij}(t) \equiv \mathbf{h}(t) : \hat{\mathbf{n}}_{ij}(t) \otimes \hat{\mathbf{n}}_{ij}(t),
\label{eq:H_proj}
\end{equation}
where the colon denotes double contraction.
For general GW signal incorporating higher-order harmonics, we expand the 
GW tensor $\mathbf{h}$  in polarization bases and spin-weighted spherical harmonics:
\begin{equation}
\mathbf{h}(t) = \sum_{\alpha} \sum_{\ell m} h_{\ell m}(t)\, K^{\ell m}_\alpha(\iota, \varphi_{\rm ref})\, \mathbf{e}_\alpha(\lambda, \beta, \psi),
\label{eq:h_expand}
\end{equation}
with $\alpha\in\{+,\times\}$ labeling the polarization and $\ell m$ the harmonic indices.
$h_{\ell m}(t)$ denotes the waveform component of the $\ell m$ harmonic, and $\mathbf{e}_\alpha(\lambda, \beta, \psi)$ is the polarization basis in the GW source frame, where $\psi$ is  the source's  polarization angle.
These bases are related to the polarization bases in the SSB frame (dubbed $\mathbf{e}_\alpha^{\rm SSB}(\lambda, \beta)$) via
\begin{eqnarray}
\mathbf{e}_{+} &=& \cos 2\psi\; \mathbf{e}_{+}^{\rm SSB} + \sin 2\psi\; \mathbf{e}_{\times}^{\rm SSB},  \label{eq:pol_rotation_p} \\ 
\mathbf{e}_{\times} &=& -\sin 2\psi\; \mathbf{e}_{+}^{\rm SSB} + \cos 2\psi\; \mathbf{e}_{\times}^{\rm SSB}.
\label{eq:pol_rotation_x}
\end{eqnarray}
Our convention for the definitions of these  bases follows \cite{Baghi:2026fef}, where the relations between $\mathbf{e}_{+ / \times}^{\rm SSB}$ and  $(\lambda, \beta)$ are  specified.
The coefficients $K^{\ell m}_\alpha(\iota, \varphi_{\rm ref})$ are constructed from the spin-weighted spherical harmonics ${}_{-2}Y_{\ell m}$ as
\begin{align}
K^{\ell m}_+ &\equiv \frac{1}{2}\Bigl[{}_{-2}Y_{\ell m} + (-1)^\ell\,{}_{-2}Y^*_{\ell,-m}\Bigr], \\
K^{\ell m}_\times &\equiv \frac{i}{2}\Bigl[{}_{-2}Y_{\ell m} - (-1)^\ell\,{}_{-2}Y^*_{\ell,-m}\Bigr],
\label{eq:K_coefficients}
\end{align}
which depend on the inclination angle $\iota$ and the reference orbital phase $\varphi_{\rm ref}$ of the source.
We define the source-frame and SSB-frame  antenna pattern functions as 
\begin{eqnarray}
\zeta_{\alpha,ij} &\equiv& \mathbf{e}_\alpha : \hat{\mathbf{n}}_{ij} \otimes \hat{\mathbf{n}}_{ij}  \\ 
\xi_{\alpha,ij} &\equiv& \mathbf{e}^{\rm SSB}_\alpha : \hat{\mathbf{n}}_{ij} \otimes \hat{\mathbf{n}}_{ij},
\label{eq:antenna_pattern_def}
\end{eqnarray}
respectively, which encode the projection  of polarization basis  onto the laser link. 
With these  definition,  $H_{ij}$  decomposes as
\begin{equation}
      H_{ij} = \sum_{\alpha} \sum_{\ell m} h_{\ell m} K^{\ell m}_\alpha \, \zeta_{\alpha,ij}, 
\end{equation}
a general  relation that will be used in the TDI response deduction  of Appendices~\ref{app:GB} and~\ref{app:MBHB}.

The basic principle of TDI is to cancel out laser frequency noise through appropriately delaying and combining the single-link measurements~\citep{Tinto2021}.
Despite the numerous configurations of TDIs developed in the literature, all TDI combinations (or ``channels'')  can be expressed in a  unified form
\begin{equation}
\mathrm{TDI}(t) = \sum_{ij} \mathbf{P}_{ij}\, \eta_{ij}(t),
\label{eq:tdi_unified}
\end{equation}
where $ij \in \{12, 21, 23, 32, 31, 13\}$, and $\mathbf{P}_{ij}$ are polynomials of the delay operators.
For the most commonly used second-generation Michelson-$X_2$ channel, the explicit forms of the $\mathbf{P}_{ij}$ operators are
\begin{equation}
\begin{aligned}
\mathbf{P}_{12} &= 1 - \mathbf{D}_{131} - \mathbf{D}_{13121} + \mathbf{D}_{1213131}, \\
\mathbf{P}_{23} &= 0, \\
\mathbf{P}_{31} &= -\mathbf{D}_{13} + \mathbf{D}_{1213} + \mathbf{D}_{121313} - \mathbf{D}_{13121213}, \\
\mathbf{P}_{21} &= \mathbf{D}_{12} - \mathbf{D}_{1312} - \mathbf{D}_{131212} + \mathbf{D}_{12131312}, \\
\mathbf{P}_{32} &= 0, \\
\mathbf{P}_{13} &= -1 + \mathbf{D}_{121} + \mathbf{D}_{12131} - \mathbf{D}_{1312121}.
\end{aligned}
\label{eq:tdi_polynomials}
\end{equation}
Here $\mathbf{D}_{i_1 i_2 \dots i_n} \equiv \mathbf{D}_{i_1 i_2}\,\mathbf{D}_{i_2 i_3}\cdots\mathbf{D}_{i_{n-1} i_n}$ is defined as  the sequential application of $n-1$ elementary delay operators.
In the following discussion, we omit the subscript ``2'', and all TDI channels refer to the second-generation ones by default.
The expressions for  Michelson $Y, Z$ channels can be obtained by applying the cyclic index permutation $1\rightarrow 2$, $2\rightarrow3$, $3\rightarrow 1$ to that of $X$. 

The ``optimal''  $A$, $E$, and $T$ channels~\citep{Vallisneri2005,Tinto2021} are constructed from the three original  Michelson channels $\{X, Y, Z\}$ as
\begin{eqnarray}
A &\equiv& \frac{1}{\sqrt{2}}(Z - X), \nonumber \\ 
E &\equiv& \frac{1}{\sqrt{6}}(X - 2Y + Z), \nonumber \\ 
T &\equiv& \frac{1}{\sqrt{3}}(X + Y + Z),
\label{eq:AE_combinations}
\end{eqnarray}
where  $A$ and $E$  are sensitive to GWs and serve as the primary ``signal'' channels, while $T$ has a relatively  suppressed GW response and is typically used for noise characterization.
We therefore  use the $A$ and $E$ channels for source parameter estimation.
Under the assumptions of equal arm lengths  and equal instrumental  noise levels across all the SCs and laser links, the  $\{A, E, T\}$ channels are  statistically orthogonal in terms of instrumental noise~\citep{Vallisneri2005,Tinto2021}. 
In realistic detection scenarios, these assumptions are inevitably perturbed, hence breaking   the orthogonality.  
Nevertheless, although the equal-arm and equal-noise assumptions are adopted in this work, our results do not rely on them, as  the  primay focus  lies in  the confusion foreground. 

Next, we extend our discussion  to the  network of multiple detectors, and move to the frequency domain. 
For the remainder of the paper, we adopt a compact notation for the multi-detector, multi-TDI-channel data.
Let $d_I^c(f)$ denote the frequency-domain TDI output for detector $I \in \mathcal{N} \equiv \{\mathrm{LISA}, \mathrm{Taiji}, \mathrm{TianQin}\}$ and channel $c \in \{A, E\}$.
The total data is the sum of  GW signal $h_I^c(f; \boldsymbol{\theta})$, the instrumental noise $n_I^c(f)$, and the confusion foreground $g_I^c(f)$:
\begin{eqnarray}
d_I^c(f) &=& h_I^c(f; \boldsymbol{\theta}) \;+\; n_I^c(f) \;+\; g_I^c(f),
\nonumber \\ 
g_I^c(f) &\equiv& \sum_{k \in \mathcal{U}} g_{I,k}^c(f; \boldsymbol{\theta}_k),
\label{eq:data_model}
\end{eqnarray}
where $\mathcal{U}$ denotes the set of unresolved GBs, and $g_{I,k}^c(f; \boldsymbol{\theta}_k)$ is the TDI response of detector $I$ to the $k$-th GB.
Since usually treated as stochastic processes, the second-order statistics of $n_I^c$ and $g_I^c$ are specified in the following subsection.

\subsection{Instrumental noise, confusion foreground, and network covariance}
\label{sec:foreground}

We now specify the second-order statistics of the noise components introduced in Eq.~\eqref{eq:data_model}.
The CSD  between any two zero-mean stochastic processes $x_I^c(f)$ and $x_J^{c'}(f)$ observed over a duration $T_{\rm obs}$ can be calculated as  the one-sided periodogram~\citep{Creighton2011}:
\begin{equation}
C_{IJ}^{cc'}(f) \equiv \frac{2}{T_{\rm obs}}\,
\bigl\langle x_I^c(f)\, x_J^{c'*}(f) \bigr\rangle,
\label{eq:csd_def}
\end{equation}
where $\langle\cdot\rangle$ is  the ensemble average.
The PSD  is the diagonal case $I=J$, $c=c'$, for which we use the shorthand notation  $S_I^c(f) \equiv C_{II}^{cc}(f)$.
% Eq.~\eqref{eq:csd_def} serves as the unified definition for both instrumental noise and the confusion foreground; only the physical origin of the averaging differs.

According to the baseline mission  design of LISA~\citep{Babak:2021mhe}, Taiji~\citep{taiji_1,taiji_2} and TianQin~\citep{Luo2016},  we assume that  $N_{ij}$ and $\delta_{ij}$  introduced in Eq.~\eqref{eq:noise_def}  are characterized by their design PSD curves:
\begin{equation}
\begin{aligned}
S_{{\rm OMS}, I, ij}(f) &= A^2_{{\rm OMS},I,ij}\left(\frac{2\pi f}{c}\right)^{\!2}
\left[1 + \left(\frac{2\,\text{mHz}}{f}\right)^{\!4}\,\right], \\[6pt]
S_{{\rm ACC}, I, ij}(f) &= A^2_{{\rm ACC},I,ij}\left(\frac{1}{2\pi f c}\right)^{\!2}
\left[1 + \left(\frac{0.4\,\text{mHz}}{f}\right)^{\!2}\,\right] \nonumber \\
& \quad \times \left[1 + \left(\frac{f}{8\,\text{mHz}}\right)^{\!4}\,\right],
\end{aligned}
\label{eq:single_link_psd}
\end{equation}
where $A_{{\rm OMS},I,ij}$ and $A_{{\rm ACC},I,ij}$ are the noise amplitude parameters.
The factors $(2\pi f/c)^2$ and $(1/2\pi f c)^2$ convert displacement and acceleration units, respectively,    to the fractional frequency fluctuation unit used in   Eq.~\eqref{eq:eta_def}.
For a general TDI combination, the one-sided instrumental noise  PSD after TDI  is obtained by propagating the  PSDs of component noises   through the delay-operator polynomials.
Assuming statistical independence of noise contributions from different laser links and test masses, the instrumental  noise PSD for channel $c$ of detector $I$ reads:
\begin{eqnarray}
N_I^c(f) &=& \sum_{ij} \Bigl[
|\widetilde{\mathbf{P}}_{I,ij}^{\,c}(f) + \widetilde{\mathbf{D}}_{I,ji}(f)\,\widetilde{\mathbf{P}}_{I,ji}^{\,c}(f)|^2\,
S_{{\rm ACC}, I, ij}(f) \nonumber \\
&&  + |\widetilde{\mathbf{P}}_{I,ij}^{\,c}(f)|^2\, S_{{\rm OMS}, I, ij}(f) \Bigr],
\label{eq:tdi_noise_psd}
\end{eqnarray}
where $\widetilde{\mathbf{P}}_{I,ij}^{\,c}(f)$ denotes the frequency-domain representation of the TDI polynomial  of detector $I$ for channel $c$, calculated through  replacing the elementary delay operator $\mathbf{D}_{I,ij}$ by the frequency-domain time-delay factor $\widetilde{\mathbf{D}}_{I,ij}(f) = e^{-2\pi i f d_{I, ij}}$.
Under equal-arm approximation $d_{I, ij} \equiv d_I$ and  equal-noise condition $A_{{\rm OMS}, I, ij} \equiv A_{{\rm OMS}, I}$, $A_{{\rm ACC}, I, ij} \equiv A_{{\rm ACC}, I}$, the noise PSDs of the $A$ and $E$ channels are identical by construction~\citep{Vallisneri2005,Tinto2021}, and we denote them collectively  as $N_I^{\rm inst}(f) \equiv N_I^A(f) = N_I^E(f)$.

Applying the general CSD definition of Eq.~\eqref{eq:csd_def} to the confusion foreground $g_I^c(f)$, we have:
\begin{equation}
C_{IJ}^{cc'}(f) \equiv \frac{2}{T_{\rm obs}}
\sum_{k \in \mathcal{U}} \langle g_{I,k}^c(f; \boldsymbol{\theta}_k)\, g_{J,k}^{c'*}(f; \boldsymbol{\theta}_k) \rangle,
\label{eq:conf_csd}
\end{equation}
where the sum runs over the unresolved population $\mathcal{U}$, and we have exploited  the incoherence between different sources (in the sense of ensemble average). 
Considering a time interval 
% (\textit{i.e.} $T_{\rm obs}$) 
over which the detector positions can be regarded as fixed, 
Eq.~\eqref{eq:conf_csd} admits an  analytic expression  for  the foreground CSD. 
The detailed derivation is given in Appendix~\ref{app:foreground}, and  the basic idea is as follows. 
Since confusion GBs are weak in amplitude and only considerably contribute to the  noise budget  at low frequencies below 3 mHz,  
we retain only the dominant \((2,\pm2)\) harmonic mode and model the TDI response of each GB signal under  the low-frequency and equal-arm approximation. 
Each binary then contributes a quasi-monochromatic power spectrum  \(w_k\) modulated  by   frequency-domain TDI response. 
Grouping the unresolved sources within frequency bins \(\mathcal{U}_f = \{k : f_{0,k}\in[f, f+\Delta f)\}\) and taking ensemble average  over inclination, initial phase, and polarization angle of each source, Eq.~\eqref{eq:conf_csd} reduces to the form 
%% ORIGINAL TWO-COLUMN (widetext) VERSION, COMMENTED OUT.
%% Replaced by the single-column multi-line version below so that the equation
%% stays inside one column. The narrow (242pt) column cannot hold the whole
%% polarization bracket on one line (it measures ~248pt), so the 7:5 bracket is
%% broken onto its own lines.
% \begin{widetext}
% \begin{equation}
% \begin{aligned}
% C_{IJ}^{cc'}(f) &=
% \frac{\pi f^2\, d_I\,d_J}{48}
% \sum_{k\in\mathcal{U}_f} w_k\,
% e^{\,-2\pi i f\,\hat{\mathbf{k}}_k\cdot(\mathbf{R}_I-\mathbf{R}_J)/c}
% \Bigl[\,7\,Q_{I,+}^{\,c*}(f,\hat{\mathbf{k}}_k)\,Q_{J,+}^{\,c'}(f,\hat{\mathbf{k}}_k)
%       + 5\,Q_{I,\times}^{\,c*}(f,\hat{\mathbf{k}}_k)\,Q_{J,\times}^{\,c'}(f,\hat{\mathbf{k}}_k)\,\Bigr] \\[2pt]
% &= \frac{\pi f^2\, d_I\,d_J}{48}\,S_{\rm conf}(f)
% \int {\rm d} \hat{\mathbf{k}}\,P(\hat{\mathbf{k}})\, e^{\,-2\pi i f\,\hat{\mathbf{k}}\cdot(\mathbf{R}_I-\mathbf{R}_J)/c}
% \Bigl[\,7\,Q_{I,+}^{\,c*}(f,\hat{\mathbf{k}})\,Q_{J,+}^{\,c'}(f,\hat{\mathbf{k}})
%       + 5\,Q_{I,\times}^{\,c*}(f,\hat{\mathbf{k}})\,Q_{J,\times}^{\,c'}(f,\hat{\mathbf{k}})\,\Bigr],
% \end{aligned}
% \label{eq:conf_csd_explicitsum}
% \end{equation}
% \end{widetext}
\begin{equation}
\begin{aligned}
C_{IJ}^{cc'}(f) &= \frac{\pi f^2\, d_I\,d_J}{48}
\sum_{k\in\mathcal{U}_f} w_k\,
e^{\,-2\pi i f\,\hat{\mathbf{k}}_k\cdot(\mathbf{R}_I-\mathbf{R}_J)/c} \\
&\quad \times \Bigl[\,7\,Q_{I,+}^{\,c*}(f,\hat{\mathbf{k}}_k)\,Q_{J,+}^{\,c'}(f,\hat{\mathbf{k}}_k) \\
&\quad \quad + 5\,Q_{I,\times}^{\,c*}(f,\hat{\mathbf{k}}_k)\,Q_{J,\times}^{\,c'}(f,\hat{\mathbf{k}}_k)\,\Bigr] \\
&= \frac{\pi f^2\, d_I\,d_J}{48}\,S_{\rm conf}(f)
\int {\rm d} \hat{\mathbf{k}}\,P(\hat{\mathbf{k}}) \\
&\quad \times e^{\,-2\pi i f\,\hat{\mathbf{k}}\cdot(\mathbf{R}_I-\mathbf{R}_J)/c} \\
&\quad \times \Bigl[\,7\,Q_{I,+}^{\,c*}(f,\hat{\mathbf{k}})\,Q_{J,+}^{\,c'}(f,\hat{\mathbf{k}}) \\
&\quad \quad + 5\,Q_{I,\times}^{\,c*}(f,\hat{\mathbf{k}})\,Q_{J,\times}^{\,c'}(f,\hat{\mathbf{k}})\,\Bigr],
\end{aligned}
\label{eq:conf_csd_explicitsum}
\end{equation}
where we have defined 
\begin{equation}
% Q_{I,\alpha}^{\,c}(f,\hat{\mathbf{k}}) \equiv  \sum_{ij}\,\widetilde{\mathbf{P}}_{I,ij}^{\,c}(f)\,
% \mathbf{e}_\alpha^{\rm SSB}(\hat{\mathbf{k}}) : \hat{\mathbf{n}}_{I,ij}\otimes\hat{\mathbf{n}}_{I,ij},
Q_{I,\alpha}^{\,c}(f,\hat{\mathbf{k}}) \equiv  \sum_{ij}\,\widetilde{\mathbf{P}}_{I,ij}^{\,c}(f)\,
\xi_{I, \alpha, ij}(\hat{\mathbf{k}}),  
\label{eq:antenna_vector}
\end{equation} 
and $\mathbf{R}_I$ denotes the position of detector $I$'s constellation center. 
In the first line,  the sum runs over the unresolved binaries in the frequency bin \(\mathcal{U}_f\), each contributing its quasi-monochromatic power \(w_k\). 
The \(7:5\) weighting of the two polarizations follows from the average over inclination, and the baseline  phase factor \(e^{-2\pi i f\hat{\mathbf{k}}_k\cdot(\mathbf{R}_I-\mathbf{R}_J)/c}\) originates from the time delay between the constellation centers of detector $I$ and $J$. 
In the second line, the summation is  converted into a continuous integral,  obtained when the unresolved population can be  described by a sky-distribution model, where  
% \(S_{\rm conf}(f)\equiv(2/T_{\rm obs})\sum_{k\in\mathcal{U}_f}|\tilde{h}_k(f)|^2\) 
\(S_{\rm conf}(f)\equiv \sum_{k\in\mathcal{U}_f}w_k(f)\) 
is the  power spectrum  of the bin,  and \(P(\hat{\mathbf{k}})\) its the  sky distribution function, normalized via  \(\int d\hat{\mathbf{k}}\,P(\hat{\mathbf{k}})=1\). 
During the deduction of the second line, we have assumed that the spatial distribution of GB powers is frequency-independent. 
% The two lines coincide under \(S_{\rm conf}(f)\,P(\hat{\mathbf{k}})\,d\hat{\mathbf{k}}=\sum_{k\in\mathcal{U}_f\cap d\hat{\mathbf{k}}} w_k\). 
A key point in this derivation is that, owing to the anisotropic distribution of \(P(\hat{\mathbf{k}})\) for GBs, we do not adopt the conventional ORF formalism to characterize this CSD, whose standard form  relies on the isotropy assumption~\citep{Allen:1997ad}. 
For direct comparison with the   numerical simulation results, 
we adopt the first line of Eq.~\eqref{eq:conf_csd_explicitsum} and calcualte the  analytic foreground CSD via  summation over a confusion GB catalogue. 
The continuous form, on the other hand,  would be convenient  only when a distribution model \(P(\hat{\mathbf{k}})\) is prescribed. 
The PSDs of foreground are defined as  the diagonal case \(I=J\), \(c=c'\) of Eq.~\eqref{eq:conf_csd_explicitsum}, which reads
\begin{equation}
S_I^{c}(f) = \frac{\pi f^2\, d_I^2}{48}
\sum_{k\in\mathcal{U}_f} w_k\,
\Bigl[\,7\bigl|Q_{I,+}^{\,c}(f,\hat{\mathbf{k}}_k)\bigr|^2
      + 5\bigl|Q_{I,\times}^{\,c}(f,\hat{\mathbf{k}}_k)\bigr|^2\,\Bigr],
\label{eq:conf_psd_explicitsum}
\end{equation}
for which the baseline phase factor reduces to unity. 
% Because the \(A\) and \(E\) antenna vectors differ, the CSDs between distinct channels (\(c\neq c'\)), both within a detector and across detectors, are generically non-vanishing and are evaluated directly from Eq.~\eqref{eq:conf_csd_explicitsum}.
To quantify the strength of the correlation, we introduce the dimensionless  coherence
\begin{equation}
\gamma_{IJ}^{cc'}(f) \equiv \frac{C_{IJ}^{cc'}(f)}{\sqrt{S_I^{c}(f)\,S_J^{c'}(f)}},
\label{eq:coherence_cc}
\end{equation}
which satisfies \(|\gamma_{IJ}^{cc'}|\le 1\) by the Cauchy--Schwarz inequality. 
This quantity characterizes the strength  of cross-detector foreground correlation and plays a central role in assessing the impacts on analyzing resolvable GW sources.
In  Appendix~\ref{app:foreground}, 
the analytic calculation of $\gamma$ is compared against  numerical simulation  results,  providing  a cross-validation for  both approaches.

To obtain  a more compact  matrix representation, for $N$ detectors each with two ``signal'' TDI channels ($A$ and $E$), we stack the data into a joint vector of dimension $2N$ at each frequency $f$:
\begin{equation}
\mathbf{d}(f) = \big[d_1^A, d_1^E,\; d_2^A, d_2^E,\; \dots,\; d_N^A, d_N^E\big]^\top \in \mathbb{C}^{2N}.
\end{equation}
The total noise covariance matrix decomposes into statistically independent instrumental and confusion foreground contributions:
\begin{equation}
\mathbf{C}(f) = \mathbf{C}_{\rm inst}(f) + \mathbf{C}_{\rm conf}(f).
\label{eq:noise_cov_decomposition}
\end{equation}
Under the equal-arm and equal-noise assumptions, the instrumental noise matrix is diagonal, reflecting the statistical independence of different detectors and TDI channels:
\begin{eqnarray}
\mathbf{C}_{\rm inst}(f) &=& \operatorname{diag} \big( N_1^{\rm inst}(f), N_1^{\rm inst}(f), \nonumber \\ 
&&  \quad \quad \quad  \dots,  N_N^{\rm inst}(f), N_N^{\rm inst}(f)\big),
\end{eqnarray}
where $N_I^{\rm inst}(f)$ is the instrument noise PSD  of detector $I$, assumed identical for the $A$ and $E$ channels and taking  the form of  Eq.~\eqref{eq:tdi_noise_psd}.
While, the confusion foreground matrix has a block structure organized in $2\times 2$ blocks $\mathbf{B}_{IJ}$:
\begin{equation}
\mathbf{C}_{\rm conf}(f) = \begin{bmatrix}
\mathbf{B}_{11} & \mathbf{B}_{12} & \cdots & \mathbf{B}_{1N} \\
\mathbf{B}_{21} & \mathbf{B}_{22} & \cdots & \mathbf{B}_{2N} \\
\vdots & \vdots & \ddots & \vdots \\
\mathbf{B}_{N1} & \mathbf{B}_{N2} & \cdots & \mathbf{B}_{NN}
\end{bmatrix}.
\end{equation}
% The diagonal blocks ($I=J$) contain the PSDs of the individual detectors,
% The off-diagonal blocks ($I<J$) 
Each block  contains  the foreground  CSDs for a single detector ($I = J$) or  between detector pairs ($I \neq J$): 
\begin{equation}
\mathbf{B}_{IJ}(f) = \begin{bmatrix}
C_{IJ}^{AA}(f) & C_{IJ}^{AE}(f) \\
C_{IJ}^{EA}(f) & C_{IJ}^{EE}(f)
\end{bmatrix},
\end{equation}
In practical calculation, once the upper triangular blocks are obtained, the lower triangular blocks can be  determined via  Hermitian  symmetry $\mathbf{B}_{JI} = \mathbf{B}_{IJ}^\dagger$.

Regarding the computation of  foreground CSD, our adoption of  equal-arm approximation is justified as follows.
Realistic space-based  detectors will  operate with unequal, time-varying arms, whose deviations from the nominal lengths are below $\sim 1\%$ for LISA and Taiji,  and below $\sim 0.1\%$ for TianQin~\citep{LT_arm_inequality1,LT_arm_inequality2,TQ_arm_inequality}. 
Given that low-frequency TDI response scales linearly with  arm length (see \textit{e.g.} Eqs.~\eqref{eq:app_tdi_factorized} and \eqref{eq:app_T}), these deviations affect the waveform of  individual GB  only at the sub-percent level.
For low-SNR sources within the  confusion foreground, this fluctuation is even more negligible.

%---------------------------------------------------------------------
\subsection{Network matched-filter statistics and sensitivity}
\label{sec:inner_product}
%---------------------------------------------------------------------

With the $2N\times 2N$ frequency-domain noise covariance matrix $\mathbf{C}(f)$ defined in Sec.~\ref{sec:theory}, the matched filtering  inner product between two network data vectors $\mathbf{a}(f)$ and $\mathbf{b}(f)$ is~\citep{Romano:2016dpx} 
\begin{equation}
\langle \mathbf{a} \,|\, \mathbf{b} \rangle_{\mathbf{C}}
= 4\,\Delta f\,
\Re \sum_{f}\,
\mathbf{a}^\dagger(f)\,\mathbf{C}^{-1}(f)\,\mathbf{b}(f),
\label{eq:inner_product}
\end{equation}
where $\Delta f = 1/T_{\rm obs}$ is the frequency resolution of the $T_{\rm obs}$ observation, the sum runs over all frequency bins inside the signal band, and the factor $4\Delta f$ follows  the standard normalization convention for one-sided power spectra~\citep{Creighton2011,Romano:2016dpx}.

The network SNR  for a GW signal $\mathbf{h}(f; \boldsymbol{\theta})$ with parameters $\boldsymbol{\theta}$ is
\begin{equation}
\rho(\boldsymbol{\theta}) = \sqrt{\langle \mathbf{h} \,|\, \mathbf{h} \rangle_{\mathbf{C}}}\,,
\label{eq:snr_network}
\end{equation}
and the corresponding network log-likelihood under the assumption of stationary, Gaussian noise is
\begin{equation}
\ln\mathcal{L}(\boldsymbol{\theta}) = -\frac{1}{2}\,
\langle \mathbf{d} - \mathbf{h}(\boldsymbol{\theta}) \,|\,
\mathbf{d} - \mathbf{h}(\boldsymbol{\theta}) \rangle_{\mathbf{C}}\,,
\label{eq:loglike}
\end{equation}
where $\mathbf{d}(f)$ denotes the total data vector. 
In the context of this paper, $\mathbf{d}(f)$ includes resolvable GW signal,  instrumental noise, and  confusion foreground.  
% The constant $\ln\det(2\pi\mathbf{C})$ term has been omitted since it does not depend on $\boldsymbol{\theta}$ in the noise models considered here.

% In Eqs.~\eqref{eq:inner_product}--\eqref{eq:loglike} the network signal vector $\mathbf{h}(f_k; \boldsymbol{\theta})$ is obtained by applying the TDI operators $\mathbf{P}_{ij}$ of Eq.~\eqref{eq:tdi_unified} to the single-link GW response $y_{ij}$ of Eq.~\eqref{eq:single_link_td}, followed by the $A$/$E$ rotation.
% The explicit construction is
% \begin{equation}
% \mathbf{h}(f_k; \boldsymbol{\theta}) =
% \begin{bmatrix}
% h_1^A(f_k; \boldsymbol{\theta}) \\ h_1^E(f_k; \boldsymbol{\theta}) \\
% \vdots \\ h_N^A(f_k; \boldsymbol{\theta}) \\ h_N^E(f_k; \boldsymbol{\theta})
% \end{bmatrix},
% \qquad
% h_I^c(f_k; \boldsymbol{\theta}) = \mathcal{F}\bigl[\mathrm{TDI}_I^{c}(t; \boldsymbol{\theta})\bigr],
% \label{eq:h_joint}
% \end{equation}
% where $\mathcal{F}[\cdot]$ denotes the discrete Fourier transform onto the frequency grid $\{f_k\}$, and $\mathrm{TDI}_I^{c}$ labels the TDI combination (Michelson $X_2$, optimal $A$, etc.) for detector $I$ and channel $c\in\{A,E\}$.
% For Galactic binaries, the waveform model $h_{\ell m}$ in Eq.~\eqref{eq:h_expand} is a quasi-sinusoid with a slowly evolving frequency; for MBHBs, we use the \texttt{IMRPhenomHM} phenomenological waveform~\cite{London2018} whose TDI response construction is summarized in Appendix~\ref{app:MBHB}.
Complementing the matched-filtering analysis, we define the network sensitivity, a quantity crucial for theoretical analysis,  
which  characterizes the joint detection capability of multiple detectors to GW signals. 
Our definition and derivation  are extended from  \cite{Robson2017}. 
In our  derivation, the orbital information of the  detectors   is evaluated at a fixed time, so the resulting sensitivity should be regarded as the  instantaneous value. 
The orbital motion over the whole  mission life time can be  accounted for by evaluating this instantaneous sensitivity at multiple time and taking  the average, as done for the numerical results of Figure~\ref{fig:sensitivity}. 

Under fixed orbital configuration, the frequency-domain signal in  channel $c$ of detector $I$ can be factorized as
\begin{equation}
h_I^c(f) = \mathcal{R}_I^{c,+}(f;\hat{\mathbf{k}})\,\tilde{h}_+^{\rm SSB}(f) + \mathcal{R}_I^{c,\times}(f;\hat{\mathbf{k}})\,\tilde{h}_\times^{\rm SSB}(f). 
\label{eq:tdi_transfer}
\end{equation}
Here $\tilde{h}_+^{\rm SSB}(f)$ and $\tilde{h}_\times^{\rm SSB}(f)$ are the GW polarizations for the $+$ and $\times$ modes in the SSB frame, related to the source-frame polarizations $\tilde{h}_+(f)$ and $\tilde{h}_\times(f)$ via the polarization angle $\psi$:
\begin{eqnarray}
\tilde{h}_+^{\rm SSB} &=& \cos 2\psi\,\tilde{h}_+ - \sin 2\psi\,\tilde{h}_\times, \\ 
\tilde{h}_\times^{\rm SSB} &=& \sin 2\psi\,\tilde{h}_+ + \cos 2\psi\,\tilde{h}_\times, 
\label{eq:h_ssb_source}
\end{eqnarray}
in consistency with Eqs.~\eqref{eq:pol_rotation_p} and \eqref{eq:pol_rotation_x}. 
The response  functions $\mathcal{R}_{I}^{c, \alpha}$ encode the single-link response of  Eq.~\eqref{eq:single_link_td}, the TDI combination Eq.~\eqref{eq:tdi_unified}, and is transformed to the frequency domain. 
% In the closed form below, $\hat{\mathbf{n}}_{I,ij}$ denotes the unit vector along the laser link $\mathrm{SC}_j \to \mathrm{SC}_i$ of detector $I$, and $\mathbf{R}_i$, $\mathbf{R}_j$ denote the positions of the two spacecraft of detector $I$ in the SSB frame.
Under the equal-arm approximation, they take the closed form
\begin{equation}
\begin{aligned}
\mathcal{R}_{I}^{c, \alpha}(f;\hat{\mathbf{k}})
&= i\pi f d_I\, e^{-i\pi f d_I}
\sum_{ij}\widetilde{\mathbf{P}}_{I,ij}^{\,c}(f)\, \\ 
& \quad \times \mathrm{sinc}\!\bigl[\pi f d_I\,(1-\hat{\mathbf{k}}\cdot\hat{\mathbf{n}}_{I,ij})\bigr]\\ 
& \quad \times \; e^{-i\pi f\, \frac{\hat{\mathbf{k}}\cdot(\mathbf{R}_{I, i}+\mathbf{R}_{I, j})}{c}}\,
% \mathbf{e}_\alpha^{\rm SSB}(\hat{\mathbf{k}}):\hat{\mathbf{n}}_{I,ij}\otimes\hat{\mathbf{n}}_{I,ij}. 
\xi_{I, \alpha, ij}(\hat{\mathbf{k}}). 
\end{aligned}
\label{eq:pol_transfer}
\end{equation}

Stacking  Eq.~\eqref{eq:tdi_transfer} 
% the response  functions 
for all detectors and TDI  channels into joint vectors 
% $\boldsymbol{\mathcal{R}}^+(f)$ and $\boldsymbol{\mathcal{R}}^\times(f)$, the network signal vector of Eq.~\eqref{eq:tdi_transfer} becomes $\mathbf{h}(f) = \boldsymbol{\mathcal{R}}^+(f)\,\tilde{h}_+^{\rm SSB}(f) + \boldsymbol{\mathcal{R}}^\times(f)\,\tilde{h}_\times^{\rm SSB}(f)$.
and substituting into Eq.~\eqref{eq:snr_network}, the squared network SNR takes the form
\begin{equation}
\begin{aligned}
\rho^2 = 4\Delta f \sum_f\Bigl[
&|\tilde{h}_+^{\rm SSB}|^2\,\boldsymbol{\mathcal{R}}^{+\dagger}\mathbf{C}^{-1}\boldsymbol{\mathcal{R}}^+ \\
&+ |\tilde{h}_\times^{\rm SSB}|^2\,\boldsymbol{\mathcal{R}}^{\times\dagger}\mathbf{C}^{-1}\boldsymbol{\mathcal{R}}^\times \\
&+ 2\,\mathrm{Re}\bigl(\tilde{h}_+^{{\rm SSB}\,*}\tilde{h}_\times^{\rm SSB}\,
\boldsymbol{\mathcal{R}}^{+\dagger}\mathbf{C}^{-1}\boldsymbol{\mathcal{R}}^\times\bigr)
\Bigr]. 
\end{aligned}
\label{eq:snr_expanded}
\end{equation}
% where we omit the $f$-dependence for brevity.
Averaging  over the polarization angle $\psi$ yields $\langle|\tilde{h}_+^{\rm SSB}|^2\rangle_\psi = \langle|\tilde{h}_\times^{\rm SSB}|^2\rangle_\psi = \frac{1}{2}(|\tilde{h}_+|^2 + |\tilde{h}_\times|^2)$, and  the cross term vanishes.
Following the definition in  \cite{Robson2017}, we write the squared SNR averaged over polarization and sky position as 
\begin{equation}
\langle \rho^2 \rangle_{\lambda\beta\psi}
= 4\Delta f \sum_f \frac{|\tilde{h}_+(f)|^2 + |\tilde{h}_\times(f)|^2}{S_{\mathcal{N}}(f)},
\label{eq:snr_avg}
\end{equation}
where $S_{\mathcal{N}}(f)$ is the total network sensitivity.
Comparing Eqs.~\eqref{eq:snr_avg} and \eqref{eq:snr_expanded} yields the explicit expression for the network sensitivity:
\begin{equation}
\frac{1}{S_{\mathcal{N}}(f)} \equiv \frac{1}{2}\Bigl\langle
\boldsymbol{\mathcal{R}}^{+\dagger}(f)\,\mathbf{C}^{-1}(f)\,\boldsymbol{\mathcal{R}}^+(f)
+ (+\to\times)
\Bigr\rangle_{\lambda\beta}. 
\label{eq:sensitivity_net}
\end{equation}
The average over sky directions is performed via Monte Carlo simulation, \textit{i.e.}, by sampling  $10^3$ random directions and then averaging over them. 
To  obtain the characteristic strain shown in Figure~\ref{fig:sensitivity}, we further apply the conversion $S_{\mathcal{N}}(f) \rightarrow \sqrt{f S_{\mathcal{N}}(f)}$. 

% As a consistency check, for a single detector $I$ with uncorrelated $A$/$E$ noise, Eq.~\eqref{eq:sensitivity_net} reduces to the familiar per-channel sensitivity $S_I^{\rm eff}(f) = N_I^{c}(f)/\langle \mathcal{R}_I(f)^2\rangle_{\lambda\beta\psi}$, where $N_I^{c}(f) \equiv N_I^{\rm inst}(f) + S_I^{c}(f)$ is the total per-channel noise PSD and $\langle\mathcal{R}_I(f)^2\rangle = \tfrac{1}{2}\langle|\mathcal{R}_{I,+}^{A}|^2 + |\mathcal{R}_{I,\times}^{A}|^2\rangle_{\lambda\beta\psi}$ is the sky- and polarization-averaged squared response (identical for A and E by symmetry~\cite{Vallisneri2005}); the single-detector SNR then accumulates as $\rho_I^2 = (\rho_I^A)^2 + (\rho_I^E)^2$, and the corresponding characteristic strains are $h_c^{\mathcal{N}}(f) = \sqrt{f\,S_{\mathcal{N}}(f)}$ and $h_c^I(f) = \sqrt{f\,S_I^{\rm eff}(f)}$.

% The network characteristic strain $h_c^{\mathcal{N}}(f)$ is shown in Figure~\ref{fig:sensitivity} for the LISA-Taiji-TianQin network under different foreground subtraction scenarios.

%---------------------------------------------------------------------
\subsection{Analytic estimate of the SNR impact}
\label{sec:snr_impact}
%---------------------------------------------------------------------
To develop a physical insight before numerical analysis, we derive an analytic expression for the fractional change in the squared network SNR  after the inclusion of  cross-detector foreground correlations. 
For simplicity, we capture the essential physics with a two-detector model, each detector  with a single observational channel.

Let the total noise PSDs of the two detectors be \(S_1(f)\) and \(S_2(f)\), and let \(C_{12}(f)\) be their foreground CSD. The corresponding frequency-domain detector responses to the same  signal are denoted by \(h_1(f)\) and \(h_2(f)\). 
% We define the dimensionless coherence
Restricting our analysis to the regime in which the foreground dominates over the instrumental noise, the coherence $\gamma(f)$ (see Eq.~\eqref{eq:coherence_def}) can be approximately evaluated as:
\begin{equation}
\gamma(f) \approx \frac{C_{12}(f)}{\sqrt{S_1(f)\,S_2(f)}} \equiv A_\gamma(f) e^{i \theta_\gamma(f)},
\label{eq:coherence_def}
\end{equation}
with \(A_\gamma\in[0,1]\), and the noise-rescaled signals
\begin{align}
\alpha(f) &\equiv \frac{h_1(f)}{\sqrt{S_1(f)}} \equiv A_{\alpha}(f)\,e^{i\theta_\alpha(f)}, \nonumber \\ 
\beta(f) &\equiv \frac{h_2(f)}{\sqrt{S_2(f)}} \equiv A_{\beta}(f)\,e^{i\theta_\beta(f)},
\label{eq:rescaled_signals}
\end{align}
with the  phase difference in the two detector's signal  responses  defined as 
\begin{equation}
      \Delta\varphi(f) \equiv \theta_\beta(f) - \theta_\alpha(f). 
\end{equation}
The noise covariance matrix at each frequency is \(\mathbf{C}(f) = \begin{bmatrix} S_1 & C_{12} \\ C_{12}^* & S_2 \end{bmatrix}\), with inverse
\begin{equation}
\mathbf{C}^{-1}(f) = \frac{1}{\sqrt{S_1S_2}\,(1-A_\gamma^2)} \begin{bmatrix} \sqrt{S_2/S_1} & -\gamma^* \\ -\gamma & \sqrt{S_1/S_2} \end{bmatrix}.
\label{eq:cinv_2det}
\end{equation}
Substituting this expression  into Eq.~\eqref{eq:snr_network}, 
% at a single frequency point  \(f\) (postponing the summation), 
we obtain the optimal SNRs for the uncorrelated case \((\gamma=0)\) and the correlated case $(\gamma \neq 0)$:
\begin{align}
\rho_0^2 |_f &= 4\Delta f \sum_f \bigl(A_\alpha^2 + A_\beta^2\bigr), \label{eq:rho0_2det}\\[2pt]
\rho^2 |_f &= 4\Delta f \sum_f \frac{A_\alpha^2 + A_\beta^2 - 2A_\alpha A_\beta A_\gamma \cos(\theta_\gamma-\Delta\varphi)}{1-A_\gamma^2},
\label{eq:rho_2det}
\end{align}
respectively. The relative change in \(\rho^2\) at a single  frequency point $f$ is then
\begin{equation}
\frac{\rho^2 - \rho_0^2}{\rho_0^2} \bigg|_f 
= \frac{A_\gamma (A_\gamma  - \eta) }{1-A_\gamma^2},
\quad
\eta \equiv \frac{2A_\alpha A_\beta}{A_\alpha^2+A_\beta^2}\,\cos\bigl(\theta_\gamma - \Delta\varphi\bigr),
\label{eq:dsnr2_rel}
\end{equation}
where \(|\eta| \le 1\). 
The factor \(\eta\) depends on the alignment between  $\Delta \varphi$ and $\theta_\gamma$, weighted by the  relative amplitude ratio  between the  two  detector  responses. 
\(|\eta| \approx 1\) when the response  amplitudes in the two detectors are comparable and the signal phase difference aligns (or anti-aligns) with the coherence phase. 
Conversely, \(\eta \approx 0\) when the amplitudes are highly asymmetric or the phases are orthogonal.
Expanding  in $A_\gamma$, the single-frequency relative  difference in $\rho^2$ between the two covariance models is
% \begin{equation}
%       \frac{\Delta \rho^2}{\rho^2_0}\bigg|_f = -\eta A_\gamma + A_\gamma^2 - \eta A_\gamma^3 + \mathcal{O}(A_\gamma^4),
% \label{eq:dsnr_expanded} 
% \end{equation}
% to the third order in $A_\gamma$.  
% \begin{equation}
%       \frac{\Delta \rho^2}{\rho^2_0}\bigg|_f = -\eta A_\gamma + A_\gamma^2  + \mathcal{O}(A_\gamma^3),
% \label{eq:dsnr_expanded} 
% \end{equation}
\begin{equation}
      \frac{\Delta \rho^2}{\rho^2_0}\bigg|_f = -\eta A_\gamma   + \mathcal{O}(A_\gamma^2),
\label{eq:dsnr_expanded} 
\end{equation}
to the leading   order in $A_\gamma$.  
Its magnitude is bounded by \(A_\gamma\) at leading order, 
% ---equivalently \(|\Delta\rho|/\rho_0 \lesssim A_\gamma/2\) for the relative SNR change---
with the maximum being reached for \(|\eta|=1\).
% The linear term $-\eta A_\gamma$ encodes  the leading effect.  
The sign follows $\eta$, so the correlation reduces $\rho^2$ when $\eta>0$ (the correlated foreground adds coherently to the effective noise) and enhances it when $\eta<0$. 
% While the quadratic term $+A_\gamma^2$ is positive definite, which originates from the $1/(1-A_\gamma^2)$ factor of $\mathbf{C}^{-1}$,  sharpening  each detector's noise once the correlation is included and is always favorable to $\rho^2$. 

This leading-order effect can be illustrated with a simple special case. Consider two similar detectors whose constellation centers coincide and whose spacecraft phases within the constellations are nearly identical. In this case, both $\Delta\varphi$ and $\theta_\gamma$ are close to zero ($\theta_\gamma \to 0$ because Eq.~\eqref{eq:conf_csd_explicitsum} is approximately real), so $\eta$ approaches its maximum value of unity. Consequently, including the cross-detector foreground correlation always reduces the network SNR. Intuitively, this is because strong foreground coherence correlates the noise across detectors, reducing the efficiency of the network's noise suppression and thus degrading the SNR. This behavior is confirmed with the numerical  results based on orbit Config.~II (see Appendix~\ref{app:configII}).

One should also notice  that this single-frequency picture is incomplete. The time and frequency dependence  must also be taken into account for specific signal analysis. 
For instance, the foreground correlation is not necessarily present across the entire 0.1 mHz - Hz target frequency band of space-based GW detectors. 
The effect may be  observable only when the signal lies in the correlated band (see Figure~\ref{fig:csd_comparison}). For MBHBs, this typically implies that heavier systems are more affected, since their merger frequency may fall into this band. 
Besides, for continuous signals like GBs (as opposed to transient ones), the effect may be suppressed by the time variation of both $\theta_\gamma$ and $\Delta\varphi$, particularly when the variation leads to cancellation over the observation duration. 
With the above physical understanding established, we leave the analysis of specific signals to the numerical simulations presented below.

\section{Methodology}
\label{sec:method}
%=====================================================================

\subsection{Detector orbit and noise configuration}

\begin{table}[t]
\caption{\label{tab:orbits}
Orbit and instrumental noise parameters adopted for TianQin, Taiji, and LISA.
For the heliocentric Taiji and LISA constellations, \(\kappa\) denotes the initial orbital phase of the constellation guiding center and \(\lambda_0\) the initial SC  phase within the constellation plane. 
For the geocentric TianQin constellation, \(\kappa_{\oplus}\) denotes the initial orbital phase of the Earth and \(\kappa_0\) the initial SC  phase.
The two sets of initial conditions  correspond to the default-like Config.~I and the Taiji-c-like Config.~II.}
\begin{ruledtabular}
\begin{tabular}{lccc}
% \hline
Parameter & TianQin & Taiji & LISA \\
\hline
Arm length (m) & $\sqrt{3}\times 10^8$ & $3\times 10^9$ & $2.5\times 10^9$ \\
Orbit type & Geocentric & Heliocentric & Heliocentric \\
$A_{{\rm OMS},I}$ (m/$\sqrt{\rm Hz}$)   & $1\times 10^{-12}$  & $8\times 10^{-12}$  & $15\times 10^{-12}$ \\
$A_{{\rm ACC},I}$ (m/s$^2$/$\sqrt{\rm Hz}$) & $1\times 10^{-15}$  & $3\times 10^{-15}$  & $3\times 10^{-15}$ \\
\hline
\multicolumn{4}{c}{Initial Config.~I (default-like)} \\
\hline
$\kappa_0$ or $\kappa$ (rad)        & 1.2298 & 0           & $-0.6981$ \\
$\lambda_0$ or $\kappa_{\oplus}$ (rad) & $-0.3491$ & 6.1041      & 0.1280 \\
\hline
\multicolumn{4}{c}{Initial Config.~II (Taiji-c-like)} \\
\hline
$\kappa_0$ or $\kappa$ (rad)        & 4.4012 & 0           & 0 \\
$\lambda_0$ or $\kappa_{\oplus}$ (rad) & $-0.3491$ & 2.7973      & 1.8080 \\
% \hline
\end{tabular}
\end{ruledtabular}
\end{table}

We now turn to the practical implementation of the theoretical framework established in Sec.~\ref{sec:theory}.
Table~\ref{tab:orbits} lists the orbit and noise parameters adopted for the three detectors.
We employ equal-arm analytic orbit models throughout: the  heliocentric orbit model for LISA and Taiji follows \cite{LISA_EA_Orbit}, and the  geocentric orbit model for TianQin follows \cite{TianQin_EA_Orbit,Li:2023szq}.
% This approximation is well justified for the present analysis. 
% For GW  signals, the low-frequency TDI response scales  linearly with the arm length $d_{ij}$ (see \textit{e.g.} Eq.~\eqref{eq:app_link_fd}), so the unequal-arm corrections are only at the sub-percent level for LISA and Taiji~\cite{LT_arm_inequality1,LT_arm_inequality2} and below $\sim 0.1\%$ for TianQin~\cite{TQ_arm_inequality}, and their impact on low-SNR  confusion signals  would not  be considerable.
The motion of each heliocentric detector is parametrized by two initial angles: the initial phase of the constellation's  guiding center in the SSB frame $\kappa$,  and the initial phase of the SCs within the constellation plane  $\lambda_0$. 
For the geocentric TianQin orbit, the corresponding parameters are  the Earth's initial orbit  position angle $\kappa_{\oplus}$~\citep{Luo2016} and the initial SC  phases $\kappa_0$.
Config.~I represents  the  ``default'' design of  LISA-Taiji-TianQin network, with Taiji leading the Earth by 20$^\circ$ and LISA trailing the Earth by 20$^\circ$, and Config.~II corresponds to the ``Taiji-c''  design proposed by \cite{Wang2021}, where Taiji and LISA shares the same guiding center, leading the Earch by $20^\circ$ together, and  
their constellation planes are also aligned.  
For both configurations the initial SC phases ($\lambda_0$ or $\kappa_0$) are ramdomly generated. 
Instrumental noise PSDs  follow the standard model of Eq.~\eqref{eq:single_link_psd}, with  $A_{{\rm ACC},I}$ and  $A_{{\rm OMS},I}$ specified in the table.

Two noise models are compared  throughout this work. 
The ``\textbf{block-diagonal}'' model retains the foreground PSD  $S_I^c(f)$ of each detector but sets all off-diagonal blocks $\mathbf{B}_{IJ} = \mathbf{0}$ for $I \neq J$, thereby neglecting the cross-detector foreground correlations.
The ``\textbf{full-covariance}'' model retains the complete covariance matrix including all cross-detector correlation terms.

\subsection{Simulation of Galactic confusion foreground and spectrum evaluation}
\label{sec:foreground_sim}

We employ a catalog of $\sim 3 \times 10^7$ GBs from the LDC Radler dataset~\citep{Baghi:2022ucj}.
Each source is specified by eight parameters: initial GW  frequency $f_0$, initial GW  frequency derivative $\dot{f}_0$, amplitude $A$, ecliptic longitude $\lambda$, ecliptic latitude $\beta$, inclination $\iota$, polarization angle $\psi$, and initial GW  phase $\varphi_0$.
The fast frequency-domain response calculation  is performed  with the \texttt{Triangle-GB} code, as detailed in Appendix~\ref{app:GB}.

We adopt an iterative subtraction algorithm adapted from the standard procedure in   \cite{Karnesis:2021tsh,Liu2023}, to remove the resolvable GB signals   from the data and obtain a residual confusion foreground.
The procedure begins with a pre-selection step, in which an initial  candidate list is identified using  instrumental-noise-only SNR with a threshold of 5.
The subtraction then proceeds iteratively: at each iteration the PSD  and CSD are estimated from the current residuals to construct the full network covariance matrix, and every candidate whose network SNR exceeds a threshold of $\rho_{\rm th,network}$ is subtracted simultaneously from the data of all detectors.
The iterative  procedure stops when a maximum of 10 iterations is reached.
In this work we consider two choices for the  network SNR threshold    $\rho_{\rm th,network} = 7 \  {\rm and } \ 10$.

The raw periodogram of foreground data   are too noisy for direct use in SNR calculation, therefore  
at each iteration  we estimate the PSD and CSD   using   Welch's method with 0.2-day  segments, 50\% overlap. 
Each segment is multiplied by a Hann window, and the per-segment periodograms are averaged. 
This yields the smoothed PSD and CSD  on a frequency grid with resolution $\Delta f = 1/ 0.2 \ {\rm day} = 5.8\times 10^{-5} \ {\rm Hz}$.
The estimates are then  interpolated onto the target frequency grid for  SNR calculation  via  PCHIP interpolation~\citep{doi:10.1137/0905021}, with the real and imaginary parts of the complex CSD interpolated independently.

\subsection{Network Bayesian data analysis framework for MBHBs and GBs}

For the methodology of network   Bayesian analysis, we first present the FIM, which provides  computationally efficient theoretical  forecast for source  parameter uncertainties, and   then describe the parameter estimation pipeline based on simulated data, including data preparation,  likelihood model  and the settings for posterior  sampling.
The pipelines for the two target source classes  differ due to their   distinct time-frequency structures of the two signal classes: GBs are quasi-monochromatic continuous signals  analyzed in  a narrow frequency band around $f_0$, while MBHBs are transient chirping signals, with the majority of their SNRs accumulated within the final month before merger.

\subsubsection{Fisher information matrix}

As a  computationally efficient approach, we compute the FIM to forecast parameter uncertainties for given source parameters.
The  elements of FIM  are obtained from the covariance-weighted inner product as defined by  Eq.~\eqref{eq:inner_product}:
\begin{equation}
\Gamma_{ab} = \left\langle \frac{\partial \mathbf{h}}{\partial \theta_a} \, \middle| \, \frac{\partial \mathbf{h}}{\partial \theta_b} \right\rangle_{\mathbf{C}},
\label{eq:fisher}
\end{equation}
which generalizes the standard FIM to a multi-detector network.

The waveform derivatives with respect to each parameter are evaluated via central finite differencing, with step sizes optimized through an auto-test algorithm: for each parameter,  the step is iteratively halved until the relative change in the inferred $1\sigma$ uncertainty falls below a threshold of $10^{-3}$.

Once the FIM $\Gamma_{ab}$ is assembled, the parameter covariance matrix is obtained:
\begin{equation}
\Sigma_{ab} = (\Gamma^{-1})_{ab},
\qquad
\sigma_a = \sqrt{\Sigma_{aa}},
\label{eq:fisher_sigma}
\end{equation}
where $\sigma_a$ is the $1\sigma$ marginalized uncertainty on $\theta_a$.
% The correlation coefficient between parameters $\theta_a$ and $\theta_b$ follows from $\rho_{ab} = \Sigma_{ab} / \sqrt{\Sigma_{aa}\,\Sigma_{bb}}$.

The covariance matrix used in the FIM calculation is identical to that employed in the likelihood (Eq.~\eqref{eq:loglike}), ensuring consistency between the FIM predictions and the full Bayesian results.

% To assess the statistical distribution of parameter uncertainties, we generate a Monte Carlo ensemble of 100 random MBHB realizations by varying the sky location $(\lambda, \beta)$, coalescence time $t_{\rm ref}$, aligned spins $(\chi_{1z}, \chi_{2z})$, inclination $\iota$, and reference phase $\varphi_{\rm ref}$.
% For each realization, the FIM is computed under block-diagonal and full covariance, and the ratios $\sigma_i^{\rm full} / \sigma_i^{\rm block}$ are collected to quantify the impact of foreground cross-correlations on parameter estimation precision.

\subsubsection{Data generation}
\label{subsubsec:data_generation}

\begin{table}[t]
\centering
\caption{\label{tab:signal_parameters}Parameterization of the two source classes. 
The upper subtable lists the MBHB parameters and the lower subtable is for the GB parameters.}
\textbf{MBHB parameterization}\par\vspace{2pt}
\begin{ruledtabular}
\begin{tabular}{ll}
\hline
Symbol & Parameter \\
\hline
$\mathcal{M}_c$ & Chirp mass (redshifted) \\
$q$ & Mass ratio \\
$\chi_{1z}$ & Primary spin (aligned) \\
$\chi_{2z}$ & Secondary spin (aligned) \\
$t_{\rm ref}$ & Reference time (set at coalescence) \\
$\varphi_{\rm ref}$ & Reference orbital phase \\
$d_L$ & Luminosity distance \\
$\iota$ & Inclination angle \\
$\lambda$ & Ecliptic longitude \\
$\beta$ & Ecliptic latitude \\
$\psi$ & Polarization angle \\
\hline
\end{tabular}
\end{ruledtabular}
\par\vspace{10pt}
\textbf{GB parameterization}\par\vspace{2pt}
\begin{ruledtabular}
\begin{tabular}{ll}
\hline
Symbol & Parameter \\
\hline
$f_0$ & Initial GW frequency \\
$\dot{f}_0$ & Initial GW frequency derivative \\
$A$ & Amplitude \\
$\lambda$ & Ecliptic longitude \\
$\beta$ & Ecliptic latitude \\
$\iota$ & Inclination angle \\
$\psi$ & Polarization angle \\
$\varphi_0$ & Initial GW phase \\
\hline
\end{tabular}
\end{ruledtabular}
\end{table}

Both source classes share a common framework for data generation, in which 
the  GW signals  are  computed from   source parameters using the  waveform and TDI response models given in Appendix~\ref{app:GB} and~\ref{app:MBHB},  
the instrumental noise in each TDI channel is drawn as Gaussian stochastic process   from the designed  PSD of Eq.~\eqref{eq:tdi_noise_psd} with the parameters in  Table~\ref{tab:orbits}, 
and  the confusion foreground is taken  directly from the residuals of the iterative subtraction described in  Sec.~\ref{sec:foreground_sim}, which retain the intrinsic cross-detector correlations.
Consequently, the simulated data used  for analysis are all in the frequency domain.

GBs and MBHBs  differ in their time-frequency structures. 
For a GB, the    analysis is carried out over the full 4-year  observation in a narrow frequency band around $f_0$, and the foreground data is  hence sliced onto that band directly in the frequency domain. 
The width of frequency band is set to  $\Delta f_{\rm band} = 8/(3.65\,\mathrm{day}) \approx 2.54 \times 10^{-5}$~Hz,  sufficient to encompass the width of target GB signals for all three detectors, where the duration  of  3.65 days corresponds to TianQin's orbital period around the Earth. 
While for an  MBHB, a 30-day duration is sufficient to capture the detectable network SNR for the vast majority of  systems. 
Therefore   the   analysis is carried out on a $30$-day time segment,  with the  coalescence placed near its end, and the foreground data  is mapped onto that segment by transforming the full  data  to the time domain, extracting the 30-day slice, and then transforming back to the frequency domain.

The GB signal is modeled as a quasi-monochromatic source  whose  frequency-domain TDI response is calculated using the  \texttt{Triangle-GB} code,   identical to Sec.~\ref{sec:foreground_sim} (see also Appendix~\ref{app:GB}). 
The MBHB signal is modeled according to Appendix~\ref{app:MBHB} using the    \texttt{Triangle-BBH} code. 
Specifically, we employ  the \texttt{IMRPhenomHM} waveform~\citep{London2018} with 6  harmonic modes $(\ell,m) \in \{(2,2), (3,3), (4,4), (2,1), (3,2), (4,3)\}$. 
The parameterizations of the two source classes adopted in the signal generation are summarized in Table~\ref{tab:signal_parameters}.

\subsubsection{Likelihood models}

The network log-likelihood  is given by the general expression of Eq.~\eqref{eq:loglike}.
For GBs, the narrowband nature of the signal  allows the full likelihood to be evaluated on the in-band  frequencies, with no further acceleration required.
For MBHBs, we adopt the network mode-by-mode heterodyned likelihood described in  Appendix~\ref{app:MBHB} to accelerate likelihood evaluation.

In the likelihood we compare  the two noise models described above (\textit{i.e.} block-diagonal and full covariance). 
For GBs, we use the foreground noise models estimated during the subtraction procedure. 
For MBHBs, by contrast, we re-estimate the foreground noise from the 30-day data segment using Welch's method. 
On this timescale,  the foreground correlation  are effectively stationary for the LISA-Taiji pair. 
While for the TianQin-Taiji/LISA pairs, when averaged over  the 30-day duration,  TianQin's  rapid  orbital motion averages out the correlations. 
Regarding the behavior of the foreground correlation under averaging over different observation times, our results and analysis are given in Appendix~\ref{app:foreground}.

\subsubsection{Posterior sampling}

Posterior sampling is performed with the ``Nested sampling with Artificial Intelligence'' (\texttt{nessai}) sampler~\citep{nessai,Williams:2021qyt,Williams:2023ppp}, which incorporates normalising flows to accelarate computationally expensive Bayesian analyses.
We adopt the \texttt{nessai} implementation integrated into the \texttt{bilby}  suite~\citep{bilby_paper,bilby_doi}. 
For both source classes, $N_{\rm live} = 1000$ live points are used with the stopping criterion $\Delta \ln \mathcal{Z} < 0.1$.

Owing to the inclusion of higher-order modes (for MBHBs) and joint observations with  multiple detectors, the multi-modality in the posterior distributions~\citep{Marsat:2020rtl} of source parameters is significantly alleviated. 
The remaining multi-modalities are  confined to the GB parameters $\varphi_0$ and $\psi$, which arise from the symmetry of the $(2,2)$ mode. 
Thus,  except for $\varphi_0$ and $\psi$,  we  safely adopt the  $10 \sigma$ ranges around the true values as the  priors to obtain a global characterization of the posteriors, with  the confidence intervals determined via FIM.
For  $\varphi_0$ and $\psi$, we use their full prior ranges.

To assess the impact of cross-detector foreground correlation, we apply a unified test framework to the two representative source classes.
All tests employ an  injection--recovery scheme: the simulated data always contain the full physical noise, and only the covariance model entering the likelihood varies (block-diagonal versus full-covariance), so that any difference between the two models can be attributed solely to the omitted foreground correlations.
We quantify the effect of the different noise models  using   Bayesian posteriors of selected single source, and  use P-P plots to test statistical biases in parameter estimation.

%=====================================================================
\section{Results and Discussion}
\label{sec:results_discussion}
%=====================================================================

\subsection{Foreground spectra}
\label{subsec:foreground_spectra_result}

\begin{figure*}[t]
\centering
\includegraphics[width=\textwidth]{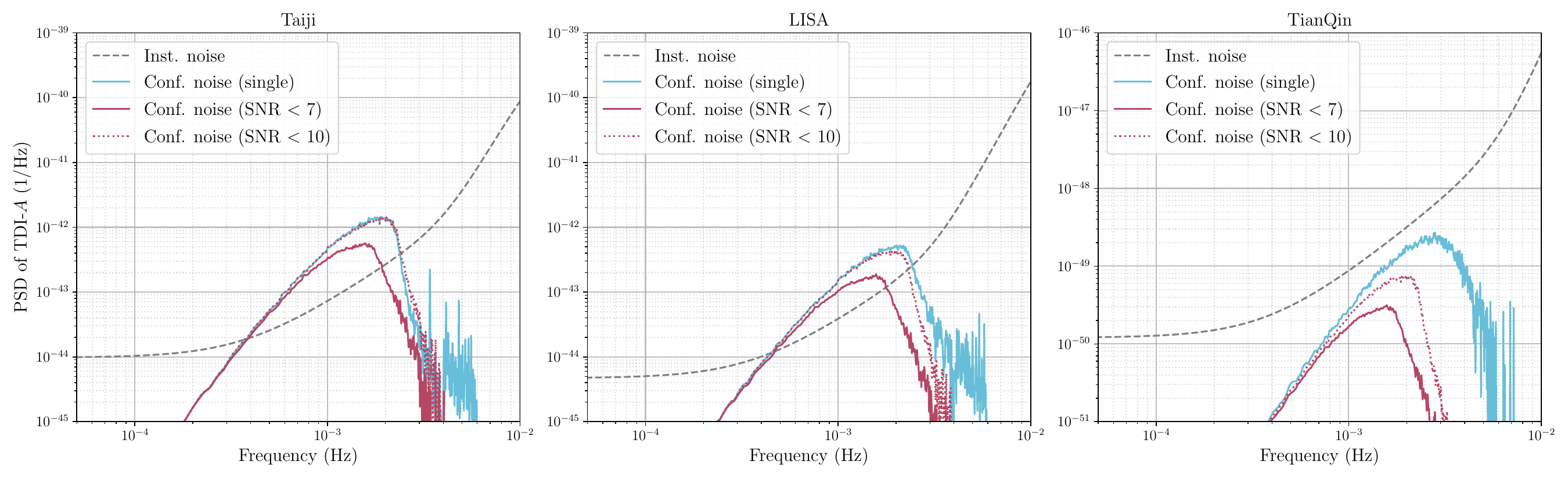}
\caption{
Four-year time-averaged noise PSDs in the TDI-\(A\) channel for Taiji (left), LISA (middle), and TianQin (right).
The grey dashed curves show the instrumental noise PSDs.
The blue solid curves show the residual foreground PSDs after  single-detector subtraction with \(\rho_{\rm th,single}=7\).
The red solid and red dotted curves show the residual foreground PSDs after network subtraction with \(\rho_{\rm th,network}=7\) and \(10\), respectively.}
\label{fig:psd_comparison}
\end{figure*}

\begin{figure*}[htbp]
\centering
\includegraphics[width=\textwidth]{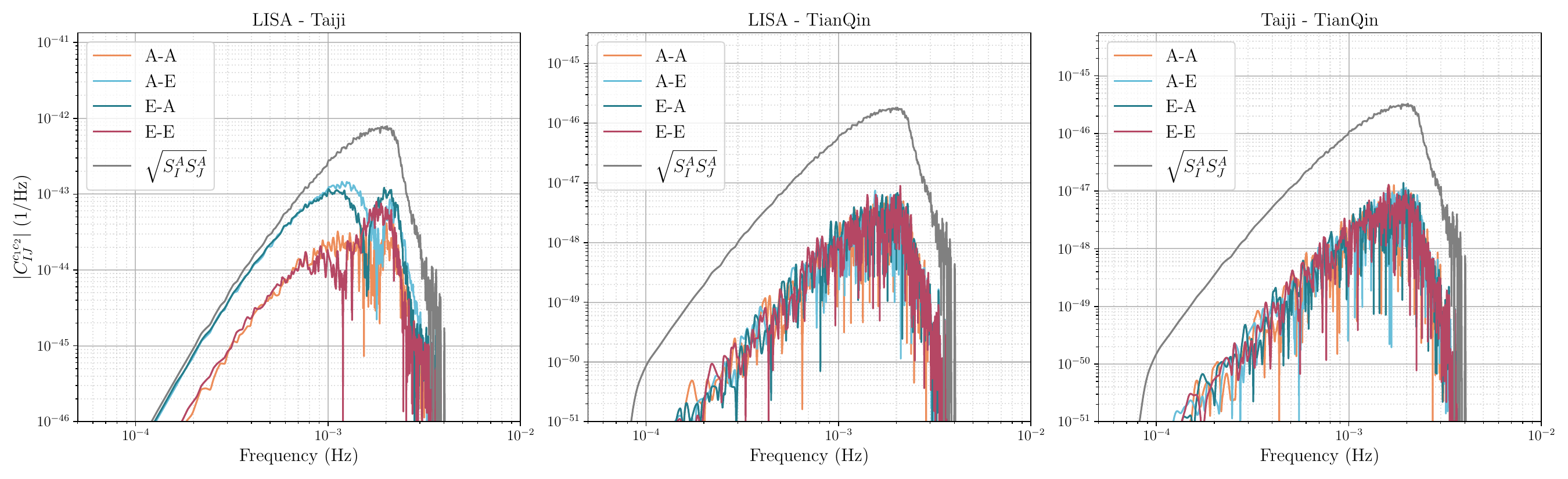}
\caption{
Four-year time-averaged magnitudes of the cross-detector foreground CSDs after network subtraction with \(\rho_{\rm th,network}=10\).
The left, middle, and right panels correspond to the LISA--Taiji, LISA--TianQin, and Taiji--TianQin  pairs, respectively.
The orange, blue, green, and red curves show the \(A\)-\(A\), \(A\)-\(E\), \(E\)-\(A\), and \(E\)-\(E\) channel combinations, where the first and second channel labels refer to the first and second detectors named in each panel.
The grey curves show  \(\sqrt{S_I^A S_J^A}\) of the foreground PSDs as  references.}
\label{fig:csd_comparison}
\end{figure*}

\begin{figure*}[htbp]
\centering
\includegraphics[width=0.7\textwidth]{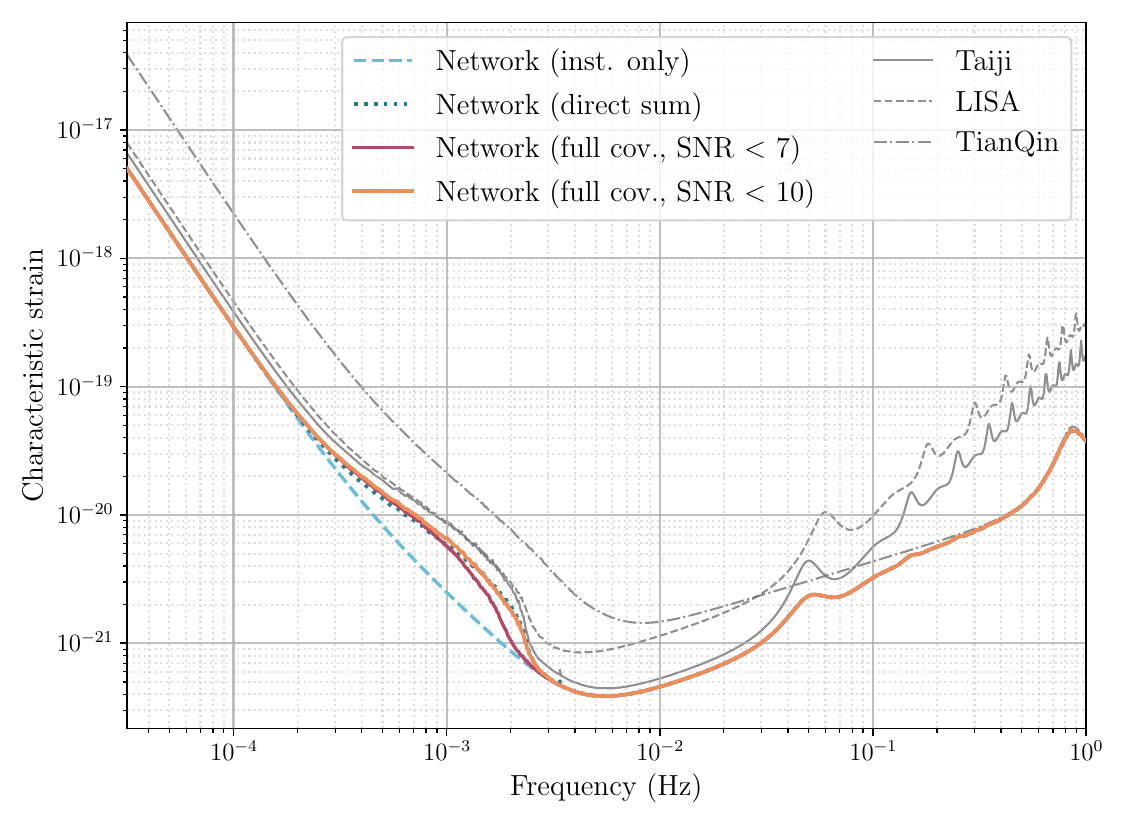}
\caption{
Sky position, polarization, and mission time averaged characteristic strain sensitivities of the LISA--Taiji--TianQin network under different settings. 
The blue dashed curve accounts for  instrumental noise only.
The green dotted curve is the direct sum of the single-detector sensitivities, including the residual foreground obtained via independent subtraction in each detector with \(\rho_{\rm th,single}=7\).
The red and orange solid curves correspond to  the  network sensitivities that use the residual foregrounds obtained with \(\rho_{\rm th,network}=7\) and \(10\), respectively, with cross-detector correlations incorporated. 
The thin grey solid, dashed, and dash-dotted curves show the individual sensitivities of Taiji, LISA, and TianQin, respectively, accounting for both  instrumental and foreground noises.
}
\label{fig:sensitivity}
\end{figure*}

Figure~\ref{fig:psd_comparison} compares the confusion foreground PSD of the TDI-$A$ channel for the three detectors, with Taiji, LISA, and TianQin shown from left to right.
In each panel, the grey dashed curve is the instrumental noise PSD, the red solid and dotted curves are the confusion foregrounds obtained with network SNR thresholds of $\rho_{\rm th,network} = 7$ and $\rho_{\rm th,network} = 10$, respectively, and the blue solid curve represents the confusion foreground obtained via applying the iterative subtraction procedure to a  single detector with  SNR threshold $\rho_{\rm th, single} = 7$ (as  adopted by \citealp{Liu2023,Karnesis:2021tsh}).
As common characteristics shared by all the resulting foregrounds, 
for Taiji and LISA, the foreground exceeds instrumental noise by roughly one order of magnitude at $f\sim1$~mHz, while  for TianQin, by contrast, the foreground is weaker than the instrumental noise in the whole frequency band.
Above $\sim4$~mHz, the foreground is dominated by instrumental noise in all cases.
Comparing the   single-detector and network results,  
the foreground amplitude of the latter is generally lower than that of the former, with the  only  exception shown in the left panel. 
Since the low-frequency sensitivity of the network is dominated by LISA and Taiji (see Figure~\ref{fig:sensitivity}), the following  qualitative interpretation can be made. 
If LISA and Taiji have  identical sensitivities,  the residual  foreground  from network subtraction with $\rho_{\rm th,network}=10$ would be equivalent to that from single-detector subtraction with  $\rho_{\rm th,single}=7$, for  $10 \approx 7\sqrt{2}$. 
However, in practice, Taiji's  low-frequency sensitivity is slightly better than LISA's due to its longer arm lengths. 
Consequently, in the Taiji panel, the  single-detector result  lies slightly below the network result with  $\rho_{\rm th,network}=10$, while the opposite holds for the LISA panel.
Besides, as expected,  the foreground amplitude  is naturally larger for a network SNR threshold of 10 than for 7. 
This paper does  not aim to prescribe an optimal threshold for bright GB subtraction, but only present results for different choices. 
For the following discussion, we adopt the foreground obtained with $\rho_{\rm th,network}=10$, so as to fully capture the foreground impact on subsequent  analysis.

In Figure~\ref{fig:csd_comparison}, we display the CSD magnitudes $|C_{IJ}^{c_1 c_2}(f)|$ for all three detector pairs and all four TDI channel combinations. 
From left to right, the LISA-Taiji, LISA-TianQin, Taiji-TianQin pairs are shown, 
and different  channel combinations are distinguished by color: orange, blue, green, and red denote the $A$-$A$, $A$-$E$, $E$-$A$, and $E$-$E$ combinations, respectively.
To illustrate   the magnitude of  correlation, we also show  $\sqrt{S_I^A S_J^A}$ with   grey curve as a reference in each panel.
The LISA-Taiji pair exhibits significant cross-correlation up to $f\sim1$~mHz, while the pairs involving TianQin exhibit negligible correlations over the whole four-year observation, for which a    detailed explanation is provided in the Appendix~\ref{app:foreground}. 
One might notice that the cross-channel correlations between LISA and Taiji appear larger than the same-channel ones. 
We argue, however, that this is an effect due to specific  initial phase of the  constellation. 
As noted in \cite{Romano:2016dpx}, the $E$ channel can be regarded as the $A$ channel rotated by $45^\circ$. 
Consequently, a rotation of the entire constellation within its plane would alter this comparative relation.

It should be noticed  that the PSDs and CSDs discussed above are the time-averaged  spectra over the full 4-year observation using the Welch method.
Nevertheless, the foreground is  not  stationary. 
As the relative positions of the detectors evolve and the antenna patterns sweep across the anisotropic Galactic sky, the foreground spectra are intrinsically time-dependent. 
This time-dependent  behavior is systematicly investigated in Appendix~\ref{app:foreground}.
The time evolution of the coherence $\gamma_{IJ}^{cc'}(f)$ at $f=1$~mHz is shown for  different detector pairs, where   the results computed from the simulated GB foreground data are compared with the theoretical  model of Sec.~\ref{sec:foreground}, and their  agreement provides a double validation of the foreground model.
The low correlation of TianQin with the other detectors is also explained there. 
Since the target GW  sources considered here (GBs and MBHBs) accumulate their SNR over timescales of  months to years, TianQin's fast geocentric orbital motion averages out its cross-correlation with the heliocentric detectors over these  periods of time.

To provide an intuitive view of the detection capability for GW  sources,  Figure~\ref{fig:sensitivity}  presents  the LISA-Taiji-TianQin network  sensitivity curves averaged over sky location, polarization angle,  and the whole  observation time,  for different  foreground  scenarios.
Notice that since $T$ channel is not ``null'' at high frequencies above 0.1 Hz~\citep{LT_arm_inequality2}, the curves  shown here   are the  total sensitivities  of $\{A, E, T\}$ channels, even though elsewhere in this paper we  consider only  $\{A, E\}$, for the sensitive  band of $T$ is not relevant in those cases.
The instrumental-noise-only sensitivity (blue dashed) is calculated considering only $\mathbf{C}_{\rm inst}$ in Eq.~\eqref{eq:noise_cov_decomposition}.
For the ``direct sum'' case (green dotted),  We compute the sensitivity for each detector using its respective foreground PSD (setting $\rho_{\rm th, single}=7$), and then combined them as $S_{\mathcal{N}} = 1 / \sum_I 1 / S_I$  
to obtain the overall sensitivity. 
This is equivalent to  assuming  that the three detectors are not analyzed jointly for GB subtraction, and that  foreground correlation are neglected  during  GW source  detection, which corresponds to the common assumption adopted in previous works.
The two solid curves are calculated with the rigorous formalism Eq.~\eqref{eq:sensitivity_net}, with red and orange colors corresponding to  $\rho_{\rm th,network}=7$  and $\rho_{\rm th,network}=10$, respectively.
Comparing the ``direct sum'' result  with that of $\rho_{\rm th,network}=10$, the figure shows that, in terms of the average sensitivity, there is no appreciable difference whether the foreground correlation is taken into account. 
For  reference, we also plot the single-detector sensitivities for LISA, Taiji and TianQin  with  thin grey  curves, and  clear improvement in detection capability from the single-detector case to the network is observed.

\subsection{Impact on MBHB analysis}
\label{sec:mbhb_results}

\begin{figure*}[t]
\centering
\includegraphics[width=\textwidth]{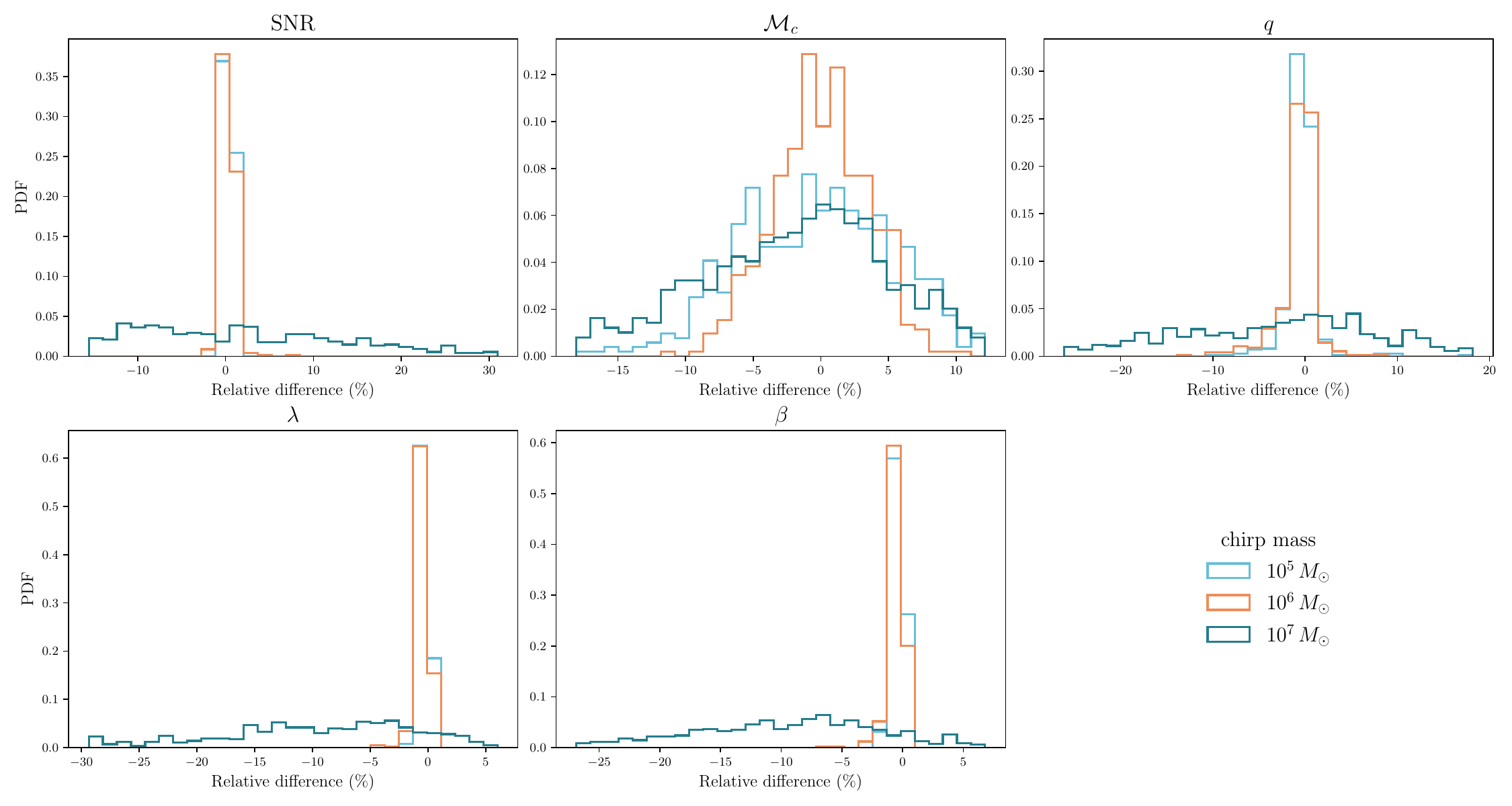}
\caption{
Normalized distributions  of 
\((X_{\rm full}-X_{\rm block})/X_{\rm block}\) ($X \in \{{\rm SNR, \sigma_{\mathcal{M_c}}, \sigma_{q}, \sigma_{\lambda}}, \sigma_{\beta}\}$), obtained from 500 MBHB realizations for each  chirp mass value in  $\mathcal{M}_c \in \{10^5, 10^6, 10^7\} M_\odot$.
% From left to right, the top row shows the network SNR \(\rho\), the chirp-mass uncertainty \(\sigma_{\mathcal{M}_c}\), and the mass-ratio uncertainty \(\sigma_q\); the bottom row shows the sky-location uncertainties \(\sigma_\lambda\) and \(\sigma_\beta\), with the shared legend in the final panel.
Each panel shows the relative deviation in  SNR or parameter uncertainty, with  the blue, orange, and green   histograms correspond to 
\(\mathcal{M}_c=10^5\,M_\odot\), \(10^6\,M_\odot\), and \(10^7\,M_\odot\), respectively.
% Here \(X=\rho\) in the SNR panel and \(X=\sigma_{\theta}\) in each parameter-uncertainty panel.
}
\label{fig:snr_mass_trend}
\end{figure*}

We compare the two noise models (\textit{i.e.} block-diagonal and full covariance) across three complementary dimensions: SNR,   FIM parameter uncertainty, and Bayesian posterior bias under noise model misspecification.

To quantify the impacts on SNR and FIM uncertainty, we perform Monte Carlo simulations for three representative chirp mass values  $\mathcal{M}_c \in \{ 10^5, 10^6, 10^7 \}\,M_\odot$. 
For each mass,  we generate 500 random source realizations at a fixed luminosity distance of $d_L = 26$~Gpc ($z\approx 3$, the  specific value of $d_L$ does not matter as it only acts as a global amplitude scaling and we are concerned solely with relative differences). 
The remaining parameters are drawn as follows: the mass ratio $q$, the aligned spins $\chi_{1z / 2z}$ are drawn from uniform distributions over $[0.1,0.99]$, $[-0.9,0.9]$, the orientation  and the sky location  of source follow isotropic distributions, 
and the reference phase $\varphi_{\rm ref}$, the polarization angle $\psi$ are uniform over $[0, 2\pi], [0, \pi]$, respectively. 
The coalescence time $t_{\rm ref}$ is drawn uniformly over the 4-year observation, with the 30-day analysis segment containing it placed fully inside the data window. 
For each realization, the network SNR and FIM  are  computed under both noise models using the identical signal. 
Figure~\ref{fig:snr_mass_trend} displays the distributions of the relative differences between two noise models $(X_{\rm full}-X_{\rm block})/X_{\rm block}$, with $X$ representing SNR or the uncertainty of a specific  parameter.

Figure~\ref{fig:snr_mass_trend} presents the histograms of relative differences  for the network SNR and four   key parameters $\{\mathcal{M}_c, q, \lambda, \beta\}$. 
All the distributions contain both positive and negative values,  
and for the SNR and  all parameters except $\mathcal{M}_c$,  the results for the  three masses follow a consistent   trend, with the high-mass $\mathcal{M}_c = 10^7\,M_\odot$ case  clearly distinct from the other two. 
Spedifically, for $\mathcal{M}_c = 10^5$ and $10^6\,M_\odot$,  the relative differences   stay at the level of a few percent, whereas for $\mathcal{M}_c = 10^7\,M_\odot$ they can reach up to  $30\%$. 
The  distribution profiles of  parameter uncertainties generally follows that of the network SNR. 
As For $\mathcal{M}_c$, by contrast, the distributions are similar across all  three masses.

%% Figures 6 and 7 are two wide (figure*) floats that are cited back to back and
%% are meant to share one page: "!p" = float page only ("!" also waives the
%% minimum-fill rule of \dblfloatpagefraction), so LaTeX collects both of them on
%% the same dedicated page instead of splitting them over two float pages.
\begin{figure*}[!p]
\centering
\includegraphics[width=0.48\textwidth]{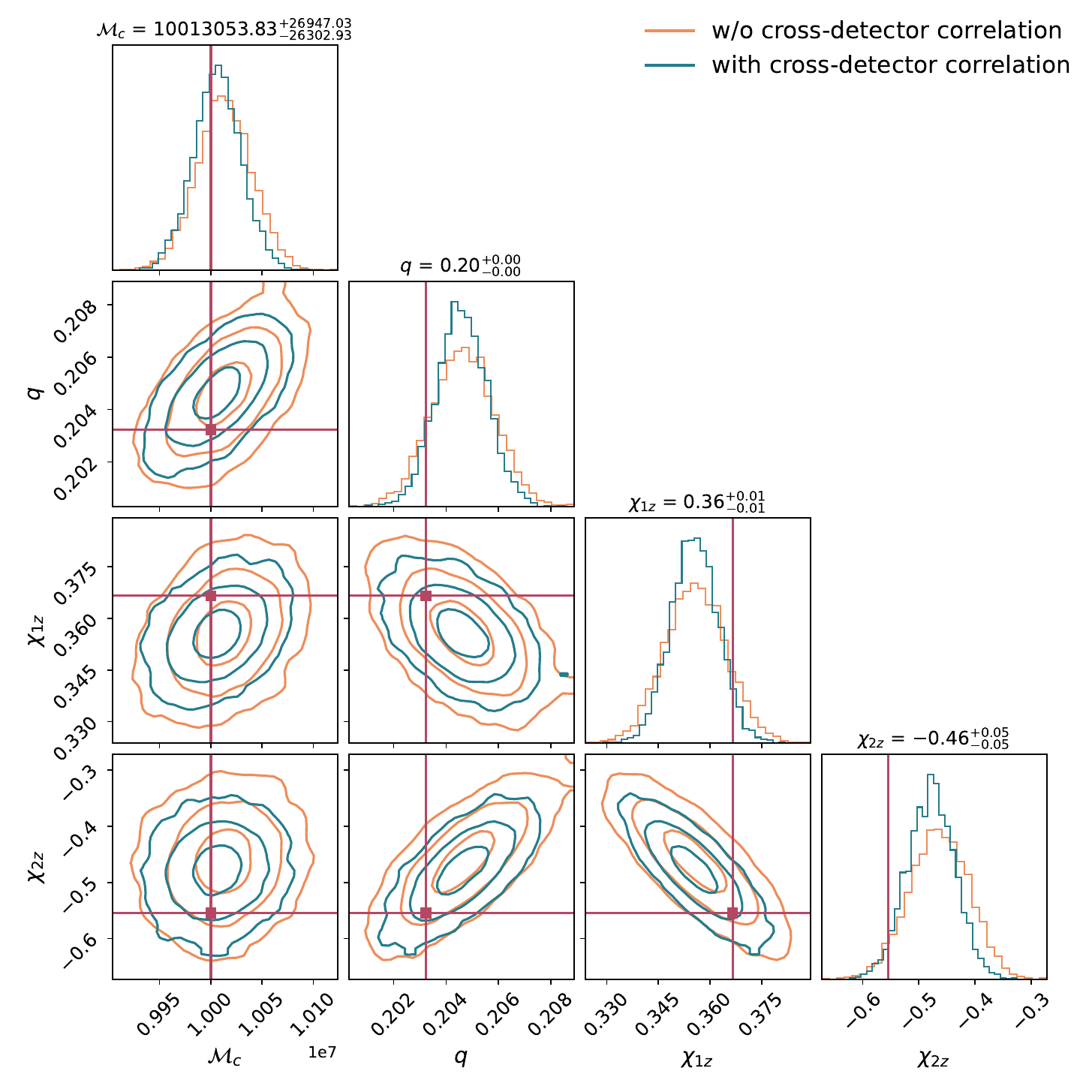}
\hfill
\includegraphics[width=0.48\textwidth]{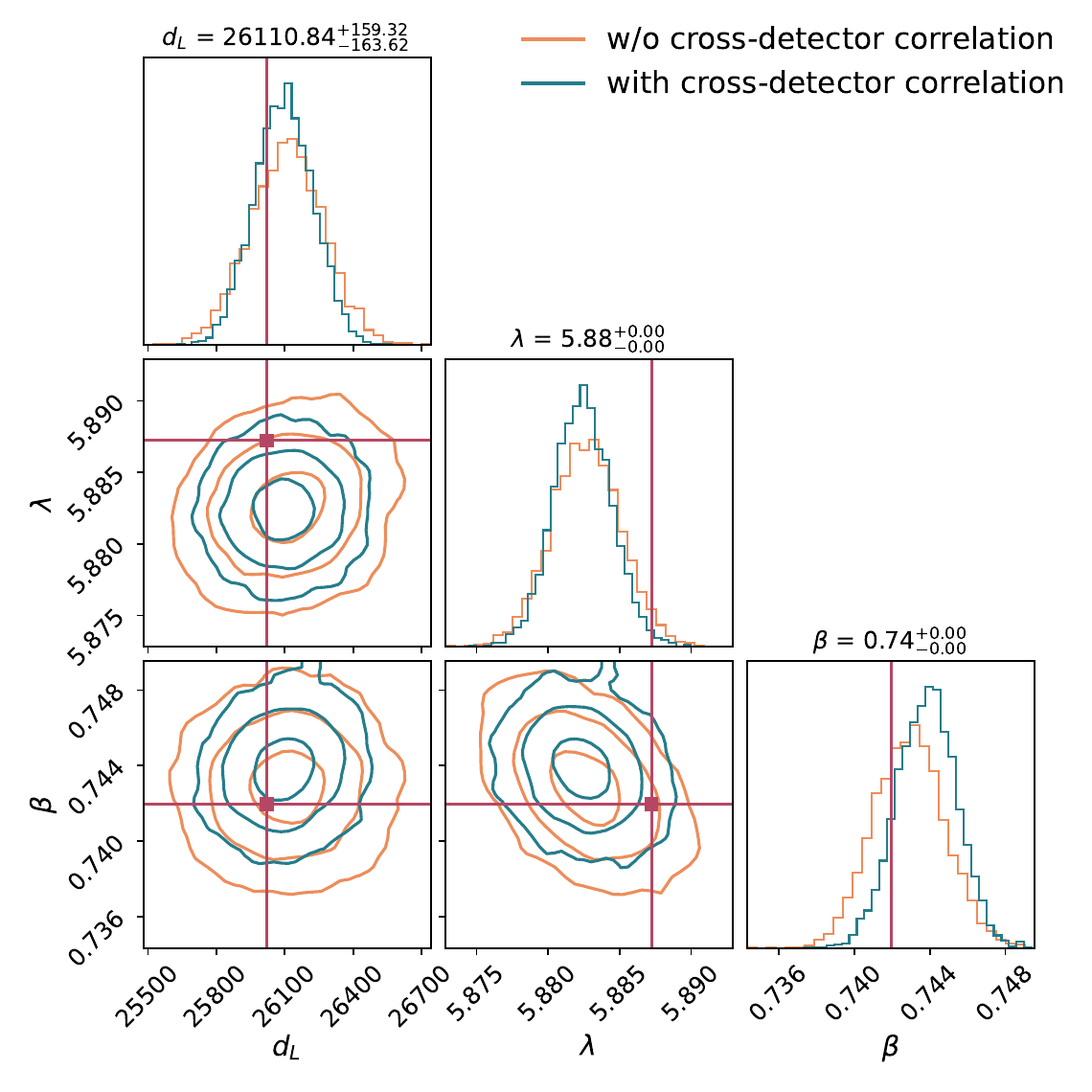}
\caption{
Posterior distributions for the selected \(\mathcal{M}_c=10^7\,M_\odot\) MBHB. 
The source is chosen since  it has  the largest relative SNR difference between the two noise models in the Monte Carlo ensemble.
The left panel shows the intrinsic parameters \(\mathcal{M}_c\), \(q\), \(\chi_{1z}\), and \(\chi_{2z}\), and the right panel shows the 3D location parameters including  luminosity distance \(d_L\) and sky coordinates \(\lambda\) and \(\beta\).
Orange and green corner plots  denote the block-diagonal and full-covariance noise models, respectively,  and  red lines mark the injected values.
}
\label{fig:posterior_mbhb}
\end{figure*}

\begin{figure*}[!p]
\centering
\includegraphics[width=0.85\textwidth]{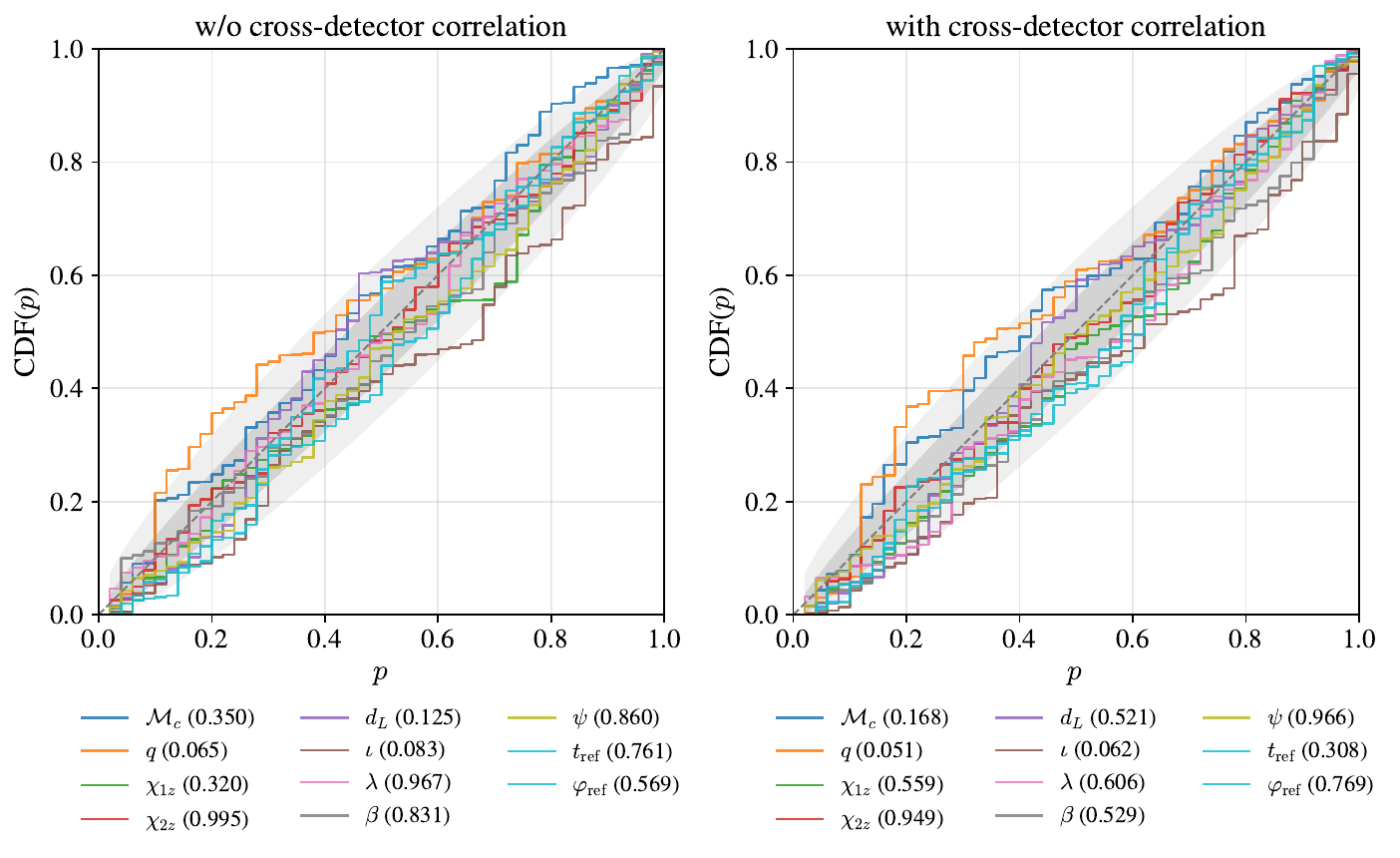}
\caption{
P--P plots for   MBHB parameter estimation from 50 random injections, performed with the  block-diagonal noise model (left) and the   full-covariance noise model (right).
Each colored  curve is the empirical cumulative distribution for one source parameter, with  
the dark- and light-grey bands indicating  the \(1\sigma\) and \(2\sigma\) regions, respectively.  
The numbers  in parentheses in the legends are the Kolmogorov--Smirnov \(p\)-values.
}
\label{fig:pp_mbhb}
\end{figure*}

These features follow directly from the theoretical analysis of Sec.~\ref{sec:snr_impact}, where we predict  that higher-mass MBHBs are more sensitive to foreground correlations. 
Since MBHB signals  are transient compared to the continuous  GB signals, the suppression in the effect of foreground  correlation due to  time averaging is relatively weak. 
We therefore attribute the correlation imprints primarily to   the frequency structure of signals.
Three  ingredients  account for the observed distributions. 
First, the leading  term in Eq.~\eqref{eq:dsnr_expanded} carries the sign of  \(\eta\), which depends on  the signal's inter-detector  phase difference $\Delta \varphi$ and the coherence phase $\theta_\gamma$. 
Since the Monte Carlo ensemble samples random sky locations and coalescence times, which determines  the  relative  positons and attitudes of detectors,  both $\Delta \varphi$ and  $\theta_\gamma$ (see Figure~\ref{fig:app_lisa_taiji_phase} for the time-dependence of $\theta_\gamma$ for orbit Config.~I) vary across realizations. 
Therefore,  \(\eta\), and hence $\Delta \rho$,  can take  both signs. 
Second, the magnitude of correlation  imprint is largely   determined by whether the signal's  frequency band  lies in the   band with significant foreground  correlation.
For \(\mathcal{M}_c = 10^7\,M_\odot\), the majority of signals, including the inspiral-merger-ringdown phases, are  accumulated  at millihertz and below, with the merger frequency  at \(\sim 0.1\)–\(0.2\) mHz. 
The entire signal  therefore falls inside the correlated regime, and for favorable sky alignments \(|\eta|\) approaches unity, allowing the fractional SNR difference to reach  \(30\%\). 
Whereas For \(\mathcal{M}_c = 10^5\) and \(10^6\,M_\odot\), since the  SNRs of MBHBs are dominated by the merger phase which  generally  lie outside the correlated  regime, thus   the deviations remain at the few-percent level.
Third, for the distribtion of parameter uncertainty difference,   \(\mathcal{M}_c\) stands out as the exception to the mass-dependent behavior. 
The constraint on $\mathcal{M}_c$ is dominated by the phase accumulation during inspiral, and  for all masses the corresponding frequencies lie in the low-frequency range  where the foreground  has significant correlation. 
The  imprint on \(\sigma_{\mathcal{M}_c}\) is therefore approximately  set by the same coherence level  regardless of the signal's total bandwidth, making it effectively mass-independent. 
By contrast,  parameters that are  constrained by the  information spread across the full signal  band should inherit the distribution profile of SNR,  and reproduce the same mass-dependent trend observed for the network SNR.

To assess whether neglecting foreground correlations would lead to biases in actual parameter estimation,  we first perform  Bayesian inference for the most extreme case in   the Monte Carlo ensemble, which is  an \(\mathcal{M}_c = 10^7\,M_\odot\) source  whose parameters are selected to maximize the relative SNR difference between the two noise  models. 
For this selected source, the network SNR is  $345$ under the block-diagonal model and $464$ under full covariance, yielding \(|\Delta\rho|/\rho_{\rm block} \simeq 35\%\).  
Bayesian posterior inference is performed under both noise models using identical simulated data.
Parameter estimation is carried out on the full 11-dimensional parameter space, and Figure~\ref{fig:posterior_mbhb} compares the resulting posteriors for both intrinsic (left panel) and localization parameters (right panel). 
Consistent with the SNR hierarchy (\(\rho_{\rm full} > \rho_{\rm block}\)), the  credible intervals under full-covariance model  are  narrower for all  parameters at  the  $\mathcal{O}(10 \%)$ level.  
Despite this, no parameter bias is statistically significant for this extreme source, which suggests that even smaller shifts would be expected for typical MBHB sources.

To validate our conclusions and inferences on a broader population, we repeat the injection–estimation  analysis on 50 MBHB realizations randomly  drawn from broad  ranges and construct P–P plots  under the two noise  models. 
For this test, 
all other parameters are drawn from the same distributions as the SNR and FIM  test above, the only differences being that $\mathcal{M}_c$ is now generated over \(10^5\)–\(10^7\,M_\odot\), and $d_L$ is drawn from luminosity  distances corresponding to  \(z\simeq1\)–\(10\).
As can be seen in Figure~\ref{fig:pp_mbhb}, the left and right panels show the P–P plots under the block-diagonal  and full-covariance  noise models, respectively, with one colored empirical CDF curve per parameter.
For both noise models, the empirical CDF curves  of all parameters lie within the \(1\sigma\)–\(2\sigma\) bands of the diagonal, with all Kolmogorov–Smirnov \(p\)-values~\citep{Karson01091968} above 0.05, showing   no evidence of  systematic bias for each parameter  under either model~\citep{Wong_2023}. 
% therefore in this sense  the two models are statistically indistinguishable at current sample size and parameter ranges. 
The absence of  appreciable population-level bias confirms that, for general MBHBs, neglecting cross-detector foreground correlations does not introduce significant systematics in parameter estimation.
As a conclusion,  whether to include the cross-detector foreground  correlation   therefore mainly  depends on the specific analysis task, \textit{i.e.} whether the at  most $\mathcal{O}(10\%)$-level difference in  the widths of  credible intervals would affect the scientific objective.

\subsection{Impact on GB analysis}
\label{sec:gb_analysis}

%% Layout: the wide GB scatter figure spans the top of the page (figure*[t]);
%% the single-column GB posterior figure then sits at the top of the left column
%% directly below it (figure[t]), with body text continuing beside and below.
%% Constraint: with the wide figure at 0.9\textwidth the top-float allowance of
%% the reduced column is ~250pt, so the column figure must stay at or below
%% 0.75\columnwidth (0.80 and larger gets deferred to the next page).
\begin{figure*}[t]
\centering
\includegraphics[width=0.9\textwidth]{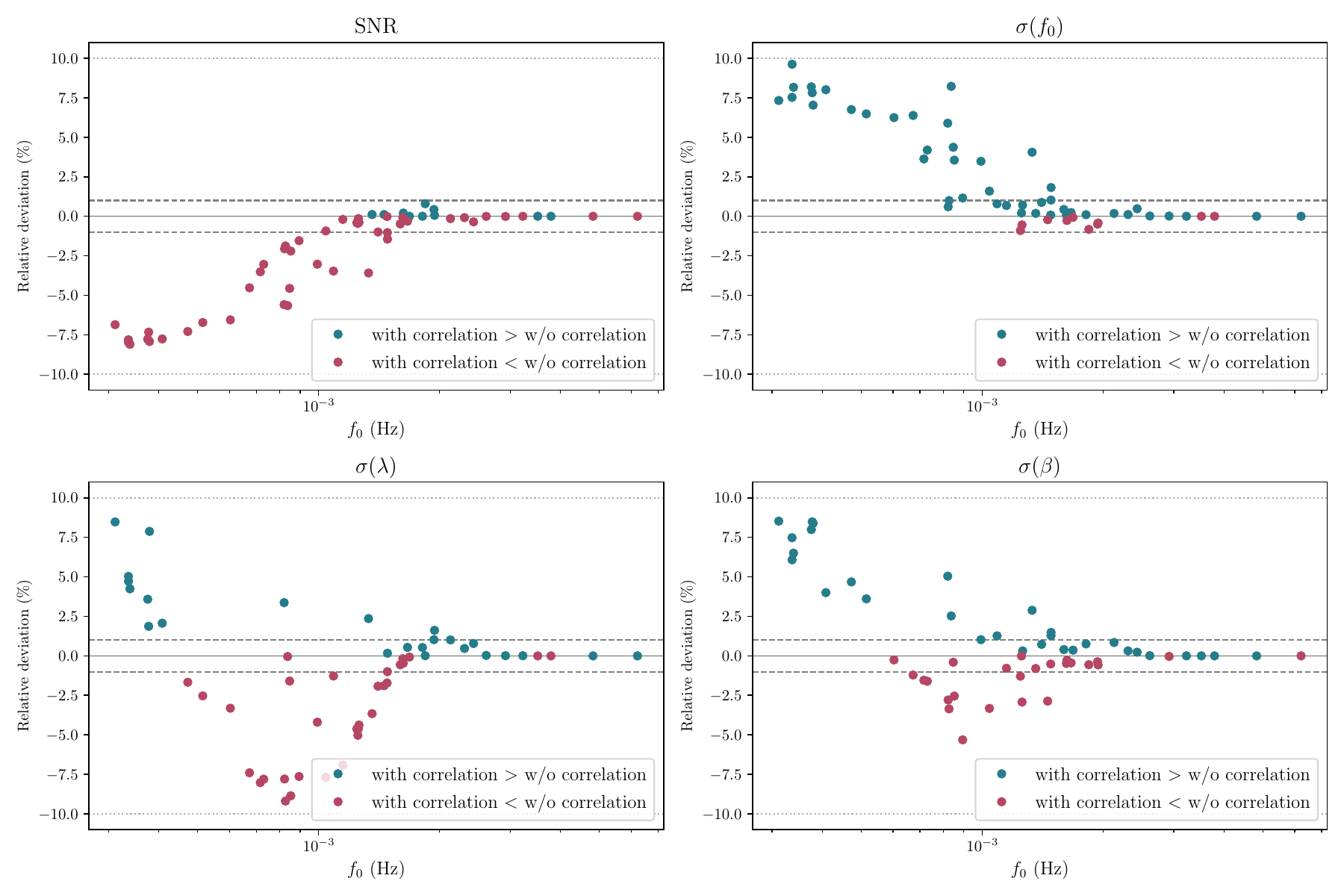}
\caption{
Scatter plots for \((X_{\rm full}-X_{\rm block})/X_{\rm block}\)  (\(X\in\{{\rm SNR}, \sigma_{f_0}, \sigma_{\lambda}, \sigma_{\beta}\}\)),   obtained from  the    catalogue of  55 VGBs,  and plotted as  functions  of the sources'  \(f_0\).
Green  and red markers denote positive and negative differences, respectively. 
The grey solid line marks zero difference, while the grey dashed and dotted lines indicate the \(\pm1\%\) and \(\pm10\%\) levels.}
\label{fig:gb_snr_comparison}
\end{figure*}

We likewise  compare the   two noise models  in terms of network SNR, FIM parameter uncertainty, and  posterior distributions.  
For test cases, rather than randomly generating the source parameters,   we  adopt  the verification Galactic binary (VGB) catalogue reported in \cite{Kupfer:2023nqx},  which includes 55  nearby compact binaries that have already been identified through electromagnetic observations.

Each panel of Figure~\ref{fig:gb_snr_comparison} presents $(X_{\rm full}-X_{\rm block})/X_{\rm block}$ as a function of the sources' initial GW  frequency $f_0$, with  $X$ representing  the network SNR or the  FIM parameter uncertainty.
The four panels show, from upper left to lower right, the relative  deviations of SNR, $\sigma_{f_0}$, $\sigma_{\lambda}$, and $\sigma_{\beta}$, respectively. 
In each panel, every VGB in the catalogue is plotted as a colored dot  at $f_0$. 
Green dots stands for  sources for which the full covariance yields a larger value than the block-diagonal model (positive deviation), while red dots mark the opposite (negative deviation).
The grey  solid line marks zero deviation, and the grey dashed and dotted lines mark the $\pm1\%$- and $\pm10\%$-level deviations, respectively. 
Two features stand out in Figure~\ref{fig:gb_snr_comparison}. 
First, the deviations are primarily concentrated at the lowest frequencies: at $f_0 \lesssim 2 \ {\rm mHz}$, where the cross-detector  foreground correlations are comparison to the foreground PSDs, as is  seen in Figure~\ref{fig:csd_comparison}. 
Second, the deviations take both positive and negative values, and the deviations of different $X$s do not point in the same direction. 
At the low frequencies the SNR deviations are predominantly negative, namely the full covariance reduces the network SNR, 
Consistently,   most of   the $\sigma_{f_0}$ deviations are  positive and nearly mirror the SNR deviations. 
% While the sky-location uncertainties $\sigma_{\lambda}, \sigma_{\beta}$ 
% scatter on both sides of zero without a preferred sign in the $f_0 \lesssim 2 \ {\rm mHz}$ range. 
However, the behavior of the sky-location uncertainties $\sigma_{\lambda}, \sigma_{\beta}$ does not fully mirror that of the SNR. 
Although they exhibit an  increase at frequencies below 0.3mHz, a decrease is observed  around 1mHz.

The relative deviations in  SNR, $\sigma_{f_0}$ for all VGBs, as well as $\sigma_{\lambda}$, $\sigma_\beta$ for the high- and low-frequency VGBs  follows directly from our theoretical expectations in  Sec.~\ref{sec:snr_impact}. 
The frequency band where these quantities show significant differences across the two noise models coincides with the region where the foreground correlation is most pronounced. 
For these cases, the results are  consistent with the general inverse relationship between SNR and parameter uncertainty. 
More intriguing  is the behavior of $\sigma_\lambda$, $\sigma_\beta$ around 1 mHz, where 
they deviate from  the trend of $\sigma_{f_0}$ (or 1/SNR). 
According to the definition of FIM in Eq.~\eqref{eq:fisher}, 
this descrepancy  can be understood by extending  Eq.~\eqref{eq:dsnr2_rel} from the signal itself to its parameter  derivatives. 
A comparative analysis between  $f_0$ and $\lambda, \beta$  is as follows. 
In terms of  $f_0$, the derivative of a quasi-monochromatic signal is approximately 
$ \partial_{f_0}h_I \simeq i \pi T_{\rm obs} h_I = \pi T_{\rm obs} h_I e^{i\pi / 2}$,
with  $h_I$ representing  the frequency-domain response of detector $I$.  
Because this scaling factor is common to all detectors,  
both the amplitude term and phase term of Eq.~\eqref{eq:dsnr2_rel} for $\Gamma_{f_0 f_0}$ are approximately the same as for SNR. 
Consequently, the change in \(\sigma_{f_0}\) generally mirrors the change in SNR. 
In contrast, for \(\{\lambda,\beta\}\), the derivatives further contain the contributions from Doppler modulations introduced via the $2 \pi f \hat{\mathbf{k}} \cdot \mathbf{R}_I / c$ term. 
Owing to the spatial separation between the detectors $(\mathbf{R}_I \neq \mathbf{R}_J)$, this term modifies  $\eta$,   hence decoupling $\sigma_\lambda, \sigma_\beta$ from the trend of $\sigma_{f_0}$ and 1/SNR. 
Below \(f_0\sim0.3\,{\rm mHz}\), however, this Doppler-induced effect is well suppressed as it scales linearly with $f$, and  the sky-location uncertainties again approximately mirror the  SNR. 
Additionally, the relative deviations in both SNR and Fisher uncertainties are generally  smaller for GBs than for MBHBs, confirming  that the  imprint correlation can be  effectively mitigated by time averaging.

\begin{figure}[t]
\centering 
% \vspace{0.1cm}
\includegraphics[width=0.9\columnwidth]{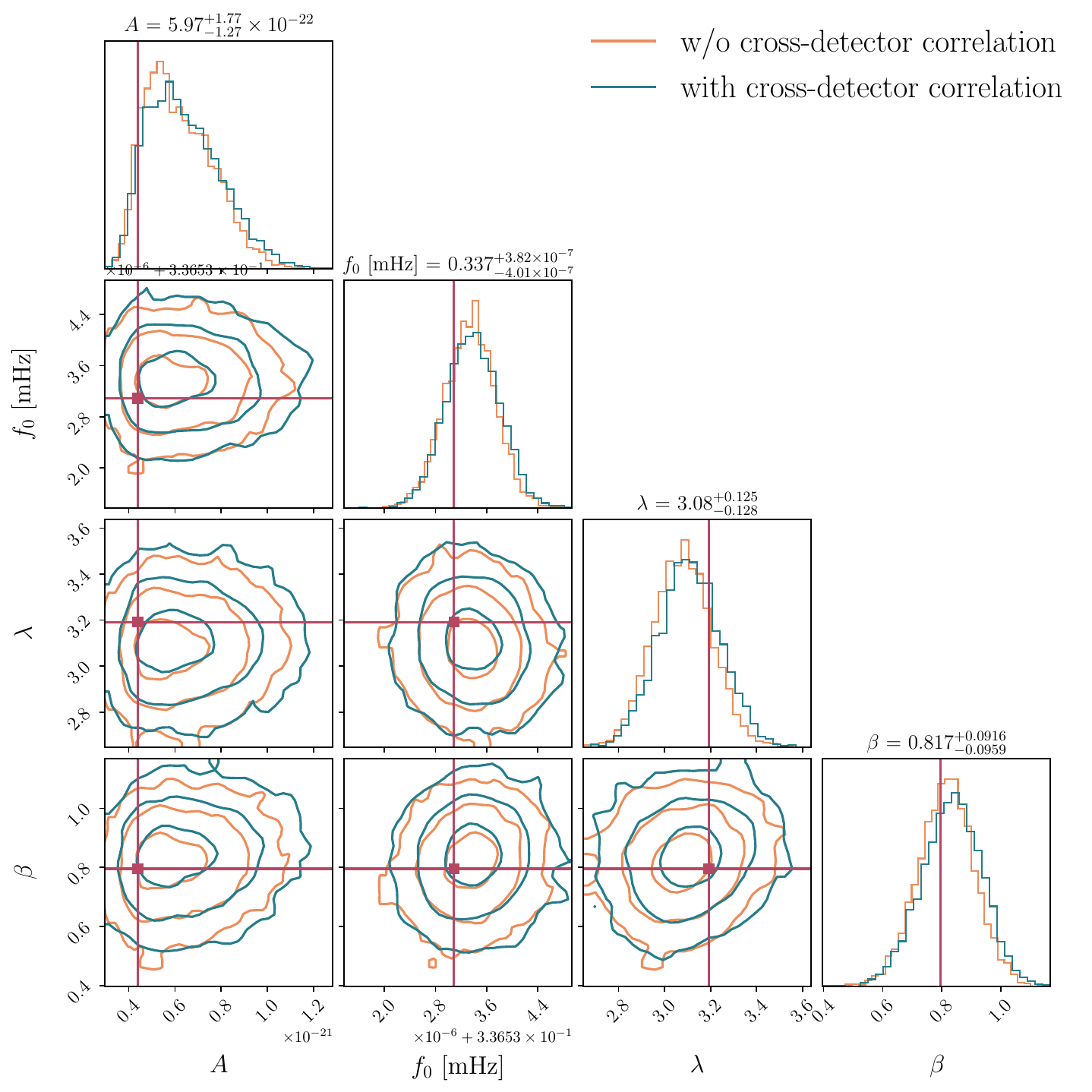}
\caption{
Posterior distributions of  four key parameters for   the  39$^{\rm th}$  VGB.  
The source   is selected  because  it exhibits  the largest relative SNR difference between the two noise models.
Orange and green corner plots  denote the block-diagonal and full-covariance noise models, respectively,   and  red lines mark the injected values.
}
\label{fig:gb_posterior}
\end{figure}

\begin{figure*}[t]
\centering
\includegraphics[width=0.85\textwidth]{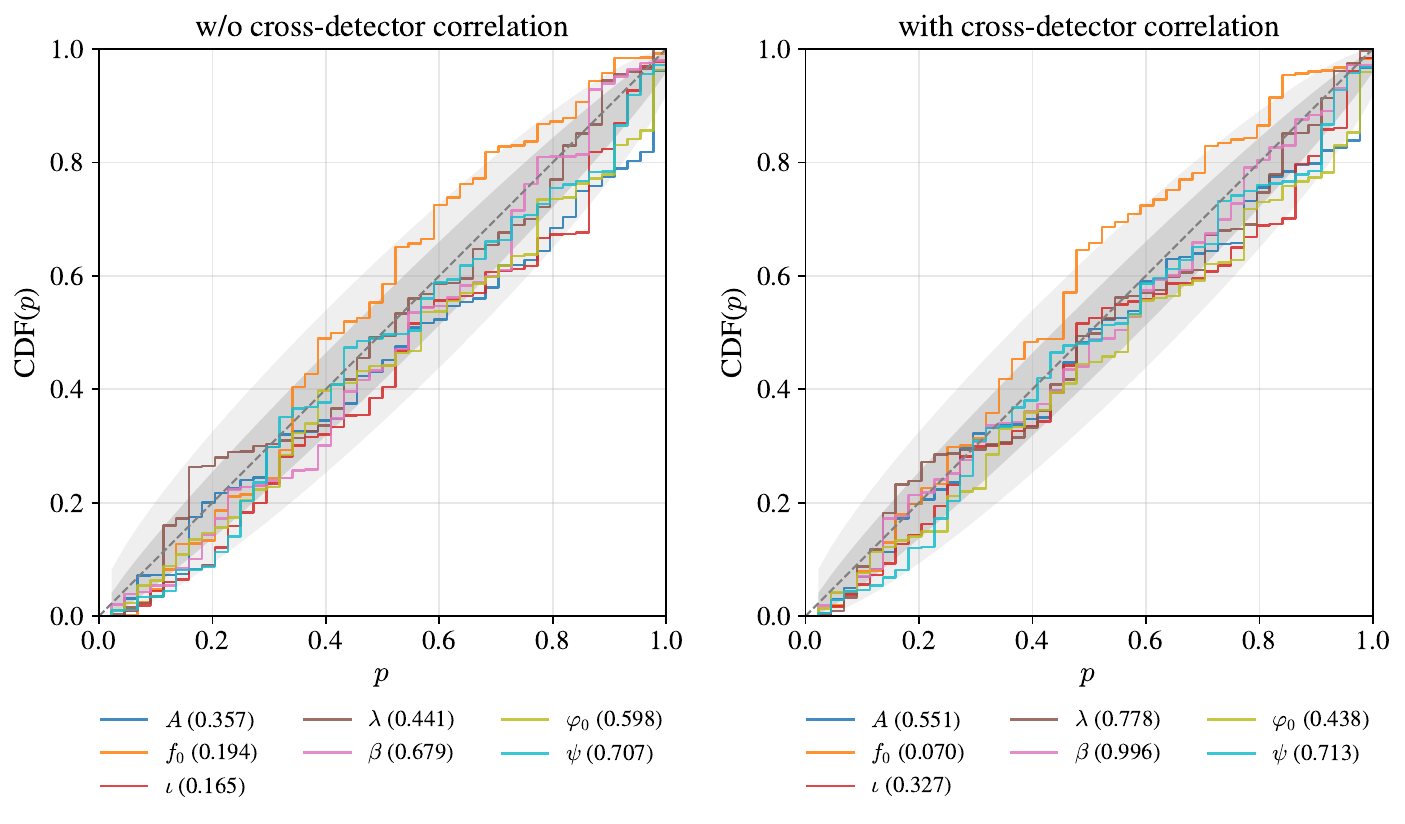}
\caption{
P--P plots for  GB  parameter estimation from 55 injection--recovery runs over the VGB catalogue, performed with the  block-diagonal noise model (left) and the   full-covariance noise model (right).
Each colored  curve is the empirical cumulative distribution for one source parameter, with  
the dark- and light-grey bands indicating  the \(1\sigma\) and \(2\sigma\) regions, respectively.  
The numbers  in parentheses in the legends are the Kolmogorov--Smirnov \(p\)-values.
The frequency derivative \(\dot f_0\) is omitted because it is well  constrained for only a small subset of the catalogue.}
\label{fig:pp_gb}
\end{figure*}

To assess whether noise model misspecification introduces biases in parameter estimation, again  we perform full Bayesian inference under both the block-diagonal and full-covariance noise models. 
For the single-source analysis, we select a representative VGB which has  the largest relative SNR differences  between the two models. 
For the population-level analysis, we carry out P-P tests over the full VGB catalogue. 
The results of these two tests are shown in Figure~\ref{fig:gb_posterior} and Figure~\ref{fig:pp_gb}, respectively.
Figure~\ref{fig:gb_posterior} compares the posteriors under  the two noise models for the selected VGB. 
The block-diagonal and full-covariance posteriors are shown in orange and green, respectively, and the injected values are marked by red lines. 
For this source, the network SNRs are \(10.49\) under the block-diagonal model and \(9.65\) under full covariance. 
As can be seen, the two posteriors are nearly indistinguishable, and  no parameter shows  a statistically significant bias. 
The credible intervals only exhibits slightly shift, with the full-covariance intervals marginally broader. 
For the P-P tests (Figure~\ref{fig:pp_gb}), we have excluded the  frequency derivative parameter  \(\dot{f}_0\)  for both models, as it is well constrained only for a small subset of the VGB catalogue. 
For all parameters shown in the figure, the empirical CDFs under both noise models lie within the \(1\sigma\)--\(2\sigma\) bands of the diagonal, with all Kolmogorov--Smirnov \(p\)-values above \(0.05\). 
These results confirm that, for GB, including or neglecting the cross-detector foreground correlations makes no appreciable difference to parameter estimation at the tested sensitivity level.

\subsection{Summary}
\label{sec:summary}
Summarizing the analysis of these  two typical resolvable source types, the impact of cross-detector foreground correlation on SNR, parameter uncertainties, and posteriors is as follows. 
For MBHBs with chirp masses at the $\mathcal{O}(10^5) \sim \mathcal{O}(10^6)$ orders,  and for GBs with frequencies  above 2 mHz, 
the two noise models yield nearly identical results: network SNR and FIM uncertainties differ by far less than 10\%.
Substantial differences arise for high-mass MBHBs $\mathcal{M}_c \sim \mathcal{O}(10^7)$,  where the network SNR and parameter uncertainties may  differ by up to $\sim 30\%$, and for low-frequency GBs, where the differences can be  up to the $10\%$ level. 
The only exception to the mass-dependent classification of MBHB  is the uncertainty of $\mathcal{M}_c$, whose difference is not affected by the source mass itself and generally remains  in  the $\pm 10\%$ range.
In terms of bias, the  choice of noise model between block-diagonal and full-covariance does not produces a statistically significant discrepancy. 
P-P tests for both the MBHB and VGB catalogues are statistically self-consistent under either model.
These conclusions primarily originate from two factors: the signal's frequency  relative to the frequency band of foreground CSD, 
and the coupling between the long observation time and the time  evolution of the coherence $\gamma$. 

%=====================================================================
\section{Conclusion and outlook}
\label{sec:conclusion}
%=====================================================================
This paper has presented a comprehensive study of the correlated Galactic confusion foreground for  the LISA--Taiji--TianQin network and its implications for the analysis  of resolvable GW sources, taking MBHBs and GBs as representative transient and continuous sources, respectively. 
We construct the full network foreground covariance matrix from a numerical simulation of $\sim 3\times 10^7$ GBs, using an iterative subtraction procedure with  multiple SNR thresholds, and derive an analytic model of the cross-detector coherence $\gamma$. 
The  consistency of this analytic model  with the simulation cross-validates both approaches. 
The resulting CSDs are non-negligible for $f \lesssim 3$~mHz, and we have characterized  their frequency and time dependence. 

As summarized in Sec.~\ref{sec:summary}, the impacts on source  analysis are  confined to specific signal regimes. 
Low- and intermediate-mass MBHBs,  along with GBs above $\sim2$~mHz,  are barely affected with differences in  SNRs and  parameter uncertainties  well below $10\%$.  
In contrast,  high-mass MBHBs ($\mathcal{M}_c\sim10^7\,M_\odot$) show network SNR and parameter uncertainty differences of up to $\sim30\%$ ($\sim35\%$ for the most extreme realization), and deviations for  the low-frequency GBs can reach the $10\%$ level. 
In terms of bias, the two noise models are statistically self-consistent: no parameter of MBHBs and GBs is significantly biased, and P-P tests over both the MBHB ensemble and the  VGB catalogue show no appreciable systematic discrepancy. 
% Thus the block-diagonal approximation is fully adequate for all GB science and for the vast majority of MBHB applications, while a full-covariance treatment is warranted only when extreme precision in sky localization or distance estimation of high-mass MBHBs is decisive. 
Results for the alternative orbit Config.~II are presented in Appendix~\ref{app:configII}.  They serve as a complement to those of   Config.~I  and verify the universality of our theoretical framework.

% We release the complete end-to-end analysis pipeline, including the fast network PE codes \texttt{Triangle-GB} and \texttt{Triangle-BBH}---which accelerate waveform and likelihood evaluations, respectively---along with the foreground covariance data and sensitivity curves for individual detectors and the network for various settings, ready for further scientific exploration with space-based detector networks.

Several limitations should be noted, indicating  directions for future research.  
First, the GB  population model carries uncertainties in the total number, spatial distribution, and binary evolution parameters. 
Alternative models (\textit{e.g.}, \citealp{Korol:2021pun}) and comparison among different catalogues merit dedicated study. 
Second,  frequency-domain analysis is widely employed in the current  state-of-the-art data anlysis pipeline prototypes~\citep{Littenberg2023,Strub:2024kbe,Katz:2024oqg,Deng:2025wgk}, while recent progress shows that it  may be complemented by a time-frequency treatment for non-stationary noise~\citep{CornishTF}. 
Extending such a framework (including computation of  SNRs~\citealp{Digman:2022jmp}, and parameter estimation~\citealp{DuTF}) to the full network with cross-detector correlations  is promising for providing a more precise and physically sound assessment, while a rigorous treatment  of residual non-Gaussianity~\citep{Racine:2007gv,Buscicchio:2024wwm} and a time-dependent covariance for year-long observations remain for future  work. 
Third, extreme mass-ratio inspiral (EMRI) systems are excluded from this work, primarily because their complicated and  computationally expensive  waveforms  make estimating the full 14- (or even higher) dimensional parameter space currently prohibitive.
While,  though AI-accelerated inference (\textit{e.g.}, simulation-based inference~\citealp{Speri:2025ucn,LB_EMRI_Research,cpl_42_8_081101}) offers a potential path forward.
Forth, we assume continuous  operation of all three detectors over 4 years, ignoring TianQin's   ``3+3'' operation  mode.  
Given that the contributions from TianQin-LISA and TianQin-Taiji cross-correlations are relatively weak, this simplification is unlikely to bias our results. 
However,  in a more general context,  data gaps are expected to frequently occur   in the data of space detectors, and a thorough treatment of them should also be  deferred to future studies.

\begin{acknowledgments}
This work is supported by National Key Research and Development Program of China
(Grant No. 2021YFC2201903, No. 2024YFC2207300, 
No. 2021YFC2201901, No. 2020YFC2200100).
\end{acknowledgments}

%=====================================================================
\appendix
%=====================================================================

\section{Frequency-domain TDI response for GBs}
\label{app:GB}
GB is  the most numerous source type  for space-based GW detection, with total population of order $\mathcal{O}(10^7)$ and detectable ones numbering in the tens of thousands. 
Efficient waveform template  is essential for data simulation and global-fit analyses. This appendix derives  the formalism of  fast frequency-domain TDI response calculation for GB signals, as adopted in our foreground generation and parameter estimation.
The acceleration scheme is based on   the  ``fast-slow decomposition'' principle proposed in  \cite{Cornish:2007if} (see also the "TDI-on-the-fly" approach in \citealp{Cornish:2025jfu}), but is generalized to accommodate arbitrary TDI combinations and  generic  orbit models with  time-varying arm lengths.

For the GB waveform in the source frame, we retain only the dominant \((2,\pm2)\) modes:
\begin{equation}
      h_{\ell m}(t) =  A_{\ell m}e^{i\Phi_{\ell m}(t)}, \quad \ell m \in \{22, 2\!-\!2\}, 
\end{equation}
where the amplitude \(A_{\ell m}\) is constant and the phase \(\Phi_{\ell m}(t)\) is quasi-monochromatic, 
described by the initial GW frequency $f_0$ and its derivatives  up to the second order: 
\begin{eqnarray}
      A_{22} = A_{2-2} &=& \sqrt{\frac{16\pi}{5}} A, \\ 
      \Phi_{22} = - \Phi_{2-2} &=& -2\pi \left(f_0 t + \frac{1}{2} \dot{f}_0 t^2 +  \frac{1}{6} \ddot{f}_0 t^3 \right), 
\end{eqnarray}   
with $\ddot{f}_0$ determined by $\ddot{f}_0 = 11 \dot{f}_0^2 / (3f_0)$. 
Substituting this into the single-link response of Eq.~\eqref{eq:single_link_td} and applying the TDI combination of Eq.~\eqref{eq:tdi_unified} yields
%% ORIGINAL TWO-COLUMN (widetext) VERSION, COMMENTED OUT.
%% Replaced by the single-column multi-line version below (the spanning form
%% cannot fit the 242pt column). Content and ordering are unchanged.
% \begin{widetext}
% \begin{equation}
% \begin{aligned}
% \mathrm{TDI}(t) &= \sum_{\ell m} \sum_{ij} \sum_{a_{ij}} \frac{K_{a_{ij}}}{2(1 - \hat{\mathbf{k}}\cdot\hat{\mathbf{n}}_{ij})} \Bigl(\sum_{\alpha} K^{\ell m}_\alpha \zeta_{\alpha,ij}\Bigr) A_{\ell m}
% \Bigl[ e^{i\Phi_{\ell m}(t - d^{\rm send}_{a_{ij}})} - e^{i\Phi_{\ell m}(t - d^{\rm recv}_{a_{ij}})} \Bigr]  \\
% &= \sum_{\ell m} \sum_{ij} \sum_{a_{ij}} \frac{K_{a_{ij}}}{2(1 - \hat{\mathbf{k}}\cdot\hat{\mathbf{n}}_{ij})} \Bigl(\sum_{\alpha} K^{\ell m}_\alpha \zeta_{\alpha,ij}\Bigr) A_{\ell m}
% \Bigl[ e^{i\Phi_{\ell m}(t - d^{\rm send}_{a_{ij}}) - i\Phi_{\ell m}(t)} - e^{i\Phi_{\ell m}(t - d^{\rm recv}_{a_{ij}}) - i\Phi_{\ell m}(t)} \Bigr] e^{i\Phi_{\ell m}(t)},
% \end{aligned}
% \label{eq:GB_tdi_td}
% \end{equation}
% \end{widetext}
\begin{equation}
\begin{aligned}
\mathrm{TDI}(t) &= \sum_{\ell m} \sum_{ij} \sum_{a_{ij}}
\frac{K_{a_{ij}}}{2(1 - \hat{\mathbf{k}}\cdot\hat{\mathbf{n}}_{ij})} \\
&\quad \times \Bigl(\sum_{\alpha} K^{\ell m}_\alpha \zeta_{\alpha,ij}\Bigr) A_{\ell m} \\
&\quad \times \Bigl[ e^{i\Phi_{\ell m}(t - d^{\rm send}_{a_{ij}})} - e^{i\Phi_{\ell m}(t - d^{\rm recv}_{a_{ij}})} \Bigr] \\
&= \sum_{\ell m} \sum_{ij} \sum_{a_{ij}}
\frac{K_{a_{ij}}}{2(1 - \hat{\mathbf{k}}\cdot\hat{\mathbf{n}}_{ij})} \\
&\quad \times \Bigl(\sum_{\alpha} K^{\ell m}_\alpha \zeta_{\alpha,ij}\Bigr) A_{\ell m} \\
&\quad \times \Bigl[ e^{i\Phi_{\ell m}(t - d^{\rm send}_{a_{ij}}) - i\Phi_{\ell m}(t)} \\
&\quad \quad - e^{i\Phi_{\ell m}(t - d^{\rm recv}_{a_{ij}}) - i\Phi_{\ell m}(t)} \Bigr] e^{i\Phi_{\ell m}(t)},
\end{aligned}
\label{eq:GB_tdi_td}
\end{equation}
with \(d^{\rm send}_{a_{ij}} = d_{a_{ij}} + d_{ij} + \hat{\mathbf{k}}\cdot\mathbf{R}_j/c\) and \(d^{\rm recv}_{a_{ij}} = d_{a_{ij}} + \hat{\mathbf{k}}\cdot\mathbf{R}_i/c\). 
Here we have decomposed each TDI polynomial \(\mathbf{P}_{ij}\) in Eq.~\eqref{eq:tdi_unified} into the linear combination of  delay operations:
\begin{equation}
\mathbf{P}_{ij} = \sum_{a_{ij}} K_{a_{ij}} \mathbf{D}_{a_{ij}}, \quad \mathbf{D}_{a_{ij}} f(t) = f(t - d_{a_{ij}}),
\label{eq:delay_operator}
\end{equation}
For example, in the Michelson \(X_2\) channel,  coefficients \(K_{a_{ij}}\) take values \(\pm 1\). 

The TDI response can be further decomposed as the product of  a  slowly  varying complex envelop and a fast varying carrier: 
\begin{equation}
\mathrm{TDI}(t) = A_{\mathrm{TDI}}(t) e^{i\Phi_{\mathrm{TDI}}(t)} = \underbrace{A_{\mathrm{TDI}}(t) e^{i\Delta\Phi(t)}}_{\text{slow envelope}} \; \underbrace{e^{i\Phi_c(t)}}_{\text{fast carrier}},
\label{eq:carrier_envelope}
\end{equation}
where we define the carrier phase as \(\Phi_c(t) = 2\pi f_0 t\).
Considering the quasi-monochromatic nature of GB waveform, 
it follows that  \(\Delta\Phi(t) = \Phi_{\mathrm{TDI}}(t) - \Phi_c(t)\) is slowly varying and hence can be absorbed into the envelope.

We then transform Eq.~\eqref{eq:carrier_envelope} to the frequency domain. For year-long data, a direct Fourier transform would require up to \(\mathcal{O}(10^6)\) time-domain  sampling points (assuming a sampling frequency of 0.1 Hz). Using the envelope–carrier separation and the convolution theorem \(\mathcal{F}[f\cdot g] = \mathcal{F}[f]*\mathcal{F}[g]\), the frequency-domain TDI response can be written as
\begin{equation}
\widetilde{\mathrm{TDI}}(f) = \frac{1}{2} \mathcal{F}\!\left[A_{\mathrm{TDI}}(t) e^{i\Delta\Phi(t)}\right] * \mathcal{F}\!\left[e^{i\Phi_c(t)}\right].
\label{eq:tdi_convolution}
\end{equation}
For the monochromatic carrier \(e^{i2\pi f_0 t}\), its Fourier transform is well approximated by a Dirac delta function, therefore 
\begin{equation}
\widetilde{\mathrm{TDI}}(f) \approx \frac{1}{2} \mathcal{F}\!\left[A_{\mathrm{TDI}}(t) e^{i\Delta\Phi(t)}\right] * \delta(f - f_0).
\end{equation}
The  envelope \(A_{\mathrm{TDI}}(t) e^{i\Delta\Phi(t)}\) varies slowly  enough so that it  requires only sparse time sampling (\textit{e.g.}, $N_{\rm sparse} = \mathcal{O}(10^2)$ points per year). 
The frequency-domain spectrum $\widetilde{\rm TDI}(f)$ can be  obtained by first taking the  fast Fourier transform  (FFT) of the envelope on this sparse grid, and then shifting the result by  \(f_0\).

The computation is divided into an offline phase and an online phase. The former is independent of source parameters and is therefore computed only once; the latter depends on the source parameters and must be repeated for each source in population simulation or parameter estimation. In the offline phase, orbit-related quantities such as \(\hat{\mathbf{n}}_{ij}\), \(\mathbf{R}_i\), \(d^{\rm send}_{a_{ij}}\), and \(d^{\rm recv}_{a_{ij}}\) are precomputed on a sparse time grid. 
While in the online phase, the amplitude \(A_{\mathrm{TDI}}(t)\) and phase \(\Phi_{\mathrm{TDI}}(t)\) are assembled on the sparse grid, and then  FFT yields the narrow-band spectrum. 
With this implementation, the data for the entire GB population can be generated within one hour on a single RTX 4080 Super GPU.

%=====================================================================
\section{Frequency-domain TDI response and network heterodyned likelihood for MBHBs}
\label{app:MBHB}

% In the millihertz band, MBHB signals are transient, with typical observation durations ranging from days to weeks. 
For the aligned-spin MBHB systems  widely considered in the literature, the frequency-domain waveform can be efficiently computed using the \texttt{IMRPhenomHM}  model~\citep{London2018}.
In this study, we utlize the \texttt{IMRPhenomHM} waveform implemented in the \texttt{WF4PY} package~\citep{Iacovelli:2022bbs,Iacovelli:2022mbg} and take into consideration 6 harmonic modes  $(\ell,m) \in \{(2,2), (3,3), (4,4), (2,1), (3,2), (4,3)\}$. 
This appendix presents the frequency-domain TDI response  of MBHB signals and the corresponding accelerated likelihood  for network data analysis.

The TDI response of MBHB signals in this work is built upon the formalism of \cite{Marsat:2018oam,Marsat:2020rtl}, and is extended to the general case with time-varying arm lengths and arbitrary TDI combinations (although in this work we restrict to constant arm lengths and the conventional  Michelson TDI combinations). 
Under stationary phase approximation (SPA)~\citep{PhysRevD.49.2658},  for a generic detector and TDI channel,   the frequency-domain response for aligned-spin MBHB signals  can be written as a sum over harmonic modes:
\begin{equation}
\widetilde{\mathrm{TDI}}(f) = \sum_{\ell m} G^{\ell m}_{\mathrm{TDI}}(f, t_{f,\ell m})\, \tilde{h}_{\ell m}(f),
\label{eq:mbhb_tdi_fd}
\end{equation}
where $\tilde{h}_{\ell m}(f) = A_{\ell m}(f) e^{-i\Phi_{\ell m}(f)}$ is the amplitude-phase decomposition of the $\ell m$ mode waveform. 
In this appendix we adopt the Fourier convention $\tilde{h}(f) \equiv \int dt\, h(t) e^{i2\pi f t}$, such that only $m>0$ modes are non-zero for $f>0$. To match the convention used elsewhere in this paper, the final frequency-domain responses derived below should be taken complex conjugate.

Following the deduction of  \cite{Marsat:2018oam}, the core idea of this formalism is  to map time-dependent orbital quantities into the frequency domain using  the time-frequency correspondence: 
\begin{equation}
t \rightarrow t_{f,\ell m} = -\frac{1}{2\pi} \frac{d\Phi_{\ell m}(f)}{df}, 
\end{equation}
and this relation should be  applied mode by mode. 
Therefore, the frequency-domain response function $G^{\ell m}_{\mathrm{TDI}}(f, t)$ for an arbitrary TDI channel is
\begin{equation}
G^{\ell m}_{\mathrm{TDI}}(f, t) = \sum_{ij} \widetilde{\mathbf{P}}_{ij}(f, t)\, G^{\ell m}_{ij}(f, t),
\label{eq:tdi_GW_response_fd}
\end{equation}
where the delay operators are converted to phase factors via the frequency-domain time-shift property $\mathbf{D}_{ij} \to \widetilde{\mathbf{D}}_{ij} = e^{i2\pi f d_{ij}(t)}$. The single-link response function $G^{\ell m}_{ij}(f, t)$ reads
% \begin{widetext}
% \begin{equation}
% G^{\ell m}_{ij}(f, t) = \; i\pi f d_{ij}(t)\,
% \mathrm{sinc}\!\left[\pi f d_{ij}(t)(1 - \hat{\mathbf{k}}\cdot\hat{\mathbf{n}}_{ij}(t))\right]
% e^{i\pi f d_{ij}(t)} e^{i\pi f\frac{\hat{\mathbf{k}}\cdot(\mathbf{R}_i(t)+\mathbf{R}_j(t))}{c}}\,
% F^{\ell m}_{ij}(t),
% \label{eq:single_link_GW_response_fd}
% \end{equation}
% \end{widetext}
\begin{eqnarray}
 G^{\ell m}_{ij}(f, t) &=& \; i\pi f d_{ij}(t)\,
\mathrm{sinc}\!\left[\pi f d_{ij}(t)(1 - \hat{\mathbf{k}}\cdot\hat{\mathbf{n}}_{ij}(t))\right] \nonumber \\ 
&& \times e^{i\pi f d_{ij}(t)} e^{i\pi f\frac{\hat{\mathbf{k}}\cdot(\mathbf{R}_i(t)+\mathbf{R}_j(t))}{c}}\,
F^{\ell m}_{ij}(t),
\label{eq:single_link_GW_response_fd}
\end{eqnarray}
Using the  antenna pattern function  $\zeta_{\alpha,ij}$ defined in Eq.~\eqref{eq:antenna_pattern_def} and the spin-weighted spherical harmonic coefficients $K^{\ell m}_\alpha$ given by Eq.~\eqref{eq:K_coefficients},   $F^{\ell m}_{ij}(t)$ can be written compactly as
\begin{equation}
F^{\ell m}_{ij}(t) \equiv \sum_{\alpha=+,\times} K^{\ell m}_\alpha(\iota, \varphi_{\rm ref})\;\zeta_{\alpha,ij}\bigl(\hat{\mathbf{k}}, \psi, t\bigr). 
\label{eq:F_ellm_def}
\end{equation}

In the context of space GW detector network,  parameter estimation for MBHBs  faces substantial computational challenges, arising from the high dimensionality of the joint data vector (six TDI streams for three detectors), the inclusion of higher-order modes, and the multiplication of full noise covariance  matrix.
Evaluating the likelihood on the full frequency grid for the numerous  calls required by stochastic sampling incurs  substantial computational cost. 
To accelarate likelihood evaluation, 
we  extend the heterodyned likelihood method proposed in \cite{Cornish:2010kf,Zackay:2018qdy,Cornish2021,Leslie:2021ssu} to our interested scenario with multiple data channel and  full noise covariance.

For brevity, we denote the frequency-domain TDI response in Eq.~\eqref{eq:mbhb_tdi_fd} as $\tilde{h}_I(f)$ and decompose it into harmonic modes: 
\begin{equation}
\tilde{h}_I(f) = \sum_{\ell m} \tilde{h}_{I,  \ell m}(f).
\end{equation}
Since  multiple indices will be  involved later, here we use $I$ to collectively index TDI channels and detectors. 
For the $\ell m $ mode,  let $\tilde{h}^{(0)}_{I,\ell m}(f)$ denote the base waveform  at a high-likelihood reference   point $\boldsymbol{\theta}_0$ (the injection values or a preliminary search result). 
Considering the general features of  inspiral-merger-ringdown waveforms~\citep{Cornish2021}, 
we construct the slowly varying waveform ratio
\begin{equation}
\tilde{r}_{I,\ell m}(f) \equiv \frac{\tilde{h}_{I,\ell m}(f)}{\tilde{h}^{(0)}_{I,\ell m}(f)},
\label{eq:waveform_ratio}
\end{equation}
and evaluate it on a sparse frequency grid $\{f_k\}_{k=0}^{N_{\rm het}}$, with linear interpolation within each interval $[f_k, f_{k+1}]$:
\begin{equation}
\tilde{r}_{I,\ell m}(f) \approx \alpha_{k,I,\ell m} + \beta_{k,I,\ell m}\,(f - f_k), 
\end{equation}
where $\alpha_{k,I,\ell m} \equiv \tilde{r}_{I,\ell m}(f_k)$ and $\beta_{k,I,\ell m} \equiv [\tilde{r}_{I,\ell m}(f_{k+1}) - \tilde{r}_{I,\ell m}(f_k)] / (f_{k+1} - f_k)$.
The inner products in the likelihood can be then  evaluated as
\begin{eqnarray}
\langle \tilde{h} | \tilde{h} \rangle_{\mathbf{C}} &\approx& \Re \sum_{k < N_{\rm het}} \sum_{I,J} \sum_{\ell m} \sum_{\ell' m'}  \Bigl[ \alpha_{k,I,\ell m}^* \alpha_{k,J,\ell' m'} A_{k,IJ}^{\ell m,\ell' m'} \nonumber \\
&& \quad + \beta_{k,I,\ell m}^* \beta_{k,J,\ell' m'} B_{k,IJ}^{\ell m,\ell' m'} \nonumber  \\
&& \quad + \bigl( \alpha_{k,I,\ell m}^* \beta_{k,J,\ell' m'} + \beta_{k,I,\ell m}^* \alpha_{k,J,\ell' m'} \bigr) C_{k,IJ}^{\ell m,\ell' m'} \Bigr],   \nonumber  \\ \\ 
% && \quad \quad \times C_{k,IJ}^{\ell m,\ell' m'} \Bigr]  \\
\langle \tilde{d} | \tilde{h} \rangle_{\mathbf{C}} &\approx& \Re \sum_{k < N_{\rm het}} \sum_{I,J} \sum_{\ell m}  \Bigl[ \alpha_{k,I,\ell m} D_{k,IJ}^{\ell m} + \beta_{k,I,\ell m} E_{k,IJ}^{\ell m} \Bigr], \nonumber \\
\end{eqnarray}
where the offline coefficients $A$, $B$, $C$, $D$, $E$ are precomputed once per reference point $\boldsymbol{\theta}_0$ (hence the term ``offline''):
\begin{equation}
\begin{aligned}
A_{k,IJ}^{\ell m,\ell' m'} &\equiv \sum_{f_k \le f < f_{k+1}} \bigl(\tilde{h}^{(0)}_{I,\ell m}\bigr)^*\, \mathbf{C}^{-1}_{IJ}\, \tilde{h}^{(0)}_{J,\ell' m'}, \\[2pt]
B_{k,IJ}^{\ell m,\ell' m'} &\equiv \sum_{f_k \le f < f_{k+1}} \bigl(\tilde{h}^{(0)}_{I,\ell m}\bigr)^*\, \mathbf{C}^{-1}_{IJ}\, \tilde{h}^{(0)}_{J,\ell' m'}\, (f - f_k)^2, \\[2pt]
C_{k,IJ}^{\ell m,\ell' m'} &\equiv \sum_{f_k \le f < f_{k+1}} \bigl(\tilde{h}^{(0)}_{I,\ell m}\bigr)^*\, \mathbf{C}^{-1}_{IJ}\, \tilde{h}^{(0)}_{J,\ell' m'}\, (f - f_k), \\[2pt]
D_{k,IJ}^{\ell m} &\equiv \sum_{f_k \le f < f_{k+1}} \tilde{d}_I^*\, \mathbf{C}^{-1}_{IJ}\, \tilde{h}^{(0)}_{J,\ell m}, \\[2pt]
E_{k,IJ}^{\ell m} &\equiv \sum_{f_k \le f < f_{k+1}} \tilde{d}_I^*\, \mathbf{C}^{-1}_{IJ}\, \tilde{h}^{(0)}_{J,\ell m}\, (f - f_k).
\end{aligned}
\label{eq:het_coefficients}
\end{equation}
The inverse covariance $\mathbf{C}^{-1}(f_k)$ is also precomputed and cached at each frequency, eliminating the cost of matrix inversion during parameter updates.
In summary, the acceleration factor relative to a full-grid evaluation is approximately $N_f / N_{\rm het}$, since only the slowly varying ratios $\alpha_k$ and $\beta_k$ need to be updated during the stochastic exploration of parameter space. 
As demonstrated in the literature (\textit{e.g.} \citealp{Leslie:2021ssu}),  $N_{\rm het}$
is typically of the $\mathcal{O}(10^2)$ order, regardless of  the original length of data, and  
the P-P plots  shown  in Figure~\ref{fig:pp_mbhb} serve as a validation of  statistical unbiasedness for our choice of $N_{\rm het}=256$. 

%=====================================================================
\section{Derivation and validation of the foreground CSD}
\label{app:foreground}
%=====================================================================
This appendix derives the analytic expression  for the confusion foreground CSD in Eq.~\eqref{eq:conf_csd_explicitsum}. The derivation follows the outline given in Sec.~\ref{sec:foreground},  and is validated against the numerical simulations in Figures~\ref{fig:app_lisa_taiji_amp} and~\ref{fig:app_lisa_taiji_phase}.

\subsection{Derivation of the foreground CSD}

We start from the definition in Eq.~\eqref{eq:conf_csd}. 
Since each unresolved GB  carries an independent random phase (random in the ensemble sense), the cross terms between distinct sources  vanish under  ensemble average.  
The CSD therefore reduces to a sum over the unresolved population $\mathcal{U}$.

The GBs  constituting the  confusion  foreground are weak in amplitude and contribute to  the total noise budget  mainly at low frequencies. 
We therefore model each GB waveform at leading order, retaining only the dominant $(2,\pm2)$ harmonic,  and evaluating the TDI response in the low-frequency, equal-arm approximation. 
We adopt the same Fourier convention $\tilde{h}(f) \equiv \int dt\, h(t)\,e^{i2\pi f t}$ as in Appendix~\ref{app:MBHB}, so that  positive frequencies correspond to $m>0$ modes, and hence  only the $(2,2)$ mode is  kept. 
Since this convention is opposite to the $e^{-2\pi i f t}$ convention used elsewhere, the final results derived below are to be complex-conjugated.

For a time interval over which the detector positions can be regarded as fixed,  in the low-frequency limit ($f d \ll 1$), the frequency-domain TDI response simplifies to
\begin{equation}
\tilde{g}_{I,k}^{c}(f) =  e^{2\pi i f\,\hat{\mathbf{k}}_k\cdot \mathbf{R}_I/c}\; \mathcal{T}_I^{\,c}(f)\;\tilde{h}_{22}(f),
\label{eq:app_tdi_factorized}
\end{equation}
where $g_{I,k}^c(f)$ is the TDI $c$ channel response of detector $I$ to the $k$-th GB as defined in Eq.~\eqref{eq:data_model}. The transfer function $\mathcal{T}_I^c$ is given by
% \begin{eqnarray}
% && \mathcal{T}_I^{c}(f;\hat{\mathbf{k}}_k)  \nonumber \\ 
% &\equiv& i\pi f d_I \sum_{\alpha=+,\times} K^{22}_\alpha(\iota_k,\varphi_k)  \sum_{ij}\widetilde{\mathbf{P}}_{I,ij}^{\,c}(f)\,\zeta_{\alpha,ij}(\hat{\mathbf{k}}_k, \psi_k).
% \label{eq:app_T}
% \end{eqnarray}
\begin{eqnarray}
\mathcal{T}_I^{c}(f;\hat{\mathbf{k}}_k, \psi_k, \iota_k, \varphi_k)  &\equiv&  \sum_{\alpha} \sum_{ij} i\pi f d_I  K^{22}_\alpha(\iota_k,\varphi_k)   \widetilde{\mathbf{P}}_{I,ij}^{\,c}(f) \nonumber \\ 
&& \times \zeta_{I,\alpha,ij}(\hat{\mathbf{k}}_k, \psi_k),
\label{eq:app_T}
\end{eqnarray}
with $\alpha \in \{+, \times\}$ and $ij \in \{12, 23, 31, 21, 32, 13\}$. 
This transfer function depends on the extrinsic source parameters: the sky direction $\hat{\mathbf{k}}_k$ and polarization angle $\psi_k$ enter through $\zeta_{\alpha, I, ij}$, while the inclination $\iota_k$ and phase $\varphi_k$ appear via $K^{22}_\alpha$. 
In principle, the ensemble average in the definition of CSD indicates averaging over the  source parameters. 
The parameters over which we can   average analyticly  include $\theta_k\equiv\{\psi_k,\iota_k,\varphi_k\}$. 
Given that we may not have an analytic model for the anisotropic  sky  distribution of confusion GBs, we exclude sky location parameters  from this analytic  averaging and  instead  average over  a specific source catalogue (for this work the  catalogue of confusion GB is obtained via the iterative subtraction procedure of Sec.~\ref{sec:foreground_sim}). 

Grouping unresolved binaries into frequency bins $\mathcal{U}_f=\{k:f_{0,k}\in[f,f+\Delta f)\}$, the CSD at frequency $f$ reads
\begin{equation}
C_{IJ}^{cc'}(f) = \sum_{k\in\mathcal{U}_f} w_k\;
e^{\,2\pi i f\,\hat{\mathbf{k}}_k\cdot(\mathbf{R}_I-\mathbf{R}_J)/c}\;
\bigl\langle \mathcal{T}_I^{c}(f)\,\mathcal{T}_J^{c'*}(f)\bigr\rangle_{\theta_k},
\label{eq:app_csd_sum}
\end{equation}
where $\langle\cdot\rangle_{\theta_k}$ denotes the ensemble average over $\theta_k$.
We perform this averaging in two steps.
First, the $\psi$-averaged response product is
\begin{equation}
\bigl\langle \mathcal{T}_I^{c}\,\mathcal{T}_J^{c'*}\bigr\rangle_{\psi_k}
= \pi^2 f^2 d_I d_J
\sum_{\alpha} |K^{22}_\alpha|^2\,
Q_{I,\alpha}^{\,c}(f,\hat{\mathbf{k}}_k)\,Q_{J,\alpha}^{\,c'*}(f,\hat{\mathbf{k}}_k), 
\label{eq:app_T_psi}
\end{equation}
with $Q_{I, \alpha}^c$ defined in Eq.~\eqref{eq:antenna_vector}. 
Second, the average over $\iota$ and $\varphi$ acts on $K^{22}_\alpha$.
For the $(2,2)$ harmonic, we have
\begin{eqnarray}
K^{22}_{+} &=& \sqrt{\frac{5}{64\pi}}\,(1+\cos^2\iota)\,e^{2i\varphi},\nonumber \\ 
K^{22}_{\times} &=& i\sqrt{\frac{5}{16\pi}}\,\cos\iota\,e^{2i\varphi}.
\label{eq:app_K22}
\end{eqnarray}
Averaging $|K^{22}_{\alpha}|^2$ over isotropic inclination (uniform in $\cos\iota$) and phase $\varphi$ yields
\begin{equation}
\bigl\langle |K^{22}_{+}|^2\bigr\rangle = \frac{7}{48\pi},\qquad
\bigl\langle |K^{22}_{\times}|^2\bigr\rangle = \frac{5}{48\pi}.
\label{eq:app_Kavg}
\end{equation}
Inserting Eq.~\eqref{eq:app_Kavg} into Eq.~\eqref{eq:app_T_psi} and taking the complex conjugate of the result,  we  finally obtain the expression for  CSD  in Eq.~\eqref{eq:conf_csd_explicitsum}.

\subsection{Time dependence and  numerical validation}

The analytic CSD model derived above is validated against the simulated foreground data obtained using the method of Sec.~\ref{sec:foreground_sim}, and the time dependence of the foreground covariance is characterized in this subsection. 
Both  are performed in terms of the coherence $\gamma$, which plays a crucial role in quantifying the impacts of foreground correlation.

\begin{figure*}[t]
\centering
\includegraphics[width=\textwidth,height=0.95\textheight,keepaspectratio]{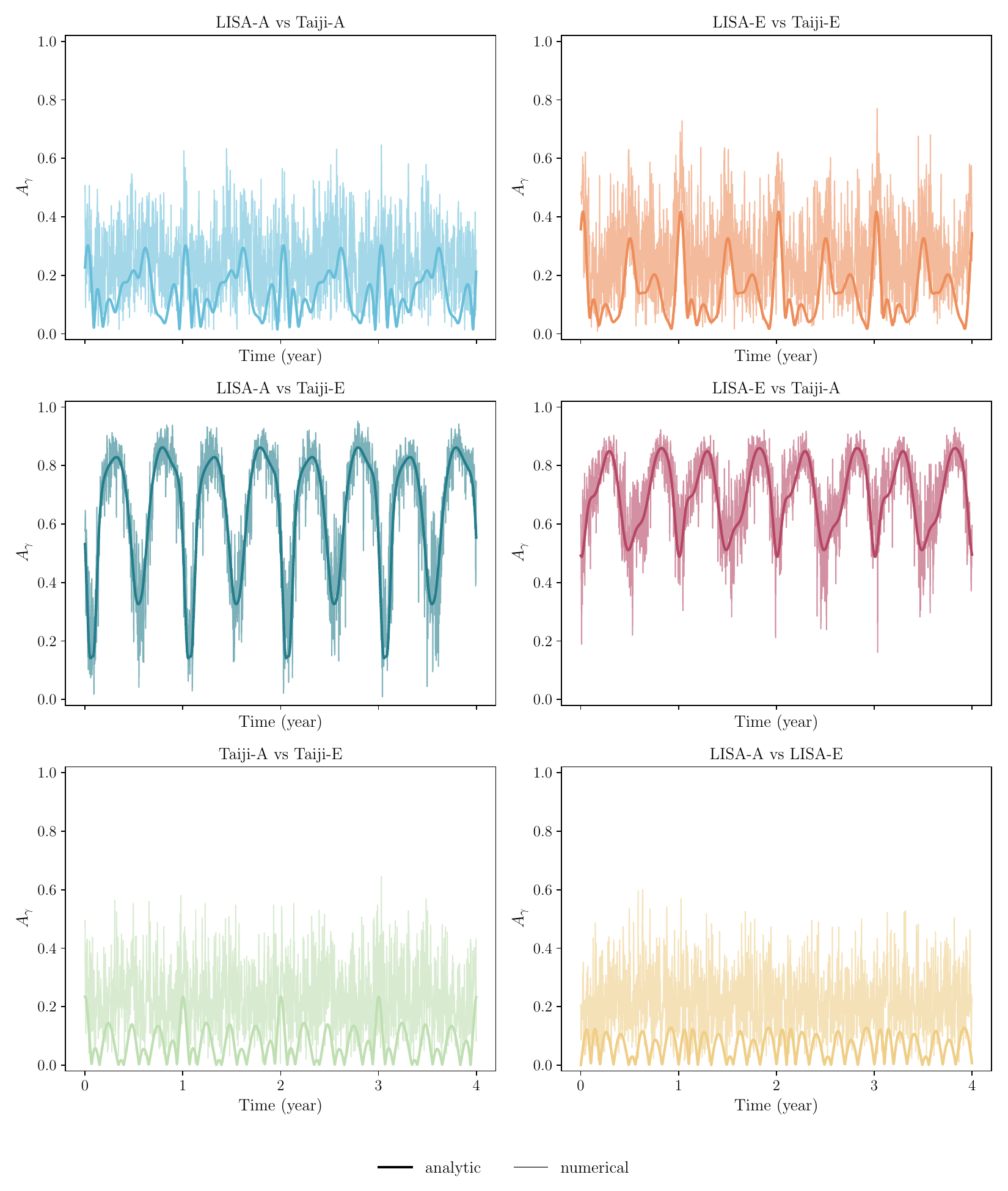}
\caption{
Coherence amplitudes \(A_\gamma(t)\equiv|\gamma(t)|\) of the confusion foreground for the LISA-Taiji detector pair  at \(f=1\)~mHz over the four-year observation.
The top row shows the LISA--Taiji \(A\)-\(A\) and \(E\)-\(E\) channel pairs, the middle row the \(A\)-\(E\) and \(E\)-\(A\) pairs, and the bottom row the same-detector \(A\)-\(E\) pairs for Taiji and LISA.
In each panel, the thick curve is the theoretical estimate from Eq.~\eqref{eq:conf_csd_explicitsum}, and the thin curve of the same color is estimated via the   Welch method  from numerical simulation.}
\label{fig:app_lisa_taiji_amp}
\end{figure*}

\begin{figure*}[t]
\centering
\includegraphics[width=0.95\textwidth]{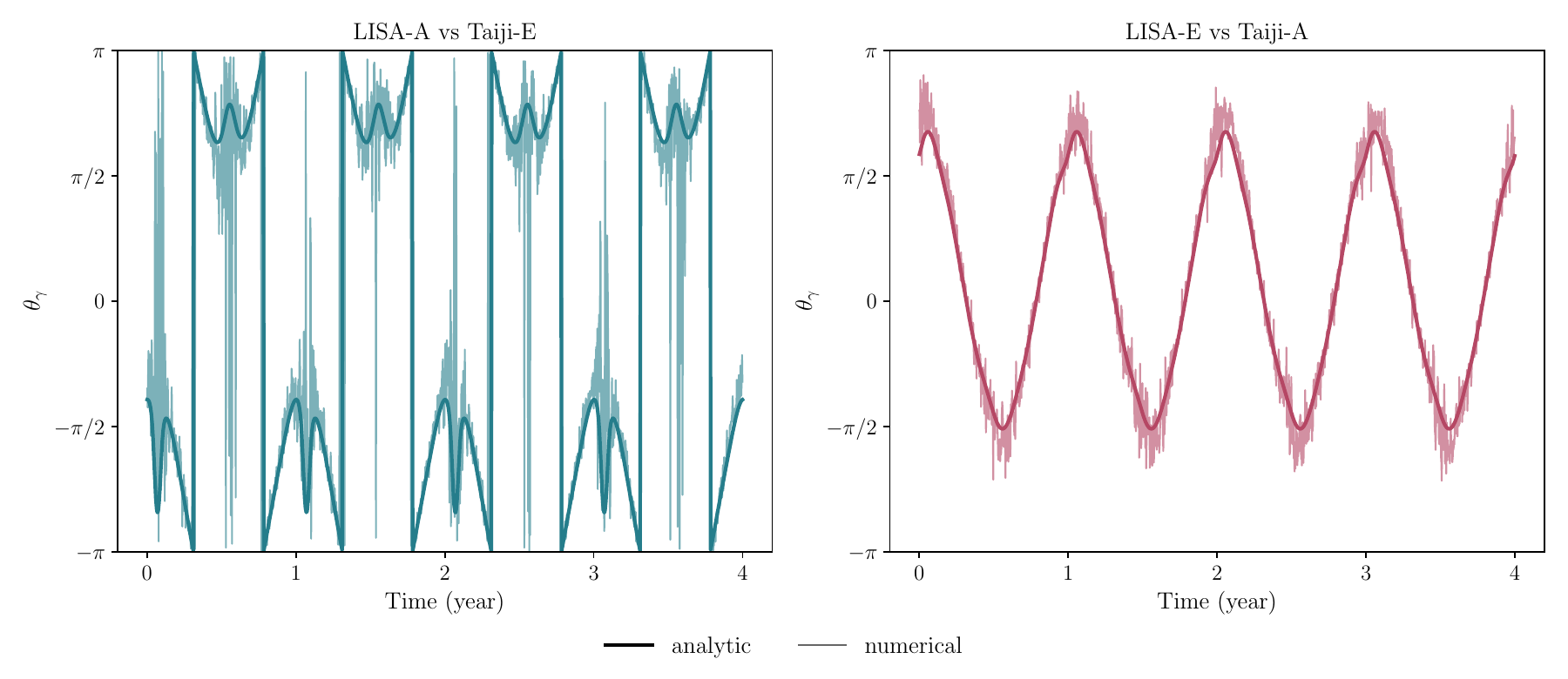}
\caption{
Coherence phases \(\theta_\gamma(t)\equiv\arg\gamma(t)\) of  confusion foreground for the LISA-Taiji detector pair  at \(f=1\)~mHz over the four-year observation.
The left and right panels show the LISA-\(A\)--Taiji-\(E\) and LISA-\(E\)--Taiji-\(A\) channel pairs, respectively.
The thick curves are the theoretical estimates from Eq.~\eqref{eq:conf_csd_explicitsum}, and the thin curves of the same colors are estimated  from numerical simulation.
Here we only present the TDI channel pairs that exhibit the strongest correlations. 
}
\label{fig:app_lisa_taiji_phase}
\end{figure*}

% \begin{figure*}
% \centering
% \includegraphics[width=\textwidth]{apjs_fig/compare_coherence_amplitude_at_freqs.pdf}
% \caption{
% Theoretical coherence amplitudes \(A_\gamma(t)\equiv|\gamma(t)|\) of the LISA--Taiji confusion foreground over the four-year observation.
% The left and right panels show the LISA-\(A\)--Taiji-\(E\) and LISA-\(E\)--Taiji-\(A\) channel pairs, respectively.
% Purple, blue, green, and yellow curves correspond to \(f=0.2\), \(0.3\), \(0.6\), and \(1.0\)~mHz, respectively.}
% \label{fig:app_lisa_taiji_amp_at_freqs}
% \end{figure*}

For the  LISA-Taiji pair, 
Figure~\ref{fig:app_lisa_taiji_amp} and Figure~\ref{fig:app_lisa_taiji_phase} respectively show the coherence amplitude $A_\gamma \equiv |\gamma|$ and coherence phase $\theta_\gamma \equiv {\rm arg}(\gamma)$ as functions of mission time, at the frequency of 1 mHz where the foreground is  most prominent.  
In each panel, each   thick   curve represents the  theoretical estimate for a pair of data channels, calculated  according to the first line of  Eqs.~\eqref{eq:conf_csd_explicitsum}, while  the thin curve plotted in  the same color shows the corresponding numerical simulation result.  
For consistency, 
the catalogue employed in the theoretical calculation  is also the one obtained  from the subtraction procedure of Sec.~\ref{sec:foreground_sim}. 
In both theoretical and numerical calculations, the whole mission time of 4 year is sliced to 1-day segments. 
To reduce the fluctuation in the CSD estimated  from  simulation, we adopt the Welch method,   with the average time  set to 0.1 days (which is adequate for 1 mHz), along with a Hann window and 50\% overlap.

\begin{figure*}
\centering
\includegraphics[width=0.9\textwidth]{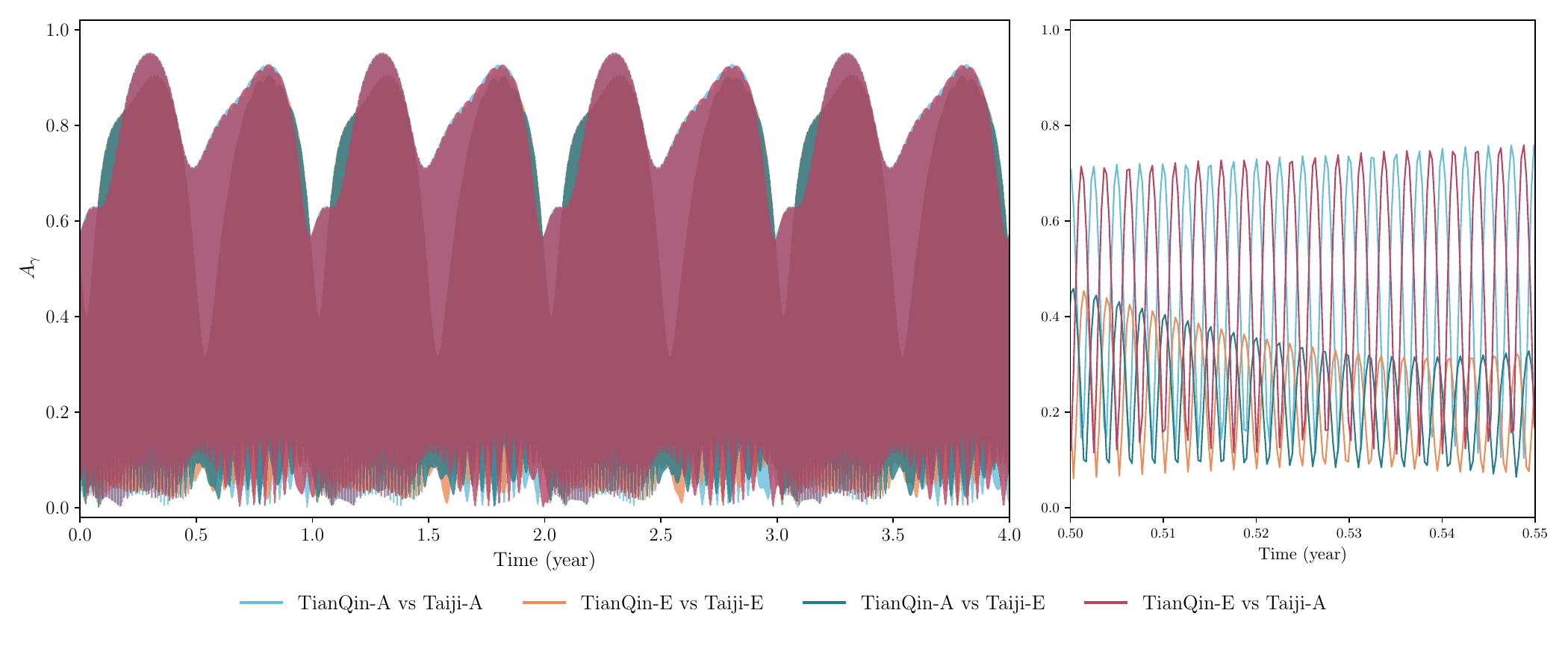}\hfill
\includegraphics[width=0.9\textwidth]{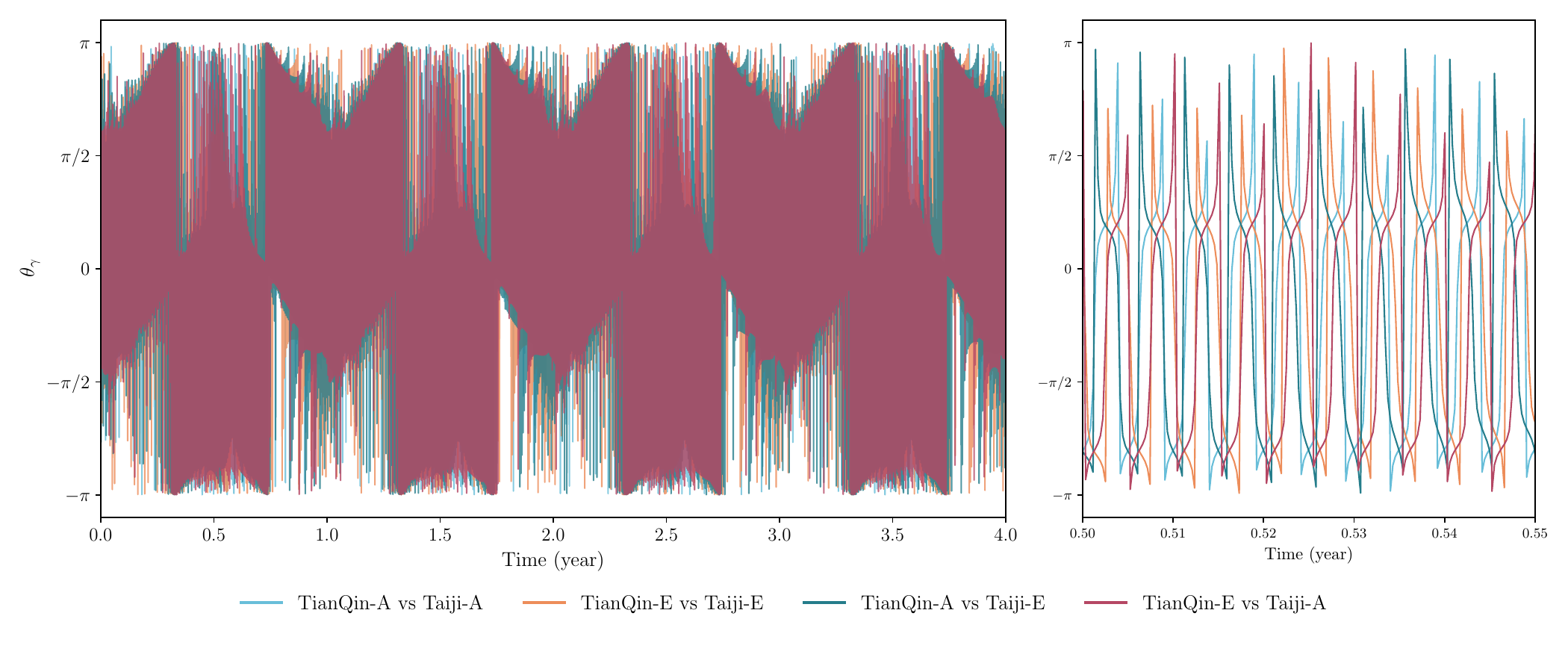}\\[8pt]
\includegraphics[width=0.8\textwidth]{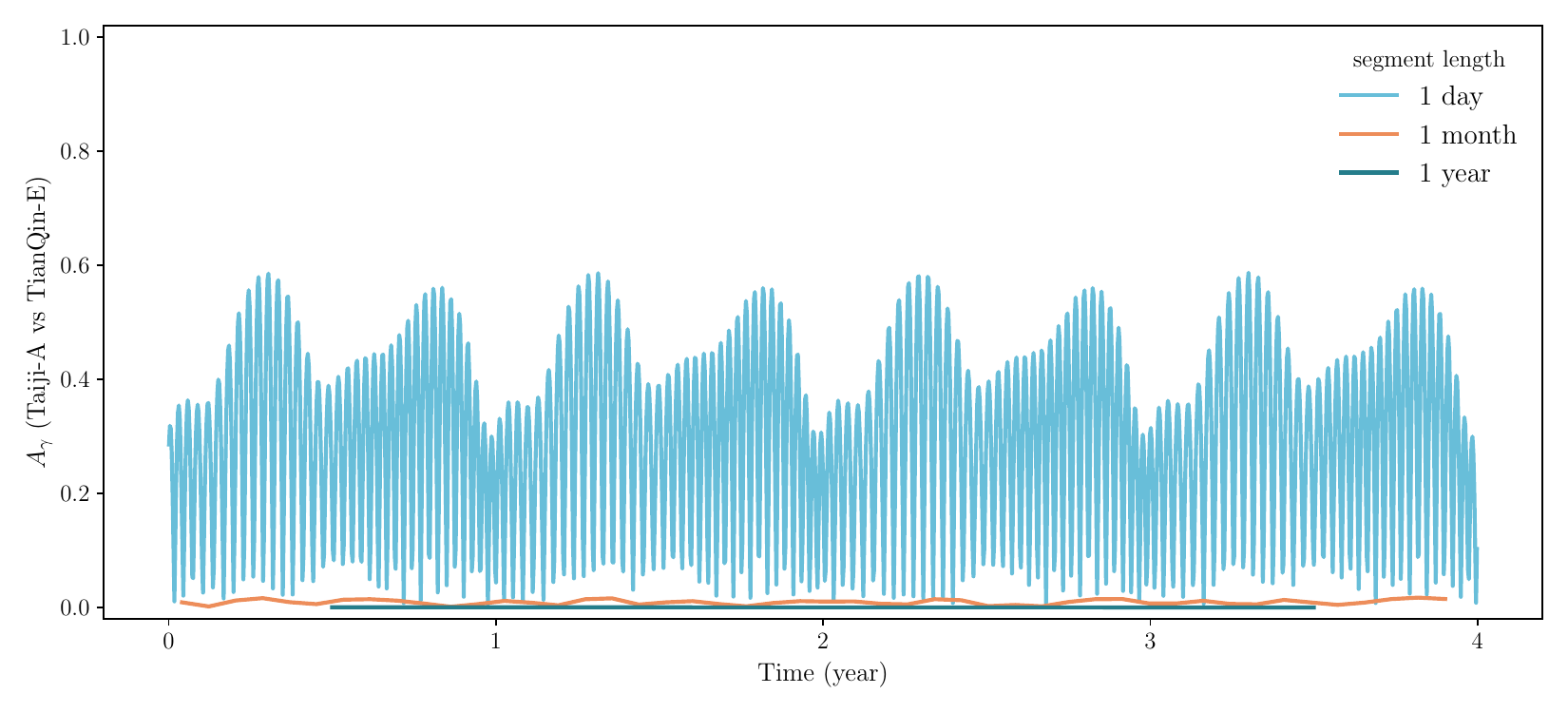}
\caption{
Theoretical TianQin--Taiji confusion foreground coherence $\gamma$ at \(f=1\)~mHz.
The first and second rows show the coherence amplitude \(A_\gamma(t)\equiv|\gamma(t)|\) and phase \(\theta_\gamma(t)\equiv\arg\gamma(t)\), respectively, calculated with 0.1-day segments, with  the left panel in each row  covering  the four-year observation and the right panel showing  the interval from 0.50 to 0.55~yr.
In these rows, light blue, orange, green, and red denote the TianQin-\(A\)--Taiji-\(A\), TianQin-\(E\)--Taiji-\(E\), TianQin-\(A\)--Taiji-\(E\), and TianQin-\(E\)--Taiji-\(A\) channel pairs, respectively.
The bottom row shows \(A_\gamma\) for the Taiji-\(A\)--TianQin-\(E\) pair after averaging the CSDs over 1-day (light blue), 1-month (orange), and 1-year (green) windows.
}
\label{fig:app_tq_taiji_time}
\end{figure*}

As can be seen from Figure~\ref{fig:app_lisa_taiji_amp},  
the same-channel coherences ($AA$, $EE$) between LISA and Taiji are less than 0.5 for the whole mission time, while we have clarified  in Sec.~\ref{subsec:foreground_spectra_result} that  these  results are sensitive to initial  conditions of the detector orbits. 
The cross-channel coherences ($AE$, $EA$) between LISA and Taiji are systematically larger, with the magnitudes ranging from 0 to near 0.9. 
Besides, we also show  the  same-detector cross-channel coherences  in the bottom panels.  
As stated  in \cite{AdamsCornish2010,Smith:2019wny}, these coherences would  vanish for isotropic SGWBs, and we show that for the  anisotropic Galactic foreground, they remain close to zero throughout.
All curves show a clear annual modulation set by the detectors'  relative motions, and   
in all panels the theoretical curves  track the Welch estimates  within the statistical fluctuations of the data, cross-validating both approaches. 
One may also notice that the smaller the magnitude of $A_\gamma$, the more easily the Welch spectrum is dominated by random fluctuations.

The coherence  serves as an important metric for quantifying the impact of correlation on data analysis. 
For the pairs of data channels with the strongest coherence, 
Figure~\ref{fig:app_lisa_taiji_phase} further displays the time dependence of their phases $\theta_\gamma$. 
The annual modulation is again observed, along with consistency between the numerical and theoretical  results. 
% As a supplement, Figure~\ref{fig:app_lisa_taiji_amp_at_freqs} shows the time evolution of the coherence amplitude $A_\gamma(t)$ over the four-year observation at the frequencies $f=0.2$--$1$~mHz, obtained via theoretical calculation; this band essentially covers the range where the foreground is most prominent. 
% On top of the quasi-periodic time dependence shared with the 1~mHz curves, the amplitude exhibits a clear frequency dependence: its time-averaged level decreases with increasing frequency, and the dips within each year become markedly deeper, so that the higher-frequency curves intermittently drop to much smaller coherence values whereas the lower-frequency ones keep a higher average level with comparatively mild modulation. 
% This frequency dependence of the amplitude, together with the phase behavior discussed above, sets the effective strength of the cross-detector foreground correlation across the band, and is used in Sec.~\ref{sec:gb_analysis} to explain the frequency dependence of the VGB sky-localization uncertainties.

For the TianQin-Taiji pair, the first and second rows of Figure~\ref{fig:app_tq_taiji_time} show the time evolution of $A_\gamma$ and $\theta_\gamma$, respectively. 
Still, the target frequency is set to 1 mHz. 
In each row, the left panel displays the variation over the full mission duration, while the right panel presents a zoom-in over a 0.05-year period. 
Since TianQin  rotates around the Earch  in the constellation plane with a period of  3.65 days, the time dependence of $\gamma$ exhibits fast oscillations relative to the LISA-Taiji pair. 
To capture this, we reduce  the segment length used in the theoretical calculation to 0.1 day. 
Given the short segment duration, the agreement between theory and simulation can only be interpreted statistically, so we omit the numerical simulation results in this figure.  
It is readily seen that the  time-dependence of $\gamma$ is  generally  consistent with the  overlap reduction function  obtained in \cite{Liang:2024tgn}, which adopts a segment length of 1 hour.  
While,  
for the signals of interest in this work, the time scales are generally longer than 0.1 day. 
\textit{e.g.}, for MBHBs, we use a 30-day window to cover the timescale over which their SNR is predominantly accumulated, and  for GBs, we directly use the full 4-year data. 
To illustrate how the time window of  analyzing  a certain signal affects the  averaged CSD within it,   we show in the third row of Figure~\ref{fig:app_tq_taiji_time} the $A_\gamma$ variations obtained from  CSDs  averaged over 1 day, 1 month, and 1 year. 
It is evident that a window of one day is already sufficient to reduce the correlation to below 0.6, while windows of one month or longer essentially lead to vanishing correlation. This explains  why the correlations between TianQin and other detectors are generally negligible in the CSDs averaged over 4 years (Figure~\ref{fig:csd_comparison}), and also why we mainly focus on the LISA-Taiji pair when studying the impacts of correlations on resolvable  signals.

\section{Results  for orbit Config.~II}
\label{app:configII}

\begin{figure*}[t]
\centering
\begin{minipage}{0.49\textwidth}
\centering
\includegraphics[width=\linewidth]{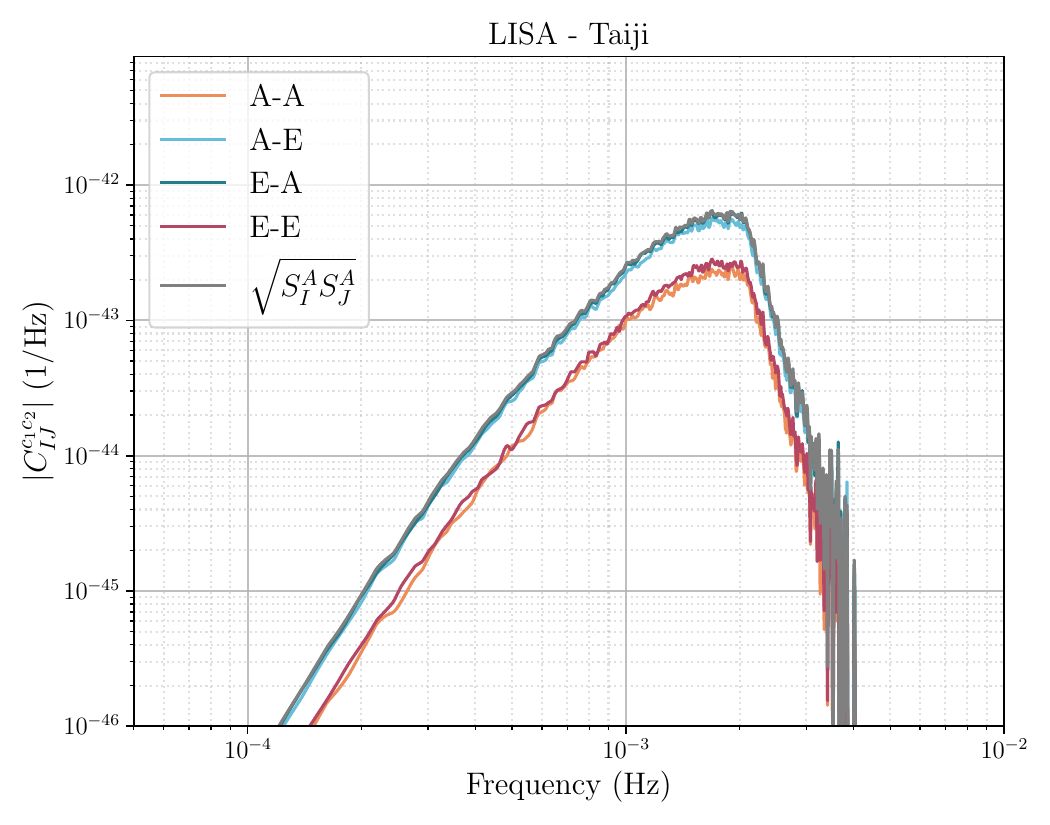}\\[2pt]
\textbf{(a)}
\end{minipage}\hfill
\begin{minipage}{0.49\textwidth}
\centering
\includegraphics[width=\linewidth]{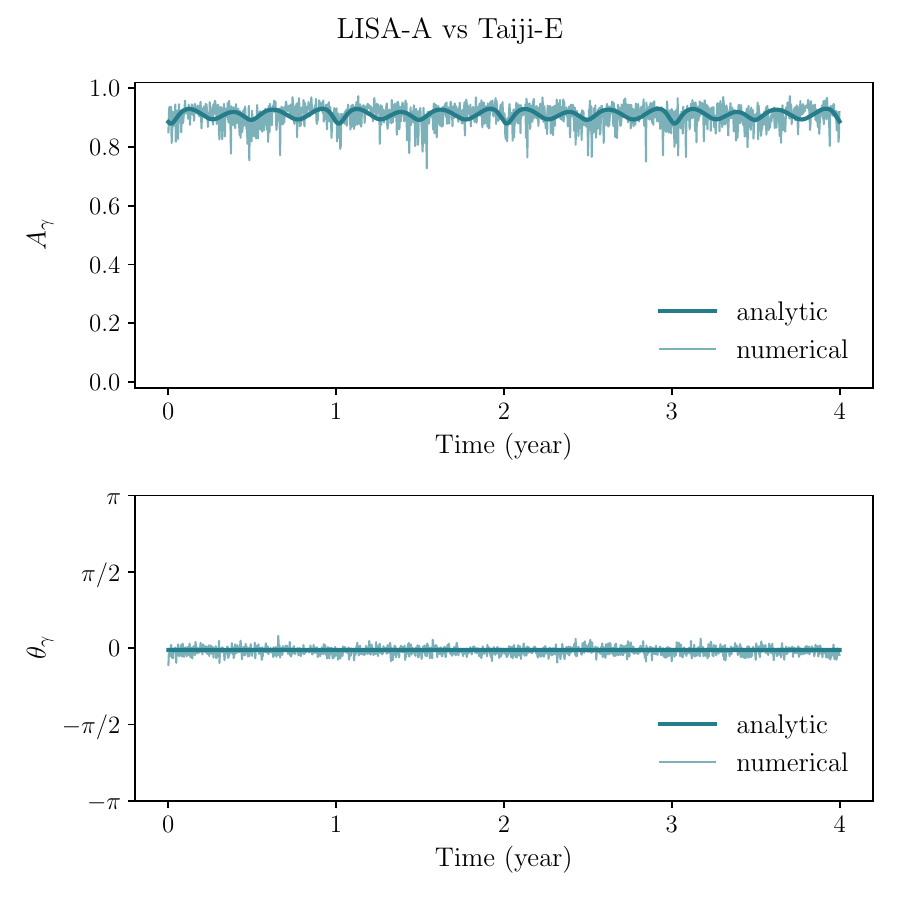}\\[2pt]
\textbf{(b)}
\end{minipage}\\[6pt]
\begin{minipage}{0.49\textwidth}
\centering
\includegraphics[width=\linewidth]{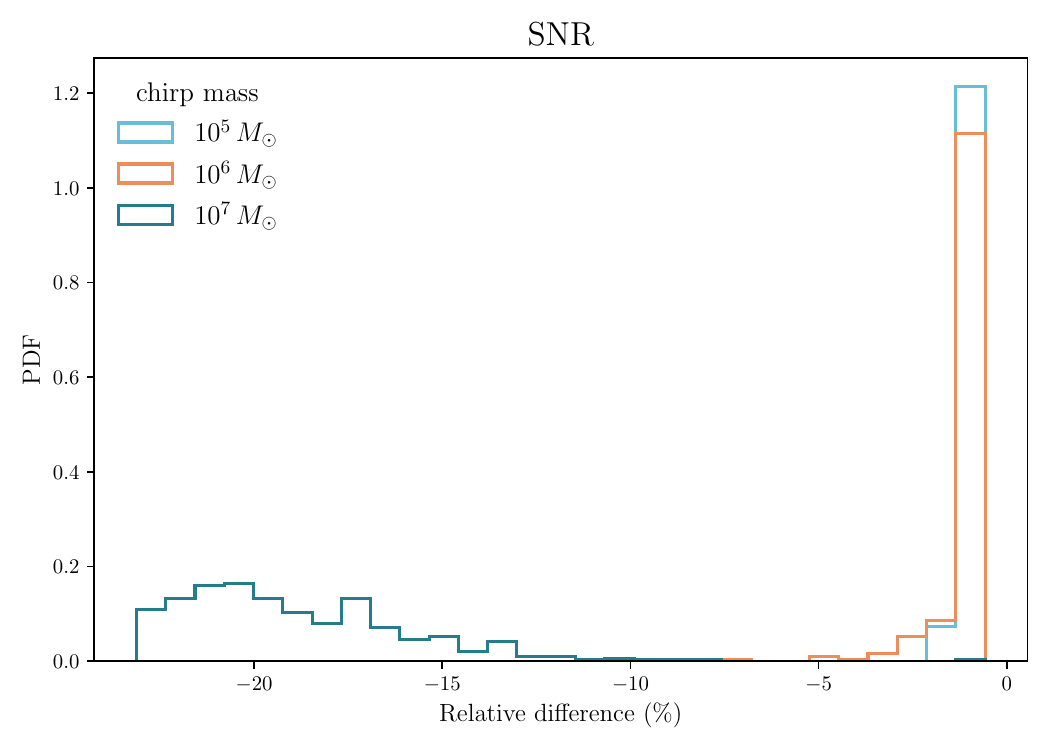}\\[2pt]
\textbf{(c)}
\end{minipage}\hfill
\begin{minipage}{0.49\textwidth}
\centering
\includegraphics[width=\linewidth]{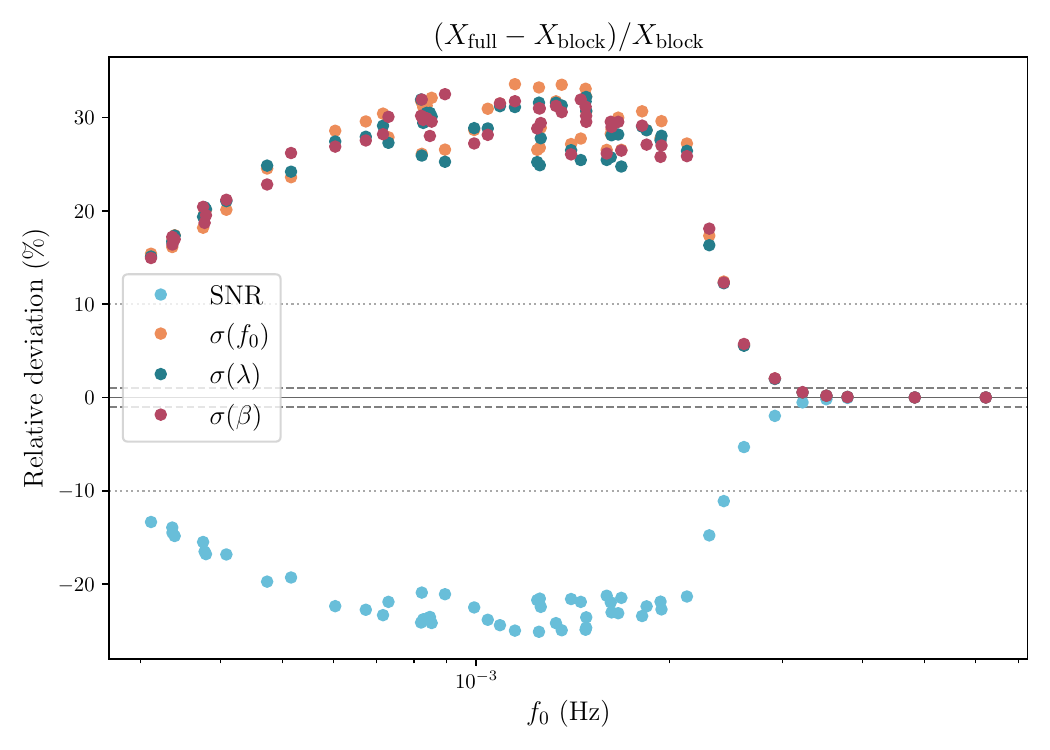}\\[2pt]
\textbf{(d)}
\end{minipage}
\caption{
Representative results for  Config.~II.
\textbf{(a)} Four-year time-averaged magnitudes of the LISA--Taiji cross-detector foreground CSDs after network subtraction with \(\rho_{\rm th,network}=10\), with the orange, blue, green, and red curves showing the \(A\)-\(A\), \(A\)-\(E\), \(E\)-\(A\), and \(E\)-\(E\) channel pairs, respectively, and the grey curve showing the reference \(\sqrt{S_I^A S_J^A}\).
\textbf{(b)} Coherence amplitude \(A_\gamma(t)\) (top) and phase \(\theta_\gamma(t)\) (bottom) for the LISA-\(A\)--Taiji-\(E\) pair at \(f=1\)~mHz over the four-year observation, with the thick and thin curves showing the theoretical estimates  and the  Welch estimates, respectively.
\textbf{(c)} Normalized distributions of \(({\rm SNR}_{\rm full}-{\rm SNR}_{\rm block})/{\rm SNR}_{\rm block}\) for 500 MBHB realizations for each chirp mass, with the light-blue, orange, and green unfilled histograms denoting \(\mathcal{M}_c=10^5\), \(10^6\), and \(10^7\,M_\odot\), respectively.
\textbf{(d)} Signed relative differences \((X_{\rm full}-X_{\rm block})/X_{\rm block}\) for the 55 resolved VGBs, plotted against the initial GW frequency \(f_0\), with the blue, orange, green, and red markers denoting \(X={\rm SNR}\), \(\sigma_{f_0}\), \(\sigma_\lambda\), and \(\sigma_\beta\), respectively, and the grey solid, dashed, and dotted lines marking zero, \(\pm1\%\), and \(\pm10\%\).}
\label{fig:configII}
\end{figure*}

Orbit Config.~II given in Table~\ref{tab:orbits} does not represent the default design for the LISA and Taiji orbits, but rather an alternative configuration explored in the literature~\citep{Wang2021}. 
In this configuration, the centers of LISA and Taiji coincide, with both detectors leading Earth by $20^\circ$. 
Their constellation planes are also aligned, with the initial phases of the SCs within each plane  randomly generated. 
We investigate this configuration here for two purposes. 
First, it demonstrates the dependence of the foreground correlation, including its amplitude, phase, and time dependence, on the relative orbital geometry. 
Consequently, the impact on the analysis of GW signals is also configuration dependent. 
Second, by repeating the analysis for this alternative realization, it complements the results for Config.~I and tests the universality of the theoretical framework and analysis developed in the main text.

Representative results are shown in Figure~\ref{fig:configII}. 
Panel (a) presents the CSDs between different data channels for the LISA-Taiji pair (corresponding to the left panel of Figure~\ref{fig:csd_comparison}). 
Because LISA and Taiji share the same constellation plane in this configuration, their antenna patterns are much similar  than in Config.~I. 
Consequently, the overall correlation level across the entire frequency band is also higher. For the $AE$ and $EA$ channel pairs, the CSDs are very close to the PSD reference levels, indicating a coherence amplitude of $A_\gamma \rightarrow 1$. 
Indeed, as shown in Panel (b), which displays $A_\gamma$ and $\theta_\gamma$ as functions of time at 1 mHz (corresponding to the LISA-$A$ vs. Taiji-$E$ pair in Figure~\ref{fig:app_lisa_taiji_amp} and Figure~\ref{fig:app_lisa_taiji_phase}), $A_\gamma$ is markedly enhanced relative to Config.~I, and its phase remains close to zero throughout the mission period. The behavior of $\theta_\gamma$ can be understood from Eq.~\eqref{eq:conf_csd_explicitsum}, since the baseline phase factor is always unity.
For both the amplitude and the phase, the fluctuation amplitudes over time are much smaller, and they remain at a roughly constant level throughout the entire mission period. 

The impacts of foreground CSDs on the analyses of MBHBs and GBs are illustrated in Panel (c) and Panel (d), respectively. 
For MBHBs, Panel (c)  shows the distribution of the relative SNR difference $({\rm SNR}_{\rm full}-{\rm SNR}_{\rm block})/{\rm SNR}_{\rm block}$ for three chirp mass values under the block-diagonal and full-covariance noise models (corresponding to the first panel of Figure~\ref{fig:snr_mass_trend}).   
For GBs, Panel (d) presents the $(X_{\rm full} - X_{\rm block}) / X_{\rm block}$ values for all the 55 VGBs, with $X$ representing network SNR, $\sigma_{f_0}$, $\sigma_\lambda$, and $\sigma_\beta$ (corresponding to Figure~\ref{fig:gb_snr_comparison}). 
Compared with Config.~I, a notable similarity is in  the frequency dependence of these deviations: for MBHBs, massive sources are  more significantly affected by correlations, while for VGBs, the deviations are also predominantly concentrated in low-frequency sources below 2 mHz.
Meanwhile, the differences between the results of the two configurations are  mainly attributed to the time dependence of coherence and relative orbital  geometry. 
The fact that $\gamma$ remains roughly constant in Config.~II gives rise to a  notable  feature that, for the transient signals of MBHBs,   the signs of $({\rm SNR}_{\rm full} - {\rm SNR}_{\rm block}) / {\rm SNR}_{\rm block}$   are all consistent.   
Additionally, following  the same logic as in  Sec.~\ref{sec:gb_analysis},   the localization uncertainties of VGBs no longer deviates from that of $\sigma_{f_0}$  as in  Config.~I.  
Because  $\mathbf{R_{\rm LISA}} \equiv \mathbf{R}_{\rm Taiji}$ in this configuration,  no alteration to $\eta$ is induced by the Doppler term. 
Consequently,  the trends of all parameter uncertainties perfectly mirror that  of the SNR.

%% Start the reference list on a fresh page: \clearpage first flushes the still
%% pending floats of Appendix D (the Config. II figure), so the figure keeps its
%% page and the bibliography begins cleanly at the top of the next one.
\clearpage
\bibliography{paper}
\bibliographystyle{aasjournalv7.1}

\end{document}